%% file: main.tex
\documentclass[pdflatex,sn-aps]{sn-jnl}

\usepackage{setspace}
\usepackage{amsfonts}
\usepackage{amsmath, bm}
\usepackage{upgreek}
\usepackage{graphicx}
\renewcommand{\thesubsection}{\Alph{subsection}}
\makeatletter
\renewcommand{\p@subsection}{}
\makeatother

\usepackage[dvipsnames]{xcolor}
\usepackage{soul}

\input{macros}

\usepackage{graphicx}%
\usepackage{multirow}%
\usepackage{amsmath,amssymb,amsfonts}%
\usepackage{amsthm}%
\usepackage{mathrsfs}%
\usepackage[title]{appendix}%

\renewcommand{\thesubsection}{\Alph{subsection}.}
\makeatletter
\renewcommand{\p@subsection}{}
\makeatother

\usepackage{xcolor}%
\usepackage{textcomp}%
\usepackage{manyfoot}%
\usepackage{booktabs}%
\usepackage{algorithm}%
\usepackage{algorithmicx}%
\usepackage{algpseudocode}%
\usepackage{listings}%
\theoremstyle{thmstyleone}%
\theoremstyle{thmstyletwo}%

\theoremstyle{thmstylethree}%

\begin{document}

\title{\bf\large Disaggregated Deep Learning via In-Physics Computing at Radio Frequency}


\author{\normalsize Zhihui Gao\textsuperscript{1}, Sri Krishna Vadlamani\textsuperscript{2}, Kfir Sulimany\textsuperscript{2}, Dirk Englund\textsuperscript{2}, and Tingjun Chen\textsuperscript{1}}

\affil[1]{\small\it Department of Electrical and Computer Engineering, Duke University, Durham, NC 27708, USA}

\affil[2]{\small\it Research Laboratory of Electronics, Massachusetts Institute of Technology, Cambridge, MA 02139, USA}

% \author[1]{\fnm{Zhihui} \sur{Gao}}\email{zhihui.gao@duke.edu}

% \author[2]{\fnm{Sri} \sur{Krishna Vadlamani}}\email{srikv@mit.edu}

% \author[2]{\fnm{Kfir} \sur{Sulimany}}\email{kfir@mit.edu}

% \author[2]{\fnm{Dirk} \sur{Englund}}\email{englund@mit.edu}

% \author[1]{\fnm{Tingjun} \sur{Chen}}\email{tingjun.chen@duke.edu}

% \affil*[1]{\orgdiv{Department of Electrical and Computer Engineering}, \orgname{Duke University}, \orgaddress{\city{Durham}, \postcode{27708}, \state{NC}, \country{USA}}}

% \affil[2]{\orgdiv{Research Laboratory of Electronics}, \orgname{Massachusetts Institute of Technolog}, \orgaddress{\city{Cambridge}, \postcode{02139}, \state{MA}, \country{USA}}}


\abstract{\input{tex/abstract}}

\maketitle

\input{tex/article}

\input{tex/methods}

\input{tex/acks}

%% BEGIN OF SUPPLEMENTARY MATERIALS
\newpage

\newpage
{

\newpage
{
\centering{\bf\large Supplementary Information: \\
Disaggregated Deep Learning via In-Physics Computing at Radio Frequency}
\vspace{2.0ex}

\centering{
Zhihui Gao\textsuperscript{1}, Sri Krishna Vadlamani\textsuperscript{2}, Kfir Sulimany\textsuperscript{2}, Dirk Englund\textsuperscript{2}, and Tingjun Chen\textsuperscript{1}}
\vspace{1.0ex}

\centering{
\textsuperscript{1}\textit{Department of Electrical and Computer Engineering, Duke University, Durham, NC 27708, USA}}

\centering{
\textsuperscript{2}\textit{Research Laboratory of Electronics, Massachusetts Institute of Technology, Cambridge, MA 02139, USA}}

}

\setcounter{section}{0}
\setcounter{page}{1}
\vspace{2.0ex}
\tableofcontents

\newpage

\setcounter{figure}{0}
\renewcommand{\thefigure}{S\arabic{figure}}
\setcounter{table}{0}
\renewcommand{\thetable}{S\arabic{figure}}
\setcounter{equation}{0}
\renewcommand{\theequation}{S\arabic{equation}}

\onehalfspacing

\input{tex/Supplementary_theory}

\newpage
\input{tex/Supplementary_experiment}

%% END OF SUPPLEMENTARY MATERIALS

\bibliography{reference}

\end{document}

%% file: macros.tex
\definecolor{lightgray}{gray}{0.9}
\definecolor{lightblue}{rgb}{0.9,0.9,1}
\definecolor{LightMagenta}{rgb}{1,0.5,1}
\definecolor{red}{rgb}{1,0,0}
\definecolor{brightgreen}{rgb}{0.4, 1.0, 0.0}

\newcommand{\remove}[1]{}

\newcommand{\autorefmethod}{\hyperref[sec:methods]{Methods}~section}

\newcommand{\autorefresults}{\hyperref[sec: main-result]{Results}~section}

\newcommand{\autoreffig}[1]{Fig.~\hyperref[#1]{\ref*{#1}}}
\newcommand{\autorefsubfig}[2]{Fig.~\hyperref[#1]{\ref*{#1}#2}}
\newcommand{\autoreftab}[1]{Table~\hyperref[#1]{\ref*{#1}}}

\newcommand{\iu}{{j}}
\newcommand{\eu}{{e}}
\newcommand{\RE}[1]{\textsf{Re}\left\{#1\right\}}
\newcommand{\IM}[1]{\textsf{Im}\left\{#1\right\}}
\newcommand{\complexity}{\mathcal{O}}
\newcommand{\sinc}{\text{sinc}}
\newcommand{\abs}[1]{\left|{#1}\right|}

\newcommand{\PAPR}[1]{\textrm{PAPR}[#1]}

\newcommand{\matInv}[1]{{#1}^{-1}}
\newcommand{\trans}[1]{{#1}^{\top}}
\newcommand{\conj}[1]{\overline{#1}}
\newcommand{\hermitian}[1]{{#1}^{*}}

\newcommand{\DAC}[1]{\textsf{DAC}\left\{#1\right\}}
\newcommand{\ADC}[1]{\textsf{ADC}\left\{#1\right\}}
\newcommand{\LPF}[1]{\textsf{LPF}\left\{#1\right\}}

\newcommand{\powerTx}{P_{x}}
\newcommand{\powerRx}{P_{y}}
\newcommand{\powerNoise}{P_{n}}

\newcommand{\energyMVM}{E_{\textrm{mvm}}}
\newcommand{\energyMVMTx}{E_{1}}
\newcommand{\energyMVMADC}{E_{2}}
\newcommand{\energyMVMDec}{E_{3}}
\newcommand{\energyMAC}{e_{\textrm{mvm}}}
\newcommand{\energyMACTx}{e_{1}}
\newcommand{\energyMACADC}{e_{2}}
\newcommand{\energyMACDec}{e_{3}}
\newcommand{\energyMACTDL}{e_{\textrm{tdl}}}
\newcommand{\energyMVMIPMult}{E'_{\textrm{mvm}}}
\newcommand{\energyMACIPMult}{e'_{\textrm{mvm}}}
\newcommand{\energyMVMIPSing}{E_{\textrm{ip}}}
\newcommand{\energyMACIPSing}{e_{\textrm{ip}}}
\newcommand{\throughput}{\Lambda}

\newcommand{\zadoffVec}{\boldsymbol{\Phi}_{\textrm{zc}}}
\newcommand{\zadoff}{\phi_{\textrm{zc}}}

\newcommand{\userNum}{U}

\newcommand{\powerMax}{P_{\textrm{max}}}

\newcommand{\sampRate}{f_{s}}
\newcommand{\sampDuration}{T_{s}}
\newcommand{\band}{B}
\newcommand{\specVec}{\mathbf{S}}
\newcommand{\spec}{S}
\newcommand{\specIdx}{k}
\newcommand{\freq}{f}
\newcommand{\freqSub}{\Delta{f}}
\newcommand{\sampVec}{\mathbf{s}}
\newcommand{\samp}{s}
\newcommand{\sampIdx}{n}
\newcommand{\sampVecCP}{\mathbf{s}^{\prime}}
\newcommand{\sampCP}{s^{\prime}}

\newcommand{\fftSize}{L}
\newcommand{\fftSizeHalf}{L/2}
\newcommand{\wave}{s}
\newcommand{\waveIdx}{t}
\newcommand{\waveLen}{T}

\newcommand{\radio}{r}

\newcommand{\inputVecNew}{\mathbf{x}^{\prime}}
\newcommand{\inputElemNew}{x^{\prime}}
\newcommand{\sampInputVecNew}{\mathbf{s}_{x}^{\prime}}
\newcommand{\sampInputNew}{s_{x}^{\prime}}
\newcommand{\weightMatNew}{\mathbf{W}^{\prime}}
\newcommand{\weightElemNew}{W^{\prime}}

\newcommand{\sampWeightNew}{s_{w}^{\prime}}
\newcommand{\outputVecNew}{\mathbf{y}^{\prime}}
\newcommand{\outputElemNew}{y^{\prime}}

\newcommand{\outputSizePad}{\Delta M}
\newcommand{\weightMatPad}{\mathbf{W}^{\prime\prime}}
\newcommand{\weightElemPad}{W^{\prime\prime}}
\newcommand{\outputVecPad}{\mathbf{y}^{\prime\prime}}
\newcommand{\outputElemPad}{y^{\prime\prime}}

\newcommand{\overheadPad}{\alpha}
\newcommand{\fftSizeCP}{\Delta L}

\newcommand{\overheadCP}{\beta}
\newcommand{\sampInputVecCP}{\mathbf{s}_{x}^{\prime}}
\newcommand{\sampInputCP}{s_{x}^{\prime}}
\newcommand{\sampWeightVecCP}{\mathbf{s}_{w}^{\prime}}
\newcommand{\sampWeightCP}{s_{w}^{\prime}}

\newcommand{\channelFunc}{h}
\newcommand{\channelMat}{\mathbf{H}}
\newcommand{\channelMatElem}{H}
\newcommand{\channelVec}{\mathbf{h}}
\newcommand{\channelVecElem}{h}

\newcommand{\specChannelVec}{\mathbf{S}_{h}}
\newcommand{\specChannel}{S_{h}}
\newcommand{\outputVecEst}{\hat{\mathbf{y}}}

\newcommand{\precodeMat}{\mathbf{V}}
\newcommand{\precodeMatElem}{V}
\newcommand{\precodeVec}{\mathbf{v}}
\newcommand{\precodeVecElem}{v}

\newcommand{\sampPrecode}{s_{v}}
\newcommand{\wavePrecode}{v}

\newcommand{\innerproduct}[2]{{\langle{#1},{#2}\rangle}}

\newcommand{\convolution}{\ast}
\newcommand{\convIdx}{\kappa}
\newcommand{\dft}{\textsf{DFT}}
\newcommand{\dftMat}{\mathbf{D}}
\newcommand{\dftElem}{d}
\newcommand{\idft}{\textsf{IDFT}}
\newcommand{\shiftMat}{\mathbf{R}}
\newcommand{\idftMat}{\mathbf{{D}}^{-1}}

\newcommand{\fourier}{\mathcal{F}}

\newcommand{\carrier}{F}

\newcommand{\inputSize}{N}
\newcommand{\inputIdx}{n}
\newcommand{\outputSize}{M}
\newcommand{\outputIdx}{m}
\newcommand{\inputVec}{\mathbf{x}}
\newcommand{\inputElem}{x}
\newcommand{\weightMat}{\mathbf{W}}
\newcommand{\weightElem}{W}
\newcommand{\outputVec}{\mathbf{y}}
\newcommand{\outputElem}{y}

\newcommand{\specInputVec}{\mathbf{S}_{x}}
\newcommand{\specInput}{S_{x}}
\newcommand{\sampInputVec}{\mathbf{s}_{x}}
\newcommand{\sampInput}{s_{x}}
\newcommand{\waveInput}{x}

\newcommand{\carrierInput}{F_{x}}

\newcommand{\specWeightVec}{\mathbf{S}_{w}}
\newcommand{\specWeight}{S_{w}}
\newcommand{\sampWeightVec}{\mathbf{s}_{w}}
\newcommand{\sampWeight}{s_{w}}
\newcommand{\waveWeight}{w}

\newcommand{\carrierWeight}{F_{w}}

\newcommand{\specOutputVec}{\mathbf{S}_{y}}
\newcommand{\specOutput}{S_{y}}
\newcommand{\sampOutputVec}{\mathbf{s}_{y}}
\newcommand{\sampOutput}{s_{y}}
\newcommand{\waveOutput}{y}

\newcommand{\carrierOutput}{F_{y}}

\newcommand{\waveInputPre}{x_{\textrm{pre}}}
\newcommand{\sampInputPre}{s_{x, \textrm{pre}}}
\newcommand{\sampInputPreVec}{\mathbf{s}_{x, \textrm{pre}}}
\newcommand{\waveWeightPre}{w_{\textrm{pre}}}
\newcommand{\sampWeightPre}{s_{w, \textrm{pre}}}
\newcommand{\sampWeightPreVec}{\mathbf{s}_{w, \textrm{pre}}}
\newcommand{\waveOutputPre}{y_{\textrm{pre}}}
\newcommand{\sampOutputPre}{s_{y, \textrm{pre}}}
\newcommand{\sampOutputPreVec}{\mathbf{s}_{y, \textrm{pre}}}
\newcommand{\sampOutputPreConj}{\bar{s}_{y, \textrm{pre}}}
\newcommand{\preamLen}{L_{\textrm{pre}}}

\newcommand{\sampRateLow}{f_{s\downarrow}}
\newcommand{\bandLow}{B_{\downarrow}}

\newcommand{\sampOutputVecLow}{\mathbf{s}_{y\downarrow}}
\newcommand{\sampOutputLow}{s_{y\downarrow}}
\newcommand{\specOutputVecLow}{\mathbf{S}_{y\downarrow}}
\newcommand{\specOutputLow}{S_{y\downarrow}}
\newcommand{\outputSizeNew}{M'}

\newcommand{\image}{\mathbf{I}}

\newcommand{\layerIdx}{l}

\newcommand{\boltzmann}{k_{B}}
\newcommand{\temperature}{T_{0}}
\newcommand{\efficiencyTX}{\eta_{\textrm{radio}}}
\newcommand{\efficiencyRX}{\eta_{\textrm{nf}}}
\newcommand{\efficiencyMixer}{\eta_{\textrm{mixer}}}
\newcommand{\efficiencyTRX}{\eta}

\newcommand{\activation}{\sigma}
\newcommand{\outputOdd}{c_f}

\newcommand{\radioInput}{r_{x}}
\newcommand{\radioOutput}{r_{y}}
\newcommand{\radioWeight}{r_{w}}

\newcommand{\name}{{\sc WISE}}
\newcommand{\namebf}{{\sc\textbf{WISE}}}
\newcommand{\smarttrx}{{\sc WISE-R}}
\newcommand{\smarttrxbf}{{\sc\textbf{WISE-R}}}

%% file: tex/abstract.tex
Modern edge devices, such as cameras, drones, and Internet-of-Things nodes, rely on deep learning to enable a wide range of intelligent applications, including object recognition, environment perception, and autonomous navigation. However, deploying deep learning models directly on the often resource-constrained edge devices demands significant memory footprints and computational power for real-time inference using traditional digital computing architectures.
In this paper, we present {\name}, a novel computing architecture for wireless edge networks designed to overcome energy constraints in deep learning inference. {\name} achieves this goal through two key innovations: disaggregated model access via wireless broadcasting and in-physics computation of general complex-valued matrix-vector multiplications directly at radio frequency. Using a software-defined radio platform with wirelessly broadcast model weights over the air, we demonstrate that {\name} achieves {95.7\%} image classification accuracy with ultra-low operation power of {6.0}\thinspace{fJ/MAC} per client, corresponding to a computation efficiency of {165.8}\thinspace{TOPS/W}. This approach enables energy-efficient deep learning inference on wirelessly connected edge devices, achieving more than two orders of magnitude improvement in efficiency compared to traditional digital computing.
% \zhihui{200 words.}

%% file: tex/article.tex
\section*{Introduction}
\label{sec: main-introduction}

%% figure begins
\begin{figure*}[!t]
    \centering
    \includegraphics[width=1.0\textwidth]{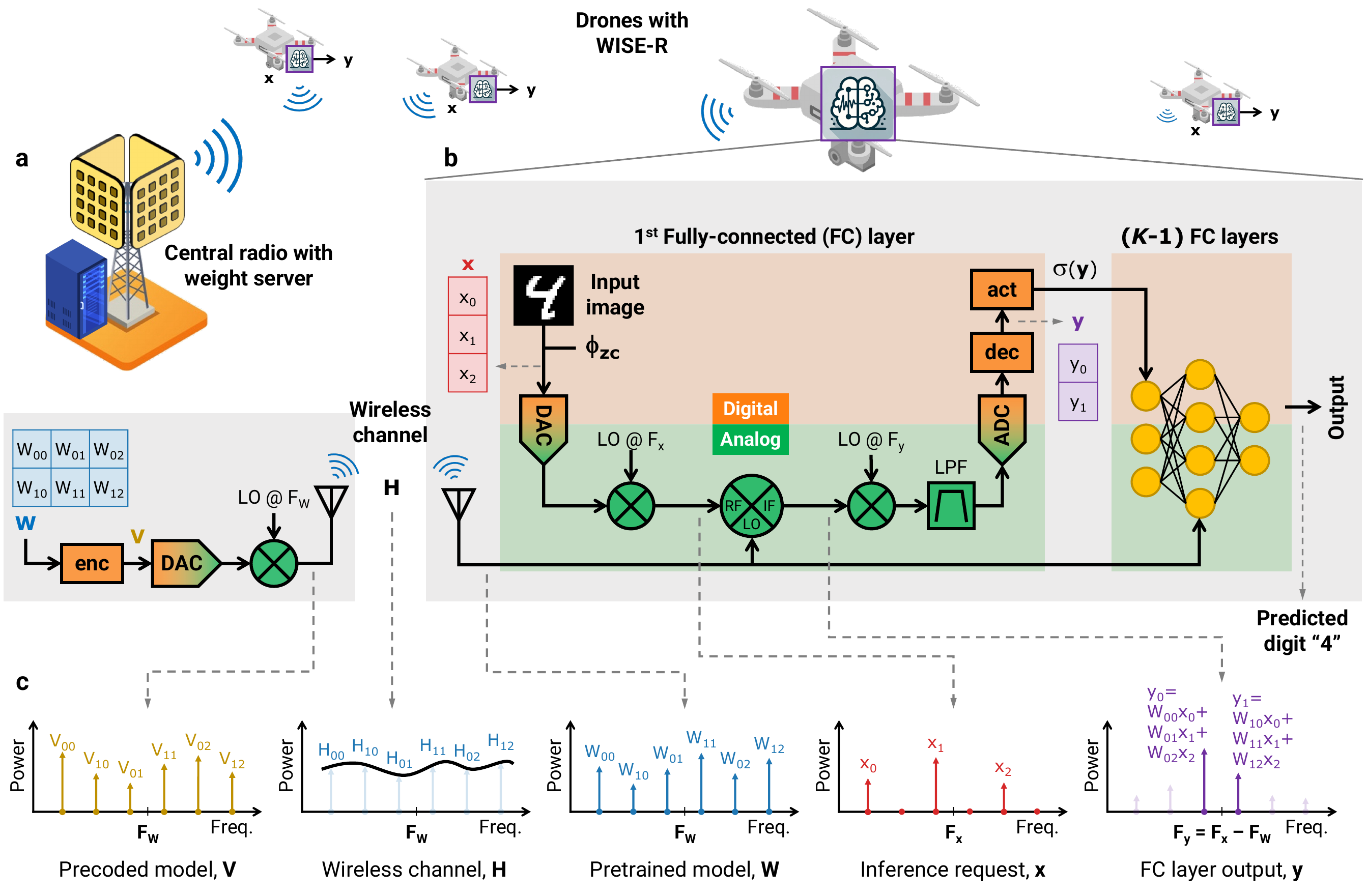}
    \caption{\textbf{The {\namebf} architecture enables disaggregated model access and energy-efficient deep learning (DL) to multiple clients in wireless edge networks.}
    \textbf{a}, A central radio broadcasts frequency-encoded model weights, $\weightMat$, onto a radio-frequency (RF) signal at the carrier frequency $\carrierWeight$, which is precoded to $\precodeMat$ to mitigate the distortion introduced during propagation over the wireless channel, $\channelMat$.
    \textbf{b}, Each client equipped with a {\smarttrx} encodes the inference request $\inputVec$ at the carrier frequency $\carrierInput$, and performs local DL inference for $\outputVec$ at the carrier frequency $\carrierOutput$, where the matrix-vector multiplications (MVM), or essentially the fully connected (FC) layers, are realized using a passive frequency mixer.
    \textbf{c}, Illustration of the in-physic MVM computation during frequency down-conversion with frequency-encoded $\weightMat$, $\inputVec$, and $\outputVec$.
    }
    \label{fig:figure-concept}
\end{figure*}
%% figure ends

Deep learning (DL) has revolutionized modern computing, enabling breakthroughs across a wide range of applications, including the Internet-of-Things (IoT), computer vision, and large language models (LLMs)~\cite{lecun2015deep, guo2016deep, vinyals2019grandmaster, rokhsaritalemi2020review, brown2020language}. 
As models now scale to billions of parameters~\cite{brown2020language, touvron2023llama}, the primary energy efficiency bottleneck is no longer just the raw computation efficiency, but also the energy cost of data movement between the local memory and processing units~\cite{horowitz20141}.
Moreover, retrieving DL model weights on demand from the cloud requires significant wireless bandwidth, while offloading DL inference to the cloud introduces potential privacy concerns~\cite{sulimany2024quantum}.
At the same time, the theoretical lower bound for irreversible computation is set by Landauer's principle at ${2.9}\thinspace\textrm{zeptojoules (zJ)}$ per bit operation~\cite{landauer1961irreversibility, miller2017attojoule, hamerly2019large} at room temperature. In comparison, modern digital computing application-specific integrated circuits (ASICs) operate at energy efficiency in the picojoule range~\cite{horowitz20141}.
Bridging this gap calls for fundamentally different computing paradigms, including in-physics computing architectures that perform computing using continuous quantities (e.g., waves) with minimum data movement.

Recent works have explored a variety of in-physics computing approaches to overcome the memory wall, leveraging integrated photonic and optical waveguides~\cite{shen2017deep, zuo2019all, xu202111, zhang2021optical, feldmann2021parallel, wright2022deep, wang2023image, chen2023all, ma2025quantum, hamerly2019large}, memristor-based crossbars with analog weight storage~\cite{agarwal2016energy, marinella2018multiscale, ankit2019puma, sebastian2020memory, ambrogio2023analog, jung2022crossbar}, and reconfigurable metasurfaces~\cite{liu2022programmable, sanchez2022airnn, reus2023airfc, tong2024sensor, cotrufo2024reconfigurable}.
While these approaches have demonstrated promising energy efficiency gains~\cite{ambrogio2023analog, chen2023all, hamerly2019large}, they often require specialized photonic or electronic hardware, limiting their scalability and practicality for large-scale deployments.
In contrast, radio-frequency (RF) systems~\cite{ross2023multilayer} present a compelling alternative by enabling wireless broadcast of model weights to edge devices, especially given that modern edge devices rely on wireless connectivity (e.g., cellular or wireless local area networks) for control signaling, data transfer, and Internet access.

In this work, we present {\name} (\underline{WI}reless \underline{S}mart \underline{E}dge networks), the first edge computing architecture designed for disaggregated and energy-efficient DL via in-physics computing directly at RF (\autoreffig{fig:figure-concept}).
In {\name}, a central radio broadcasts RF signals that encode model weights ($\weightMat$) and leverages the shared wireless channel to provide simultaneous, disaggregated model access to multiple edge clients (\autorefsubfig{fig:figure-concept}{a}).
Each edge client, equipped with a {\name} radio ({\smarttrx}), performs inference on local data ($\inputVec$) upon receiving the broadcast RF signals and obtains the matrix-vector multiplication (MVM) result, $\outputVec = \weightMat \cdot \inputVec$, as part of the DL inference (\autorefsubfig{fig:figure-concept}{b}).
Both model weights and inference requests are frequency-encoded and I/Q modulated to an RF carrier, and in-physics MVM computation is realized using a passive frequency mixer (\autorefsubfig{fig:figure-concept}{c}).
For example, a drone can execute object detection and image classification tasks on its captured images without the need to store the DL model locally. Our analysis shows that this computing paradigm, in the ideal case, achieves an energy efficiency approaching the thermodynamic limit (TDL) of analog hardware as the problem size scales, even exceeding the Landauer bound~\cite{landauer1961irreversibility} for irreversible digital computation.

To enable ultra-low-power inference, each client requires minimum active hardware--primarily for analog-to-digital conversion (ADC) and lightweight digital signal processing--while offloading the most computationally intensive MVMs to the analog domain. This is achieved by exploiting RF electronics, such as mixers, that inherently perform signal multiplication and are widely used in modern edge devices. In addition, the encoding of both model weights and inference requests is optimized for spectral efficiency, drawing inspiration from modern wireless communication systems employing orthogonal frequency-division multiplexing (OFDM) and I/Q (de)modulation. A channel estimation and calibration process is integrated to mitigate signal distortions during wireless transmission. 

We evaluate the energy efficiency of {\name} for general inner-product (IP) computation and DL model inference tasks. Experimental results on a software-defined radio (SDR) platform with over-the-air transmissions demonstrate that {\name} achieves an energy efficiency of {6.0}\thinspace{fJ/MAC}, measured as energy per multiply-and-accumulate (MAC) operation, for {95.7\%} classification accuracy on the MNIST dataset~\cite{lecun1998gradient}. This corresponds to a computation efficiency of {165.8}\thinspace{TOPS/W} (Tera MAC operations per second per Watt). The energy efficiency can be further improved to {4.6}\thinspace{fJ/MAC} ({216.4\thinspace{TOPS/W}}) with a slightly reduced classification accuracy of {90\%}. These represent a two to three orders of magnitude improvement compared to state-of-the-art digital computing ASICs operating at {1}\thinspace{pJ/MAC}~\cite{horowitz20141}.
Detailed analysis and comprehensive experiments show that {\name} has the potential to transform the landscape of wireless edge networks with embedded intelligence and to offer enhanced energy efficiency in a myriad of real-world applications.

%%%%%
%%%%%
\section*{Results}
\label{sec: main-result}

%% figure begins
\begin{figure*}[!t]
    \centering
    \includegraphics[width=0.98\textwidth]{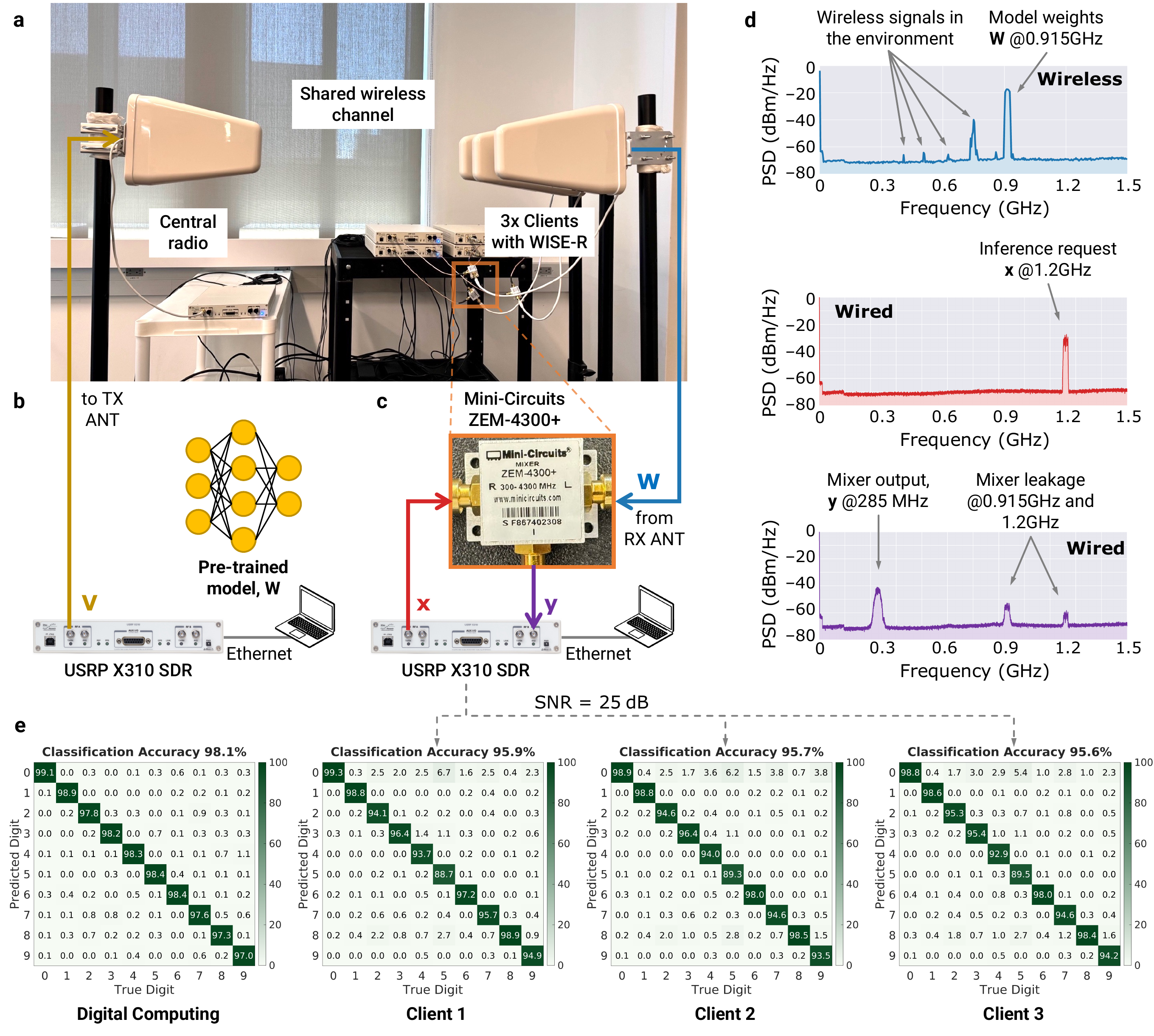}
    \caption{
    \textbf{{\namebf}'s workflow with one central radio and multiple clients.}
    \textbf{a}, Experimental setup for {\name} using a software-defined radio (SDR) platform: 
    \textbf{b}, A central radio simultaneously provides disaggregated deep learning (DL) model access to three edge clients, each equipped with a {\smarttrx}.
    \textbf{c}, On each client, the computing mixer performs general matrix-vector multiplications (MVMs) in-physics using the wirelessly received model weights ($\weightMat$) and local inference request ($\inputVec$).
    \textbf{d}, The model weights $\weightMat$ is modulated at $\carrierWeight={0.915}\thinspace\textrm{GHz}$ over a wireless channel, and the inference request $\inputVec$ is modulated at $\carrierInput={1.2}\thinspace\textrm{GHz}$; after down-conversion, the MVM result $\outputVec$ is located at {0.285}\thinspace{GHz}.
    \textbf{e}, {\name} achieves classification accuracies of {97.1\%--97.4\%} across the three clients on the MNIST dataset using the LeNet-300-100 model, which is comparable to the accuracy of {98.1\%} achieved by traditional digital computing but with significantly improved energy efficiency.
    }
    \label{fig:figure-experiment}
\end{figure*}
%% figure ends

%%%%%
%%%%%
\subsection*{Central Radio and {\smarttrxbf}}

In {\name}, a central radio wirelessly broadcasts model weights to a set of clients for local inference. The complex-valued model parameters and inference requests are encoded in the frequency domain of two RF waveforms via I/Q modulation. These signals are subsequently passed into an RF ``computing'' mixer, which naturally performs the time-domain multiplication (or frequency-domain convolution) of the two input waveforms during frequency mixing. The resulting output signal carries the desired analog computing results. Essentially, {\name} effectively realizes the computation of complex-valued fully-connected (FC) layers, represented by $\outputVec = \weightMat \cdot \inputVec$, with $\inputVec \in \mathbb{C}^{\inputSize}$, $\outputVec \in \mathbb{C}^{\outputSize}$, and $\weightMat \in \mathbb{C}^{\outputSize \times \inputSize}$, directly in the analog domain.

\autorefsubfig{fig:figure-experiment}{a} shows the experimental setup of {\name}'s implementation with three edge clients using an SDR platform (see details in Supplementary Section~\ref {sec:supplementary-experiment-setup}).
Specifically in \autorefsubfig{fig:figure-experiment}{b}, the central radio encodes the DL model weights in the $\layerIdx$-th layer $\weightMat^{(\layerIdx)}$ onto a complex-valued waveform $\waveWeight^{(\layerIdx)}(\waveIdx)$ with bandwidth $\band$, which is then I/Q modulated to a time-domain waveform $\radioWeight^{(\layerIdx)}(\waveIdx) = \RE{\waveWeight^{(\layerIdx)}(\waveIdx) \cdot e^{\iu 2\pi \carrierWeight t}}$ at frequency $\carrierWeight$ for broadcasting.
As shown in \autorefsubfig{fig:figure-experiment}{c}, each {\smarttrx} consists of three main components: 
a transmitter (TX), which encodes the input to the $\layerIdx$-th layer $\inputVec^{(\layerIdx)}$ onto a complex-valued waveform $\waveInput^{(\layerIdx)}(\waveIdx)$, which is then I/Q modulated to $\radioInput^{(\layerIdx)}(\waveIdx) = \RE{\waveInput^{(\layerIdx)}(\waveIdx) \cdot e^{\iu 2\pi \carrierInput t}}$ at $\carrierInput$;
a passive frequency mixer as the analog MVM (or IP) engine for computing $\radioOutput^{(\layerIdx)}(\waveIdx) = \radioWeight^{(\layerIdx)}(\waveIdx) \cdot \radioInput^{(\layerIdx)}(\waveIdx)$; and a receiver (RX), which I/Q demodulates, filters, and samples the mixed signal $\radioOutput^{(\layerIdx)}(\waveIdx)$ at $\carrierOutput$ using the minimal required sampling rate, and decodes the output of the $\layerIdx$-th layer, $\outputVec^{(\layerIdx)} = \weightMat^{(\layerIdx)} \cdot \inputVec^{(\layerIdx)}$.
Note that when the computing mixer is used for frequency down-conversion, the carrier frequencies satisfy $\carrierOutput = \carrierInput - \carrierWeight$, and a spectrum example of $\radioInput^{(\layerIdx)}(\waveIdx)$, $\radioWeight^{(\layerIdx)}(\waveIdx)$ and $\radioOutput^{(\layerIdx)}(\waveIdx)$ is shown in \autorefsubfig{fig:figure-experiment}{d}.
An activation function involving the absolute value function and a Zadoff-Chu (ZC) phase sequence~\cite{chu1972polyphase} is then applied to $\outputVec^{(\layerIdx)}$ to generate the input to the next layer, $\inputVec^{(\layerIdx+1)} = \activation(\outputVec^{(\layerIdx)}) = |\outputVec^{(\layerIdx)}| \cdot \bm{\upphi}_{\textrm{zc}}$. The use of $\bm{\upphi}_{\textrm{zc}}$ converts the real absolute values into complex, which ensures that the power of $\waveInput^{(\layerIdx+1)}(\waveIdx)$ is evenly distributed across frequency. See Supplementary Section~\ref{sec:supplementary-experiment-case-study-three-layer} for a detailed workflow example with a three-layer DL model, and Supplementary Section~\ref{sec:supplementary-experiment-ml-regression} for the single-layer linear regression model without this ZC-phased activation function in digital.
{\name} also accounts for signal distortion caused by the wireless channel by incorporating channel state information (CSI), $\channelMat$, into the encoding of $\weightMat$ at the central radio, as detailed in {\autorefmethod} and Supplementary Section~\ref{sec:supplementary-theory-weight-precoding-scheme}.
\autorefsubfig{fig:figure-experiment}{e} illustrates an example of the {\name}'s in-physics computation on the DL-based image classification task (MNIST) on the three clients, respectively.
Note that the CSI precoding can also be applied on $\inputVec$ on each client (see Supplementary Sections~\ref{sec:supplementary-theory-input-precoding-scheme} and \ref{sec:supplementary-experiment-ml-calibration}), or eliminated for a wired channel (see Supplementary Sections~\ref{sec:supplementary-theory-basic-scheme} and \ref{sec:supplementary-experiment-ml-wired}).

%% figure begins
\begin{figure*}[!t]
    \centering
    \includegraphics[width=0.98\textwidth]{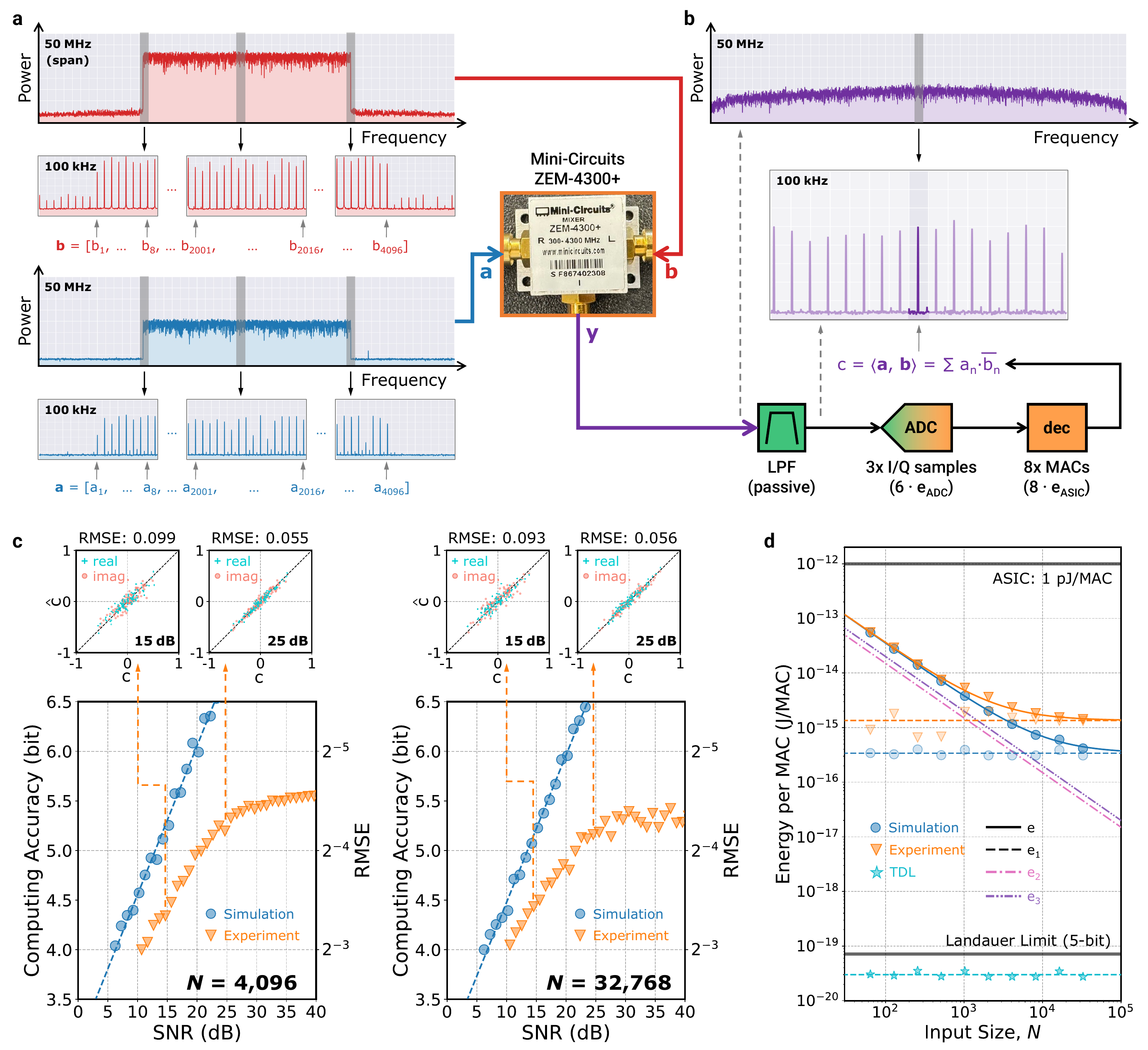}
    \caption{\textbf{Benchmarking general complex-valued inner-product (IP) computation: computing accuracy and energy efficiency.}
    \textbf{a}, Complex-valued IP computation of two length-$\inputSize$ vectors, $c = \innerproduct{\mathbf{a}}{\mathbf{b}} = \sum_{\inputIdx=1}^{\inputSize} a_{\inputIdx} \cdot \overline{b_{\inputIdx}}$, where $\mathbf{a}$ and $\mathbf{b}$ are frequency encoded onto $\inputSize$ (4,096) subcarriers across a bandwidth of $\band$ ({25}\thinspace{MHz}).
    \textbf{b}, Decoding of the IP result, $c$, after the in-physics IP computation, low-pass filtering, and sampling using an analog-to-digital converter (ADC).
    \textbf{c}, IP computing accuracy achieved by {\name} as a function of the signal-to-noise ratio (SNR) for $\inputSize = 4,096$ and $\inputSize = 32,768$.
    \textbf{d}, Energy efficiency of {\name}, $\energyMAC$ (J/MAC), required to achieve $\textsf{RMSE} < 0.0625$ (equivalent to 5-bit computing accuracy~\cite{sludds2022delocalized, davis2022frequency}) as a function of the IP size, $\inputSize$.
    }
    \label{fig:figure-inner}
\end{figure*}
%% figure ends

\subsection*{General IP Computation and Scalability}

We benchmark {\name}'s analog computing performance for the complex-valued IP of two length-$\inputSize$ vectors, $c = \innerproduct{\mathbf{a}}{\mathbf{b}} = \sum_{\inputIdx=1}^{\inputSize} a_{\inputIdx} \cdot \overline{b_{\inputIdx}}$, where $\overline{b_{\inputIdx}}$ denotes the complex conjugate of $b_{\inputIdx}$. 
Compared to MVM, $\inputVec$ is replaced by $\mathbf{a}$ produced by the client, and $\weightMat$ is replaced by $\mathbf{b}$ broadcast by the central radio.
This IP computation involves $\inputSize$ complex-valued MACs, equivalent to $4\inputSize$ real-valued MACs.
In particular, the amplitude and phase of $a_{\inputIdx}$ and $b_{\inputIdx}$ are drawn from independent uniform distributions $\mathcal{U}[0, 1]$ and $\mathcal{U}[0, 2\pi]$, respectively.
$\inputSize$ subcarriers (excluding padded zero-subcarriers) are placed in the frequency domain when generating $\waveInput(\waveIdx)$ and $\waveWeight(\waveIdx)$, and a single subcarrier is captured after the LPF on the $\waveOutput(\waveIdx)$ (\autorefsubfig{fig:figure-inner}{a--b}).

\autorefsubfig{fig:figure-inner}{c} shows the experimental IP computing accuracy, measured by the root mean squared error (RMSE) of the IP obtained by in-physics computing ($\widehat{c}$) compared to the ground truth ($c$), with a normalization factor of $1/\sqrt{\inputSize}$ and under varying SNR values (see detailed definition in Supplementary Section~\ref{sec:supplementary-experiment-ml-calibration}).
The normalization ensures consistent distributions of the IP results across different problem sizes ($\inputSize$).
{\name} achieves an RMSE of 0.055 at {25}\thinspace{dB} SNR with $\inputSize = 4,096$, equivalent to a computing accuracy of $-\log_{2} (\textsf{RMSE}/2) \approx 5\thinspace\textrm{bit}$~\cite{sludds2022delocalized, davis2022frequency}, sufficient for various ML inference tasks~\cite{choi2018pact, garg2022dynamic}.
Simulation results demonstrate slopes of {6.7}\thinspace{dB/bit} computing accuracy for {4,096}-point and {32,768}-point IP.
A similar trend is observed from the experiments in the low SNR regime ($\textsf{SNR} < {25}\thinspace\textrm{dB}$).
In the high SNR regime ($\textsf{SNR} > {25}\thinspace\textrm{dB}$), the computing accuracy is no longer limited by the thermal noise but by the imperfect channel estimation and computing mixer that inherently operates using on-off switching instead of performing the ideal multiplication.

\autorefsubfig{fig:figure-inner}{d} plots the energy efficiency of a {\smarttrx} to achieve $\textsf{RMSE}<0.0625$ (i.e., 5-bit computing accuracy) for IP across varying values of $\inputSize$. For $\inputSize = 4,096$, {\smarttrx} achieves an energy efficiency of {2.4}\thinspace{fJ/MAC} ({421.9}\thinspace{TOPS/W}) in experiments.
As the IP dimension increases, the energy consumption of I/Q sampling and digital FFT is amortized, leading to energy efficiency asymptotically approaching $\energyMACTx$ for the waveform generation and I/Q (de)modulation. 
Experimental results further validate this trend, demonstrating an energy efficiency of {1.4}\thinspace{fJ/MAC} ({699.3}\thinspace{TOPS/W}) with $\inputSize = 32,768$, nearly three orders of magnitude lower than the state-of-the-art ASICs operating at {1}\thinspace{pJ/MAC}~\cite{horowitz20141, abari201427, jouppi2017datacenter}.
In simulations with ideal hardware, $\energyMACTDL$ required for achieving 5-bit computing accuracy is projected to be {28.7}\thinspace{zJ/MAC} ({34.8}\thinspace{ExaOPS/W}), surpassing the Landauer limit~\cite{landauer1961irreversibility} for MAC computation with 5-bit accuracy.

%%%%%
%%%%%
\subsection*{ML for Image/Audio Classification}

%% figure begins
\begin{figure*}[!t]
    \centering
    \includegraphics[width=0.98\textwidth]{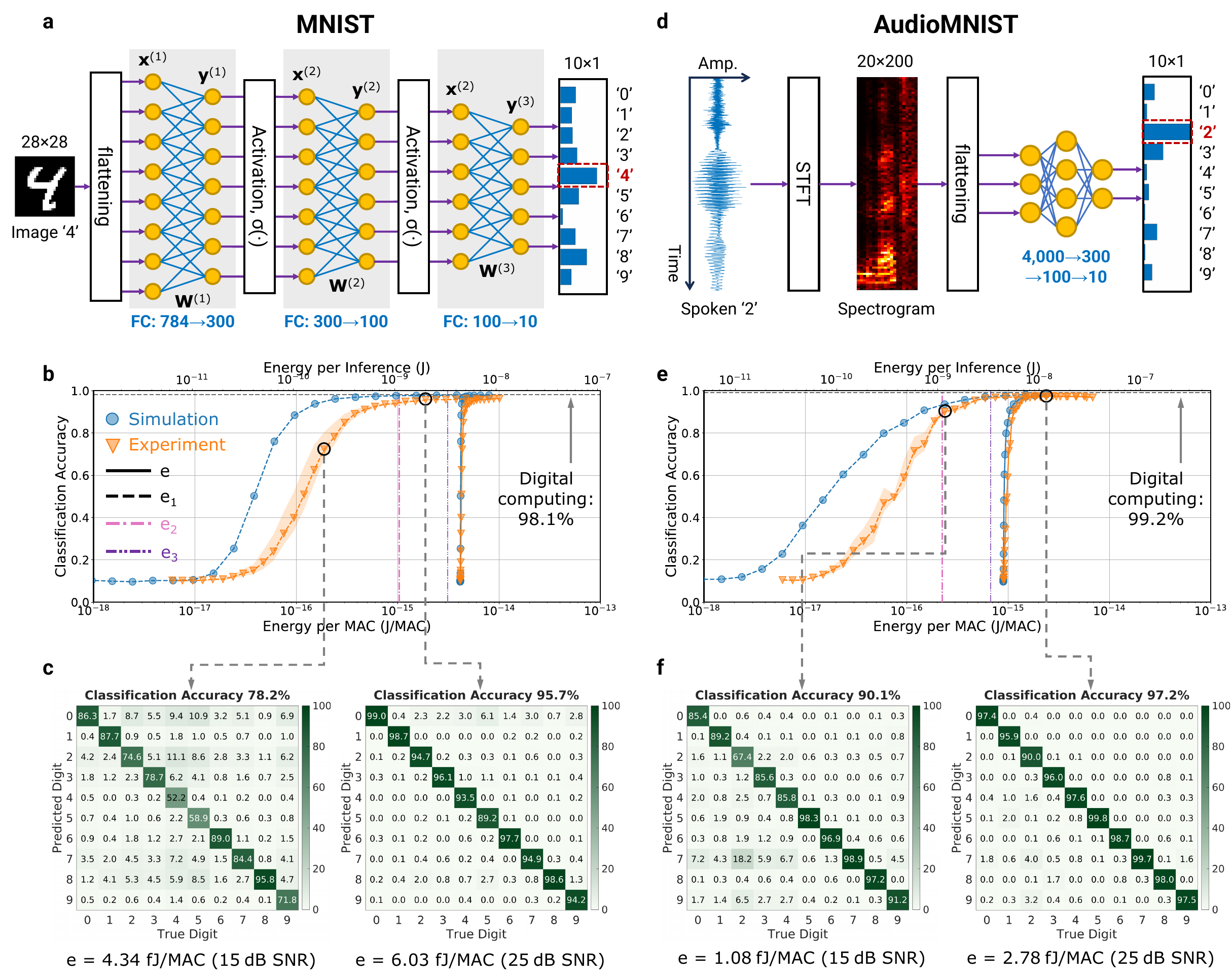}
    \caption{\textbf{{\namebf} for energy-efficiency deep learning (DL) inferences.}
    \textbf{a}/\textbf{d}, Deployment of {\name} for DL tasks using complex-valued three-layer models: classification of handwritten digits on the MNIST dataset (\textbf{a}) and spoken digits on the AudioMNIST dataset (\textbf{d}). 
    \textbf{b}/\textbf{e}, Experimental classification accuracy achieved by {\name} on the MNIST (\textbf{b}) and AudioMNIST (\textbf{e}) datasets over different energy efficiency (J/MAC), and the corresponding energy consumption per inference.
    \textbf{c}/\textbf{f}, Confusion matrices for classification accuracy at {10}\thinspace{dB} and {20}\thinspace{dB} SNR on the MNIST (\textbf{b}) and AudioMNIST (\textbf{e}) datasets.
    }
    \label{fig:figure-dataset}
\end{figure*}
%% figure ends

We deploy {\name} for two DL inference tasks, where the central radio wirelessly broadcasts model weights to three clients equipped with {\smarttrx}: image classification on the MNIST dataset~\cite{lecun1998gradient}, and audio signal classification on the AudioMNIST dataset~\cite{audiomnist2023}.
Both tasks employ a complex-valued model following LeNet-300-100~\cite{lecun1998gradient} with three FC layers.
The complex-valued model exploits the absolute function combined with a pre-defined Zadoff-Chu phase sequence as the activation function after each of the first two FC layers; for the last layer, only the absolute function is applied before the output layer.
The models are trained on an NVIDIA A100 GPU using cross-entropy loss; during the training process, the models with the highest testing accuracy by digital computing are recorded and used.

Each {\smarttrx} performs local inference upon receiving the three-layer model broadcast by the central radio. Each FC layer is formulated as an MVM, which is naturally realized during down-conversion as $\waveInput^{(\layerIdx)}(\waveIdx)$ and $\waveWeight^{(\layerIdx)}(\waveIdx)$ pass through the computing mixer. The mixer output is subsequently low-pass filtered, digitized, and decoded before applying the activation function. 
In the experiment, the MVMs corresponding to individual FC layers are divided into smaller MVMs with $\outputSizeNew = 6$,  $\overheadPad=0.33$, and $\overheadCP=0.25$.
This process is repeated for each layer, and the final classification results are obtained from the output $\outputVec^{(3)}$.

For the MNIST dataset, the images are gray-scaled with dimensions of {28}$\times${28} pixels, which are flattened into $\inputVec^{(1)} \in \mathbb{C}^{784}$ as inputs to the DL model.
The three-layer FC model has an architecture of 784--300--100--10 following LeNet-300-100~\cite{lecun1998gradient} (\autorefsubfig{fig:figure-dataset}{a}), consisting of {0.27} million complex-valued parameters and requiring {1.06} million real-valued MACs.
By digital computing, the classification accuracy of the pre-trained three-layer FC model is {98.1}\% across a testing set of {10,000} images.
\autorefsubfig{fig:figure-dataset}{b} shows the averaged classification accuracy across three clients by {\name}'s in-physics computing at varying different SNR levels.
To achieve {90\%} classification accuracy, the experimental energy efficiency is {4.6}\thinspace{fJ/MAC} ({216.5}\thinspace{TOPS/W}) at {11.8}\thinspace{dB} SNR, with a breakdown of {0.5}\thinspace{fJ/MAC}, {1.0}\thinspace{fJ/MAC}, and {3.1}\thinspace{fJ/MAC} for $\energyMACTx$, $\energyMACADC$, and $\energyMACDec$, respectively.
Simulations validate this trend, demonstrating an energy efficiency of {4.2}\thinspace{fJ/MAC} ({236.1}\thinspace{TOPS/W}) for achieving {90\%} classification accuracy.
\autorefsubfig{fig:figure-dataset}{c} shows the detailed confusion matrices across three clients at {15}\thinspace{dB} and {25}\thinspace{dB} SNR, with experimental classification accuracies of {78.2\%} and {95.7\%}, respectively.

AudioMNIST is a dataset of audio signals containing spoken digits from `0' to `9'. Each audio clip is converted into a spectrogram using the short-time Fourier transform (STFT), which is then concatenated as a vector with Zadoff-Chu phases, $\inputVec^{(1)} \in \mathbb{C}^{4,000}$.
The pre-trained AudioMNIST model consists of {1.23} million complex-valued parameters across three FC layers, involving {4.92} million real-valued MACs (\autorefsubfig{fig:figure-dataset}{d}).
Such a three-layer model achieves a digital computing accuracy of {99.2\%} on the AudioMNIST's testing set with {3,000} audio clips.
As shown in \autorefsubfig{fig:figure-dataset}{e}, an experimental classification accuracy of {90\%} requires the experimental energy consumption of {1.1}\thinspace{fJ/MAC} ({885.0}\thinspace{TOPS/W}), including the energy efficiency breakdown of {0.2}\thinspace{fJ/MAC}, {0.2}\thinspace{fJ/MAC}, and  {0.7}\thinspace{fJ/MAC} for $\energyMACTx$, $\energyMACADC$, and $\energyMACDec$, respectively.
Simulations under this accuracy level reveal an energy efficiency of {1.0}\thinspace{fJ/MAC} ({1.0}\thinspace{PetaOPS/W}).
The discrepancy between experimental and simulation results in both DL tasks is mainly due to imperfect wireless channel estimation and calibration.
\autorefsubfig{fig:figure-dataset}{f} shows the confusion matrices with {15}\thinspace{dB} and {25}\thinspace{dB} SNR. Under {25}\thinspace{dB} SNR, the average experimental accuracy across the three clients is {97.2}\%, with an energy efficiency of {2.8}\thinspace{fJ/MAC} ({359.7}\thinspace{TOPS/W}).

%%%%%
%%%%%
\section*{Discussion}
\label{sec: main-conclusion}

We presented {\name}, a novel computing paradigm that enables disaggregated and energy-efficient DL inference simultaneously on multiple edge clients equipped with {\smarttrx}. Leveraging wireless delivery of DL models broadcast by a central radio, each {\smarttrx} utilizes a (passive) frequency mixer to perform IP or MVM computation directly at RF. 
Through comprehensive theoretical analysis and simulations, we show that {\name} achieves energy efficiency approaching the thermodynamic limit as the problem size $\inputSize \to +\infty$, surpassing the Landauer bound of conventional digital computing.
Extensive experiments demonstrate that {\name} achieves over 5-bit computing accuracy for IPs up to $\inputSize = 32,768$. For DL tasks involving large-scale MVMs, {\name} achieves a classification accuracy of 95.7\% and 97.2\% using the MNIST and AudioMNIST dataset, respectively, at energy efficiencies of {6.0}\thinspace{fJ/MAC} and {2.8}\thinspace{fJ/MAC}, corresponding to computation efficiencies of {165.8}\thinspace{TOPS/W} and {359.7}\thinspace{TOPS/W}.
This represents two to three orders of magnitude of energy efficiency improvement compared to digital computing using state-of-the-art ASICs. 
{\name} can also be adapted to various MVM-based DL tasks, including convolutional neural networks~\cite{lecun1998gradient, lecun2015deep, xu202111, feldmann2021parallel} and transformers~\cite{vaswani2017attention, touvron2023llama}.

Taking one step further, the energy efficiency of {\name} can be further improved through an all-analog architecture, where energy consumption is primarily attributed to analog waveform generation.
Supplementary Section~\ref {sec:supplementary-experiment-ml-regression} demonstrates the effectiveness of a single-layer analog model, and multi-layer analog models can be realized by integrating electronics that inherently perform non-linear activation functions based on their physical properties, such as transistors or diodes~\cite{davis2022frequency, wright2022deep, zuo2019all, chen2023all}. 
The gap between practical energy efficiency and the theoretical limit can be further narrowed using advanced hardware and ASICs~\cite{zhang20190, choi202419, rashed2024scalable}.
Beyond outdoor deployments constrained by limited spectrum, {\name} is also applicable to indoor compute clusters performing DL inference in a shielded environment, where directional antennas mounted on top of server racks~\cite{ghobadi2016projector} can stream model weights to clients with increased bandwidth.
Moreover, a central radio equipped with large-scale antenna arrays~\cite{shepard2012argos} can accelerate DL inference for a single broadcast task or serve multiple models to multiple clients, exploiting the spatial multiplexing gain of the wireless channel.
The physical separation between the central radio (hosting DL models) and edge clients (generating local inference requests) offers an additional privacy benefit by mitigating the risk of information leakage~\cite{sulimany2024quantum}, where the inherent ``noisy'' nature of the wireless channel can be harnessed for model weight precoding.
By integrating pervasive RF signals into the in-physics computing ecosystem, {\name} unlocks large-scale DL deployment on ubiquitous edge devices at orders of magnitude lower power consumption and complexity.

%% file: tex/methods.tex
\section*{Methods}
\label{sec:methods}

%%%%%
%%%%%
\subsection*{Energy Efficiency}

The energy consumption of {\smarttrx} is minimized via the wireless broadcast of disaggregated model weights from a central radio. The energy consumed by a {\smarttrx} to perform an MVM consists of three parts:
$\energyMVMTx$ for the generation of $\waveInput(\waveIdx)$ and I/Q modulation,
$\energyMVMADC$ for the I/Q sampling of $\outputVec$ from $\waveOutput(\waveIdx)$ using two ADCs after I/Q demodulation, and
$\energyMVMDec$ for the digital FFT operation to decode $\outputVec$, i.e.,
\begin{align}
    \energyMVM
    & = \energyMVMTx + \energyMVMADC + \energyMVMDec \nonumber \\
    & = (1+\overheadPad) (1+\overheadCP) \cdot \inputSize\outputSize \cdot \efficiencyTRX^{-1} \cdot \textsf{SNR} \cdot \boltzmann \temperature 
    + (1+\overheadPad) \cdot 2\outputSize \cdot e_{\textrm{adc}} 
    + (1+\overheadPad) \cdot 2\outputSize \log_2 \left((1+\overheadPad) \outputSizeNew\right) \cdot e_{\textrm{dig}}.
\end{align}
Here, $\boltzmann\temperature = -{174}$\thinspace{dBm/Hz} is the thermal noise power spectrum density at room temperature of $\temperature ={300}\thinspace\textrm{K}$. $\efficiencyTRX \in (0,1]$ is the overall loss of the {\smarttrx} hardware including the energy efficiency of the TX, insertion loss of the computing mixer, and noise floor of the RX.
To ensure robust performance, $\overheadPad > 0$ is the overhead coefficient of the zero-subcarriers to overcome the LPF's roll-off effect, and $\overheadCP$ is the overhead coefficient of the cyclic prefix for a better timing synchronization tolerance (see Supplementary Section~\ref{sec:supplementary-experiment-setup}).
$\textsf{SNR}$ refers to the signal-to-noise ratio (SNR) measured at the RX.
Moreover, $e_{\textrm{adc}}$ is the energy consumed per sample by an ADC, and $e_{\textrm{dig}}$ is the energy consumed by an ASIC per real-valued MAC operation in digital computing.

Since each complex-valued MVM involves $4\inputSize\outputSize$ real-valued MACs, the \emph{energy efficiency} of MVM computation, $\energyMAC$, measured by energy per real-valued MAC (J/MAC), is given by
\begin{align}
    \energyMAC
    & = \frac{\energyMVM}{4\inputSize\outputSize} = \energyMACTx + \energyMACADC + \energyMACDec \nonumber \\
    & = \frac{(1+\overheadPad)(1+\overheadCP)}{4} \cdot \efficiencyTRX^{-1} \cdot \textsf{SNR} \cdot \boltzmann\temperature + \frac{1+\overheadPad}{2\inputSize} \cdot e_{\textrm{adc}}
    + \frac{1+\overheadPad}{2\inputSize} \cdot \log_2 \left((1+\overheadPad)\outputSizeNew\right) \cdot e_{\textrm{dig}}.
    \label{eqn:emac-mvm}
\end{align}
It can be seen that $\energyMAC$ significantly improves for large values of $\inputSize$ since $\energyMACADC$ and $\energyMACDec$ scale as $\complexity\left(1/\inputSize\right)$, e.g., $\inputSize = 11,008$ in emerging LLMs such as Llama-2-7b~\cite{touvron2023llama}.
With $\outputSizeNew = 1$, equation~\eqref{eqn:emac-mvm} is reduced to $\energyMACIPSing$ for IP computation (See Supplementary Sections~\ref{sec:supplementary-theory-weight-precoding-scheme} and \ref{sec:supplementary-experiment-ml-inner}).
With ideal hardware ($\efficiencyTRX = 1$, $\overheadPad = \overheadCP = 0$), $\energyMAC$ approaches its thermodynamic limit (TDL) as $\inputSize \to +\infty$,
\begin{align}
    \energyMACTDL :=
    \lim_{\inputSize \to +\infty} \energyMAC
    = \lim_{\inputSize \to +\infty} \energyMACIPSing
    = \textsf{SNR} \cdot \boltzmann \temperature/4.
    \label{eqn:emac-tdl}
\end{align}
We define the corresponding \emph{computation efficiency} for each {\smarttrx} as the reciprocal of energy per MAC, $(\energyMAC)^{-1}$, measured by the number of (real-valued) MAC operations per second per Watt (OPS/W).
See Supplementary Sections~\ref{sec:supplementary-theory-basic-scheme}--\ref{sec:supplementary-theory-input-precoding-scheme} for more details on the energy efficiency and overhead analysis.

\subsection*{Computation Throughput}

The disaggregated setup of {\name} treats the shared wireless medium as a channel for the central radio to deliver DL model parameters for energy-efficient inference at each client, which is different than conventional communication systems for data delivery.
Hereby, in the context of DL inference, we define the computation throughput as the number of (real-valued) MAC operations per second (OPS). We define the \emph{computation throughput} of this channel over $\userNum$ clients as a function of $\band$, $\inputSize$, and $\outputSize$.
Consider the complex-valued MVM, $\outputVec = \weightMat \cdot \inputVec$, involving $\inputSize \outputSize$ complex-valued MACs, or $4\inputSize\outputSize$ real-valued MACs.
Waveforms $\waveInput(\waveIdx)$ and $\waveWeight(\waveIdx)$ corresponding to $\inputVec$ and $\weightMat$ last for a time duration of $\waveLen = (1+\overheadPad)(1+\overheadCP) \cdot \inputSize\outputSize/\band$. This waveform time $\waveLen$ dominates the latency of the disaggregated computation process. Across $\userNum$ clients, a total number of $\userNum \cdot 4\inputSize\outputSize$ MACs can be completed within the waveform time $\waveLen$, corresponding to a computation throughput given by
\begin{align}
    \throughput
    & = \frac{\userNum \cdot 4\inputSize\outputSize}{\waveLen} = \frac{4 \cdot \userNum \band}{(1+\overheadPad)(1+\overheadCP)}~\textrm{[OPS]},
    \label{eqn:computation-throughput-mvm}
\end{align}
which scales as a function of the available bandwidth, $\band$, and number of clients, $\userNum$. More details can be found in Supplementary Section~\ref{sec:supplementary-theory-computation-analysis}.

\subsection*{Wireless Channel Calibration}

In the wireless setting, the channel carrying DL model parameters in $\weightMat$ exhibits propagating delay and multi-path effect, therefore requiring a channel estimation and calibration process to guarantee accurate delivery of the model parameters.
The channel state information (CSI) from the central radio to a client equipped with {\smarttrx} can be represented by a complex matrix $\channelMat = [\channelMatElem_{\outputIdx, \inputIdx}] \in \mathbb{C}^{\outputSize \times \inputSize}$, which has the same dimension as $\weightMat$. Using a pre-defined signal, $\channelMat$ can be estimated by minimum mean squared error (MMSE) and nearest neighbor interpolation, as described in Supplementary Section~\ref{sec:supplementary-theory-weight-precoding-scheme}. The estimated $\channelMat$ is fed back to the central radio, and this process is only performed once as long as the wireless environment does not change significantly.
To account for signal distortion introduced by the wireless channel, we apply a precoder on $\weightMat$ to generate the transmitted signal given by $\precodeMat = [\precodeMatElem_{\outputIdx, \inputIdx}] \in \mathbb{C}^{\outputSize \times \inputSize}$, where $\precodeMatElem_{\outputIdx, \inputIdx} = \frac{\weightElem_{\outputIdx, \inputIdx}}{\channelMatElem_{\outputIdx, \inputIdx}}$. This precoding on the central radio, termed the $\weightMat$-precoding scheme, ensures that the signal received by the client contains the desired frequency-encoded model weights, $\weightMat$, which can then be used for local inference.

For multiple clients located in proximity, the same estimated $\channelMat$ can be applied to the model weight broadcast to all clients, which does not require modification of the client behavior. One alternative scheme that tolerates diverse CSI across clients with better computing accuracy is to precode $\waveInput(\waveIdx)$ on each client using $\channelMat$ estimated for individual clients. While this client-side precoding scheme incurs extra computing overhead on the client side, it achieves improved computing accuracy. More details about the wireless channel calibration and different precoding schemes can be found in Supplementary Sections~\ref{sec:supplementary-theory-input-precoding-scheme} and \ref{sec:supplementary-experiment-ml-calibration}.

\subsection*{MNIST and AudioMNIST Dataset Preparation}

Each data sample in MNIST~\cite{lecun1998gradient} is a gray-scaled image $\image \in [0, 255]^{28\times28}$ representing a handwritten digit from `0' to `9'.
Each image is first reshaped to a {784}-point vector and then element-wisely modulated by a {784}-point ZC phase sequence $\zadoffVec = [\zadoff[\outputIdx]] \in \mathbb{C}^{784}$ to generate $\inputVec \in \mathbb{C}^{784}$~\cite{chu1972polyphase}, where
\begin{align}
    \zadoff[\outputIdx] = -\frac{\outputIdx (\outputIdx + \outputOdd)}{\outputSize} \cdot \pi,\
    \text{where}~ \outputOdd = \outputSize \bmod 2,\ \forall \outputIdx = 0, 1, \dots, \outputSize-1.
\end{align}
Each audio clip in AudioMNIST~\cite{audiomnist2023} is a real-valued waveform of English spoken digits from `0' to `9' by {60} people whose native languages are English, German, Chinese, and Spanish.
Each waveform, sampled at {48}\thinspace{kHz}, lasts for about {0.5} seconds. In our implementation, each waveform is first downsampled to {8}\thinspace{kHz}, and the middle {0.5} seconds is truncated to a {4,000}-point vector.
Then, we perform a short-time Fourier transform (STFT) every {25}\thinspace{ms} for {200} non-overlapped time windows to form a spectrogram. Note that we only take the amplitudes of this spectrogram and drop the phase information. Each time window contains {20} samples, which are converted into {20} complex frequency bins by STFT.
Finally, the {200} time windows are concatenated into a vector, which is then modulated with the {4,000}-point ZC phase sequence $\zadoffVec$ to generate $\inputVec \in \mathbb{C}^{4,000}$.

%%%%%
%%%%%
\subsection*{Dataset and Complex Model Architecture}

We consider FC layers in our DL model architecture for {\name} that employ large-scale MVMs. The complex nature of the RF signals enables FC layers with complex-valued input vector, $\inputVec$, output vector, $\outputVec$, and trainable weight matrix, $\weightMat$.
We employ an activation function $\activation(\cdot)$ that applies an absolute value operation followed by a phase adjustment using the ZC phase sequence. Specifically, for each FC layer except the last, the activation function first computes the element-wise absolute value of $\outputVec$ and then adds a phase based on a $\outputSize$-point ZC sequence $\zadoffVec$,
\begin{align}
    \activation_{\outputIdx}(\outputElem_{\outputIdx})
    = \abs{\outputElem_{\outputIdx}} \cdot \eu^{\iu \zadoff[\outputIdx]} = \abs{\outputElem_{\outputIdx}} \cdot \eu^{-\iu \cdot \frac{\pi \outputIdx(\outputIdx+\outputOdd)}{\outputSize}},\
    \text{where}~ \outputOdd = \outputSize \bmod 2,\ \forall \outputIdx = 0, 1, \dots, \outputSize-1.
\end{align}
Hereby, the subscript on $\activation_{\outputIdx}$ indicates the phase shift applied to each element of $\outputElem_{\outputIdx}$. The reason behind selecting this activation function is twofold. First, it preserves the waveform power by maintaining the amplitude of each element in $\outputVec$. Second, the use of ZC phase sequence ensures that the power of the input waveform $\waveInput(\waveIdx)$ to the subsequent FC layer is evenly distributed across the spectrum.
For the last FC layer, only the absolute function $\abs{\outputElem_{\outputIdx}}$ is applied, which converts the complex-valued $\outputVec$ to a real-valued vector. 

To train the model, we employ the cross-entropy loss on $\abs{\outputVec}$, using Adam optimizer~\cite{kingma2014adam} with a learning rate of ${10}^{-3}$ over {100} epochs. In the testing phase, the predicted class is given by the maximum $\abs{\outputVec}$ after the absolute function. Among the {100} training epochs, we select the model with the highest testing accuracy. All activation functions are computed digitally. When comparing {\name} with conventional DL models performed in digital computing, we exclude the energy consumption and latency of the activation functions as they exist in both computing paradigms and are orthogonal to the MVM operations.

%%%%%
%%%%%
\subsection*{Implementation}

To demonstrate the {\name} framework, we develop a {\smarttrx} prototype using a USRP X310 SDR and a Mini-Circuits ZEM-4300+ frequency mixer~\cite{ZEM-4300+}, which function as the TX/RX and computing mixer, respectively.
% The USRP has a measured RX noise figure of {13.6}\thinspace{dB}, and the computing mixer exhibits an insertion loss of {11.4}\thinspace{dB} within the optimized operating regime.
\autoreffig{fig:figure-experiment} shows the experimental setup, where a central radio broadcasts model weights to three clients over a {25}\thinspace{MHz} channel centered at {0.915}\thinspace{GHz}, which is limited by the available unlicensed frequency spectrum in the industrial, scientific, and medical (ISM) bands between {902--928}\thinspace{MHz}~\cite{ISM}. (See Supplementary Section~\ref{sec:supplementary-experiment-ml-wired} for the wired experiments with larger bandwidth) The wireless link distance is $\approx${1}\thinspace{meter}, limited by the LO power required for the off-the-shelf diode ring-based mixer. This constraint can be relaxed to support larger link distances using integrated analog computing circuits~\cite{rashed2024scalable} or beamforming on an antenna array~\cite{shepard2012argos}.
Each client streams the I/Q modulated waveform $\waveInput(\waveIdx)$ at a carrier frequency of {1.2}\thinspace{GHz} to the mixer's RF port over an SMA cable.
In the meanwhile, $\waveWeight(\waveIdx)$ is received by the {\smarttrx}'s antenna and then mixed with the streamed $\waveInput(\waveIdx)$. The output downconverted waveform $\waveOutput(\waveIdx)$ from {0.285}\thinspace{GHz} is I/Q demodulated, low-pass filtered, and sampled by two ADCs operating at a low sampling rate of {0.2}\thinspace{MHz}. 
For the waveform generation, we place zero padding subcarriers in the frequency domain to mitigate the roll-off effect at the LPF edge, and cyclic prefix in the time domain to improve the timing synchronization tolerance.

We evaluate the energy efficiency of {\name} given by equation~\eqref{eqn:emac-mvm}, where the \textsf{SNR} values are varied by adjusting the transmit power of $\waveInput(\waveIdx)$.
The overall loss $\efficiencyTRX = 1.48 \times {10}^{-4}$ is the combination of a TX efficiency of 10\%~\cite{imai202432}, insertion loss of the computing mixer (measured at {11.4}\thinspace{dB}), and RX noise figure (measured at {16.9}\thinspace{dB}).
We also conduct simulations for {\name} by plugging in the realistic hardware parameters above, and the computing mixer performs analog multiplication. The simulations consider a frequency-flat wireless channel between the central radio and {\smarttrx} with additive Gaussian white noise (AWGN).
For both experiments and simulations, we consider ADC energy consumption ($e_{\textrm{adc}}$) of {1}\thinspace{pJ/sample}~\cite{adc_survey}, and digital computing efficiency using ASICs ($e_{\textrm{dig}}$) of {1}\thinspace{pJ/MAC}~\cite{horowitz20141, abari201427, jouppi2017datacenter}. 
See Supplementary Section~\ref{sec:supplementary-experiment-setup} for more details on the experimental setup and measurements.
The TDL of {\name} can be simulated based on equation~\eqref{eqn:emac-tdl} assuming ideal {\smarttrx} hardware with $\efficiencyTRX = 1$ and $\overheadPad = \overheadCP = 0$.

%% file: tex/acks.tex
\section*{Acknowledgments}
Z.G. and T.C. acknowledge partial support from the NSF Athena AI Institute for Edge Computing (CNS-2112562). S.K.V. and D.E. acknowledge support from the DARPA NaPSAC program. K.S. acknowledges the support of the Israeli Council for Higher Education and the Zuckerman STEM Leadership Program. D.E. acknowledges partial support from the NSF EAGER program (ECCS-2419204) and the DARPA QuANET program. The authors thank Marc Bacvanski for the useful discussion and for providing feedback on the manuscript.

\section*{Author Contributions}

D.E. and T.C. conceived the original concept and system architecture. Z.G. and T.C. designed the experiments using the SDR platform. Z.G. conducted the experiments and analyzed the results. Z.G., D.E., and T.C. wrote the manuscript. All authors contributed to the analysis and refinement of the analog computing schemes, and provided feedback on the manuscript.

\section*{Competing Interests}
The authors declare no competing interests.

\section*{Data Availability}
The data supporting the claims in this paper is available upon reasonable request.

\section*{Correspondence}
Requests for information should be directed to Tingjun Chen (\href{mailto:tingjun.chen@duke.edu}{tingjun.chen@duke.edu}).

%% file: tex/Supplementary_theory.tex
\section*{\textbf{Supplementary Information: Theory}}

%%%%%
%%%%%
\section{Notation and Preliminaries}
\label{sec:supplementary-theory-notation}

For a complex-valued matrix $\mathbf{A} \in \mathbb{C}^{M \times N}$, let $\trans{\mathbf{A}}$, $\conj{\mathbf{A}}$, and $\hermitian{\mathbf{A}}$ denote its transpose, complex conjugate, and conjugate transpose, respectively. For a square matrix $\mathbf{A} \in \mathbb{C}^{N \times N}$ that is invertible, let $\matInv{\mathbf{A}}$ denote the inverse of $\mathbf{A}$. Let $\mathbf{I}_{N}$ be the $N \times N$ identity matrix.
For two matrices $\mathbf{A}$ and $\mathbf{B}$ with the same dimension, let $\mathbf{A} \odot \mathbf{B}$ denote the element-wise multiplication (Hadamard product), and $\mathbf{A} \oslash \mathbf{B}$ denote the element-wise division.

The discrete Fourier transform (DFT) converts a vector, $\mathbf{x} = [x_{n}] \in \mathbb{C}^{\fftSize}$, into another vector with equal length, $\mathbf{X} = [X_{k}] = \dft(\mathbf{x}) \in \mathbb{C}^{\fftSize}$, where
\begin{align}
    X_{k} = \sum_{n=0}^{\fftSize-1} x_{n} \cdot e^{-j 2\pi \frac{k}{\fftSize} n},~\forall k = 0, 1, \dots, \fftSize-1.
    \label{eq: preliminary-dft-definition}
\end{align}
The DFT operation can be written in the matrix form,
\begin{align}
    \mathbf{X} = \sqrt{\fftSize} \cdot \dftMat \cdot \mathbf{x},
\end{align}
where $\dftMat \in \mathbb{C}^{\fftSize \times \fftSize}$ is the $\fftSize$-point DFT matrix given by
\begin{align}
    \dftMat &= \frac{1}{\sqrt{\fftSize}}
    \begin{bmatrix}
    1 & 1 & 1 & \dots & 1 \\
    1 & \dftElem & \dftElem^2 & \dots & \dftElem^{\fftSize-1} \\
    1 & \dftElem^2 & \dftElem^4 & \dots & \dftElem^{2(\fftSize-1)} \\
    \vdots & \vdots & \vdots & \ddots & \vdots \\
    1 & \dftElem^{\fftSize-1} & \dftElem^{2(\fftSize-1)} & \dots & \dftElem^{(\fftSize-1)^2} \\
    \end{bmatrix},
\end{align}
where $\dftElem = \eu^{-\iu 2\pi/\fftSize}$.
Note that the DFT matrix is a unitary matrix satisfying $\dftMat \hermitian{\dftMat} = \mathbf{I}$, with $\dftMat = \trans{\dftMat}$ and $\dftMat^{-1} = \hermitian{\dftMat} = \conj{\dftMat}$.
% For brevity, we drop the subscript of a matrix indicating its dimension when the context is clear.
Symmetrically, the inverse DFT (IDFT) operation is given by $\mathbf{x} = \idft(\mathbf{X})$, where
\begin{align}
    x_{n} = \frac{1}{\fftSize} \sum_{k=0}^{\fftSize-1} X_{k} \cdot e^{\iu 2\pi \frac{k}{\fftSize} n},~\forall n = 0, 1, \dots, \fftSize-1.
    \label{eq: preliminary-idft-definition}
\end{align}
The IDFT operation can also be written in the matrix form, given by
\begin{align}
    \mathbf{x} = \frac{1}{\sqrt{\fftSize}} \cdot \idftMat \cdot \mathbf{X}.
\end{align}

The DFT and IDFT operations in equations {\eqref{eq: preliminary-dft-definition}} and {\eqref{eq: preliminary-idft-definition}} can be accelerated by the fast Fourier transform (FFT) algorithm.
Specifically, given a vector length of $\fftSize$ (assuming $\fftSize$ is the power of 2), it takes $\fftSize/2 \cdot \log_2 \fftSize$ complex-valued MACs to conduct DFT or IDFT, which is equivalent to $2 \fftSize \log_2 \fftSize$ real-valued MACs.

We define the circular shift matrix $\shiftMat_{\fftSize} \in \mathbf{R}^{\fftSize \times \fftSize}$ that, for an $\fftSize$-point vector, shifts all the elements one position to the right and puts the last element to the first position,
\begin{align}
    \shiftMat_{\fftSize} = \begin{bmatrix}
    0 & 0 & \dots & 0 & 1 \\
    1 & 0 & \dots & 0 & 0 \\
    0 & 1 & \dots & 0 & 0 \\
    \vdots & \vdots & \ddots & \vdots & \vdots \\
    0 & 0 & \dots & 1 & 0 \\
    \end{bmatrix}.
\end{align}
We can then denote a repeated shift operation applied $m$ times by $(\shiftMat_{\fftSize})^{m}$. We drop the subscript $\fftSize$ for brevity when the matrix dimension and context are clear.

\section{Digital-to-Analog and Analog-to-Digital Conversions}
\label{sec:supplementary-theory-dac-adc}

We use $\wave(\waveIdx)$ and $\samp[\sampIdx]$ to represent the continuous and discrete signal in the time domain, and $\spec(\freq)$ and $\spec[\specIdx]$ to represent the continuous and discrete spectrum of the signal in the frequency domain.
In this section, we consider real-valued $\wave(\waveIdx)$ and $\samp[\sampIdx]$ with a single DAC and ADC, and extend the scenario to complex-valued $\wave(\waveIdx)$ and $\samp[\sampIdx]$ with two DACs and ADCs for I/Q modulation in Supplementary Section~\ref{sec:supplementary-theory-iq-modulation}.

In general, a DAC reconstructs the continuous waveform $\wave(\waveIdx)$ from $\samp[\sampIdx]$ by a per-sample duration of $\sampDuration = 1/\sampRate$, or under a sampling rate of $\sampRate$. Specifically, signal construction using a DAC can be modeled by
\begin{align}
    \wave(\waveIdx)
    = \DAC{\samp[\sampIdx]}
    = \sum_{\sampIdx = -\infty}^{+\infty} \samp[\sampIdx] \cdot h_{\textrm{DAC}}\left(\frac{\waveIdx}{\sampDuration} - \sampIdx \right),
    \label{eq: digital-to-analog}
\end{align}
where $h_{\textrm{DAC}}(\cdot)$ is the reconstruction kernel of the DAC.
Note that equation~\eqref{eq: digital-to-analog} only constrains the values with integer values of $\sampIdx$. Hence, there are infinitely many reconstruction kernels from $\samp[\sampIdx]$ to $\wave(\waveIdx)$ that satisfy such reconstructions. 
For example, sinc-interpolation limits the bandwidth of the reconstructed waveform within $[-\sampRate/2, +\sampRate/2]$, which is
\begin{align}
    \wave_{\textrm{sinc}}(\waveIdx) = \textsf{DAC}_{\textrm{sinc}}\left\{ \samp[\sampIdx] \right\} &= \sum_{\sampIdx} \samp[\sampIdx] \cdot \sinc \left(\frac{\waveIdx}{\sampDuration} - \sampIdx \right),~
    \textrm{where}~\sinc(x) = \frac{\sin(\pi x)}{\pi x}.
\end{align}
In practice, a commonly used signal reconstruction for a DAC is zero-order hold (ZOH), given by
\begin{align}
    \wave_{\textrm{ZOH}}(\waveIdx) = \textsf{DAC}_{\textrm{ZOH}}\left\{ \samp[\sampIdx] \right\} &= \sum_{\sampIdx} \samp[\sampIdx] \cdot \textsf{Rect}\left(\frac{\waveIdx}{\sampDuration} - \sampIdx -\frac{1}{2} \right),~
    \textrm{where}~\textsf{Rect}(x) = 
    \begin{cases}
        1, & -\frac{1}{2} \leq x \leq +\frac{1}{2}, \\
        0, & \text{otherwise}.
    \end{cases}
\end{align}

An ADC samples the continuous waveform $\wave(\waveIdx)$ and creates the discrete sequence $\samp[\sampIdx]$ given by
\begin{align}
    \samp[\sampIdx] = \ADC{\wave(\waveIdx)} = \wave\left( \sampIdx \sampDuration \right), ~\forall \sampIdx = 0, 1, 2, \dots.
    \label{eq: analog-to-digital}
\end{align}
It can be proven that for any reconstruction kernel on DAC, as long as the DAC is synchronized with the ADC, i.e., with the shared $\waveIdx$, the original discrete sequence can be reconstructed:
\begin{align}
    \samp[\sampIdx] \propto \ADC{\DAC{\samp[\sampIdx]}}.
    \label{eq: digit-analog-constraint}
\end{align}

For the DAC, we assume a normalized power constraint on the input sequence given by
\begin{align}
    \abs{\samp[\sampIdx]}^2 \leq 1,\ \forall \sampIdx = 0, 1, 2, \dots,
\end{align}
and $\abs{\samp[\sampIdx]}^2=1$ corresponds to the DAC's peak output power, $\powerMax$. The peak-to-average power ratio (PAPR) of the sequence $\sampVec = [\samp[\sampIdx]]$ is given by
\begin{align}
    \PAPR{\sampVec} =\frac{\max_{\sampIdx} \abs{\samp[\sampIdx]}^2}{\mathbb{E}[\samp^2[\sampIdx]]},
    \label{eq: papr-discrete-definition}
\end{align}
where $\max_{\sampIdx} \abs{\samp[\sampIdx]}^2$ and $\mathbb{E}[\samp^2[\sampIdx]]$ represent the peak instant and average power of $\sampVec$, respectively.

For the ADC, we consider the constraint on the input waveform $\abs{\wave(\waveIdx)}$ that
\begin{align}
    \abs{\wave(\waveIdx)}^2 \leq 1, ~\forall \waveIdx.
\end{align}
Similarly, the PAPR of the input waveform to the ADC $\wave(\waveIdx)$ is given by
\begin{align}
    \PAPR{\wave(\waveIdx)} = \frac{\max_{\waveIdx} \abs{\wave(\waveIdx)}^2}{\frac{1}{\waveLen}\int_{0}^{\waveLen} \abs{\wave(\waveIdx)}^2\ \textrm{d}\waveIdx},
\end{align}
where $\max_{\waveIdx} \abs{\wave(\waveIdx)}^2$ refers to the peak instant power of the waveform, and $\frac{1}{\waveLen}\int_{0}^{\waveLen} \abs{\wave(\waveIdx)}^2\ \textrm{d}\waveIdx$ is the average power of the received waveform over a time period of $\waveLen$.

%%%%%
%%%%%
\section{I/Q Modulation and Demodulation}
\label{sec:supplementary-theory-iq-modulation}

%% figure begins
\begin{figure*}[!t]
    \centering
    \includegraphics[width=0.95\textwidth]{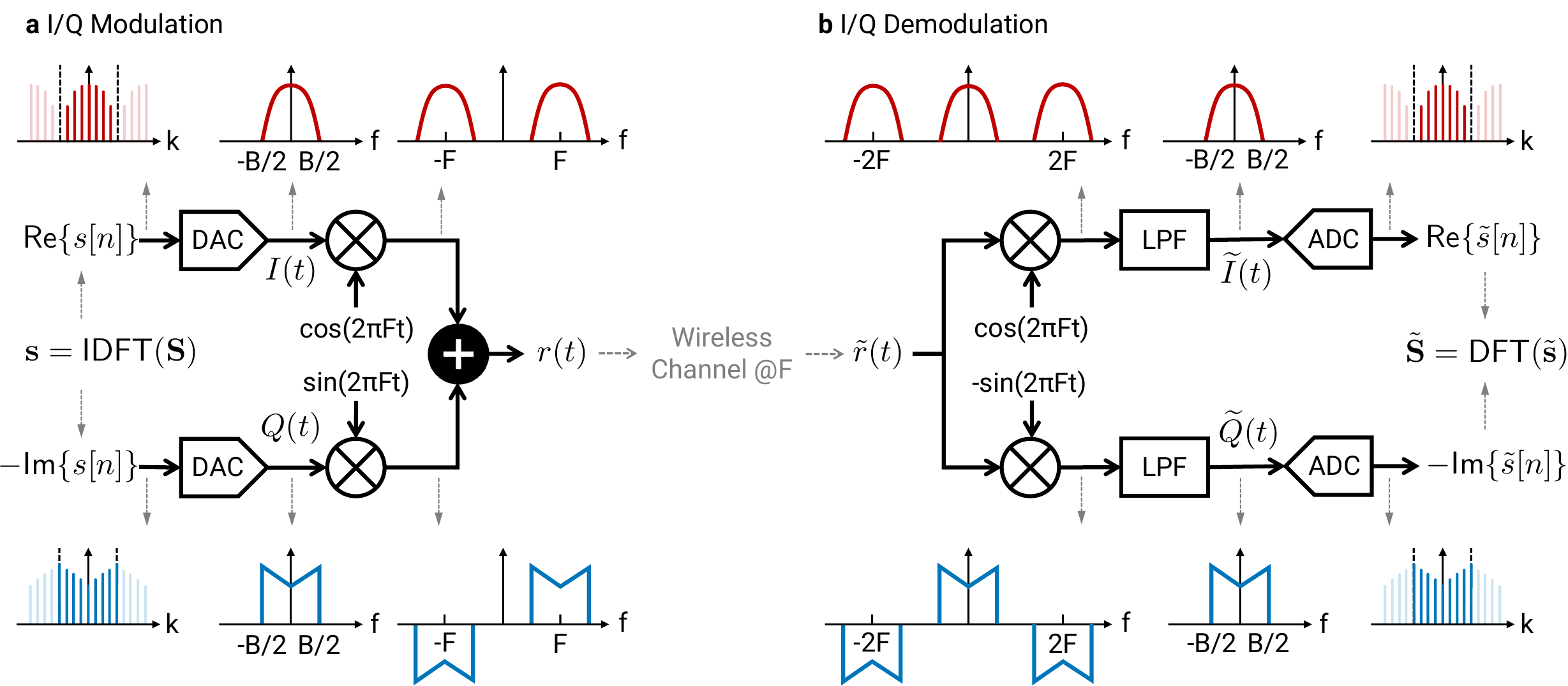}
    \caption{
    \textbf{The diagram of I/Q modulation and demodulation.}
    \textbf{a}, The I/Q modulation converts the complex-valued baseband I/Q sequence $\sampVec$ into a real-valued waveform $\radio(\waveIdx)$ modulated at carrier frequency $\carrier$. 
    \textbf{b}, The I/Q demodulation converts the received real-valued waveform $\radio(\waveIdx)$ back to the complex-valued baseband I/Q sequence $\Tilde{\sampVec}$, which is proportional to the original $\sampVec$.}
    \label{fig:supplementary-theory-iq-modulation}
\end{figure*}
%% figure ends 

In wireless communication, I/Q modulation, as illustrated in \autoreffig{fig:supplementary-theory-iq-modulation}, is a technique for efficient transmission of information over radio frequencies, which modulates data onto a carrier signal by varying its in-phase (I) and quadrature (Q) components. I/Q modulation enables complex modulation schemes by combining the I and Q components and, therefore, achieves doubled spectral efficiency by modulating information in both the amplitude and phase of the carrier signal.
Consider an I/Q modulation system with a sampling rate of $\sampRate$ (corresponding to the bandwidth $\band=\sampRate$) and a carrier frequency of $\carrier$.
The TX tasks input of a digital, complex-valued I/Q sample sequence $\samp[\sampIdx]$. The complex-valued baseband waveform with I and Q components, denoted by $I(\waveIdx)$ and $Q(\waveIdx)$, can be respectively constructed from the real and imaginary components of $\samp[\sampIdx]$ using two DACs.
Plugging in equation {\eqref{eq: digital-to-analog}}, we can formulate the waveform construction process as
\begin{align}
    \wave(\waveIdx) = \DAC{\samp[\sampIdx]} = I(\waveIdx) - \iu Q(\waveIdx),~\textrm{where}~
    I(\waveIdx) = \DAC{\RE{\samp[\sampIdx]}}~\textrm{and}~Q(\waveIdx) = -\DAC{\IM{\samp[\sampIdx]}}.
\end{align}
Here, the DAC reconstruction kernel can either be sinc interpolation or ZOH. $\wave(\waveIdx)$ is then I/Q modulated to carrier frequency $\carrier$ as shown in \autorefsubfig{fig:supplementary-theory-iq-modulation}{a}, yielding the modulate signal $\radio(\waveIdx)$ given by
\begin{align}
    \radio(\waveIdx)
    = I(\waveIdx) \cdot \cos(2\pi \carrier \waveIdx) + Q(\waveIdx) \cdot \sin(2\pi \carrier \waveIdx)
    = \RE{ s(t) \cdot \eu^{\iu 2\pi \carrier \waveIdx}}.
    \label{eq:iqmod}
\end{align}
After over-the-air transmission, the signal received by the RX, $\widetilde{\radio}(\waveIdx)$, is I/Q demodulated to recover the baseband waveform, $\widetilde{\wave}(\waveIdx)$, using an LO operating at the same carrier frequency $\carrier$ and a low-pass filter (LPF) with a cutoff frequency of $\band/2$, as shown in \autorefsubfig{fig:supplementary-theory-iq-modulation}{b}.
This I/Q demodulation process can be written as
\begin{align}
    \widetilde{\wave}(\waveIdx)
    = \LPF{\widetilde{\radio}(\waveIdx) \cdot \eu^{-\iu 2\pi \carrier \waveIdx}}
    = \widetilde{I}(\waveIdx) - \iu \widetilde{Q}(\waveIdx),~\textrm{where}~
    \widetilde{I}(\waveIdx) = \textsf{LPF}\left\{ \widetilde{\radio}(\waveIdx) \cos(2\pi \carrier \waveIdx) \right\}~\textrm{and}~ 
    \widetilde{Q}(\waveIdx) = \textsf{LPF}\left\{ \widetilde{\radio}(\waveIdx) \sin(2\pi \carrier \waveIdx) \right\}.
\end{align}
Finally, two ADCs operating at the same sampling rate $\sampRate$ convert the I and Q components of $\widetilde{\wave}(\waveIdx)$ into the discrete I/Q samples $\widetilde{\samp}[\sampIdx]$, which is
\begin{align}
    \widetilde{\samp}[\sampIdx] = \ADC{\widetilde{\wave}(\waveIdx)},
\end{align}
from which the transmitted I/Q samples $\widetilde{\samp}[\sampIdx]$ can be recovered.

%%%%%
%%%%%
\section{Frequency Mixer as Analog Multiplier}
\label{sec:supplementary-theory-frequency-mixer}

A frequency mixer is a three-port electrical circuit that produces new signals at the sum and difference of the input signal frequencies, which can be used for frequency up-conversion and down-conversion, respectively.
The three ports of a frequency mixer are labeled IF (intermediate frequency), RF (radio frequency), and LO (local oscillator). The IF and RF ports can be used as input or output, whereas the LO port always requires an input signal.
Essentially, a frequency mixer performs an analog multiplication of the two input waveforms.
Without loss of generality, assume two input waveforms $\radio_{1}(\waveIdx)$ and $\radio_{2}(\waveIdx)$ are I/Q modulated to carrier frequencies at $\carrier_{1}$ and $\carrier_{2}$, respectively, given by
\begin{align}
    \radio_1(\waveIdx) = I_1(\waveIdx) \cos(2\pi \carrier_1 \waveIdx) + Q_1(\waveIdx) \sin(2\pi \carrier_1 \waveIdx)~\textrm{and}~
    \radio_2(\waveIdx) = I_2(\waveIdx) \cos(2\pi \carrier_2 \waveIdx) + Q_2(\waveIdx) \sin(2\pi \carrier_2 \waveIdx).
    \label{eq:up-conversion}
\end{align}
Let $\wave_1(\waveIdx) = I_1(\waveIdx) - \iu Q_1(\waveIdx)$ and $\wave_2(\waveIdx) = I_2(\waveIdx) - \iu Q_2(\waveIdx)$ denote the baseband I/Q waveforms corresponding to $\radio_1(\waveIdx)$ and $\radio_2(\waveIdx)$, respectively.
The analog multiplication of the two input waveforms $\radio_1(\waveIdx)$ and $\radio_2(\waveIdx)$ yields the output waveform $\radio_o(\waveIdx)$ at the new frequency $\carrier_o = \carrier_{1} \pm \carrier_{2}$, given by
\begin{align}
    \radio_o(\waveIdx)
    \propto \radio_{1}(\waveIdx) \cdot \radio_{2}(\waveIdx)
    & = \left[ I_1(\waveIdx) \cos(2\pi \carrier_1 \waveIdx) + Q_1(\waveIdx) \sin(2\pi \carrier_1 \waveIdx) \right] \cdot
    \left[ I_2(\waveIdx) \cos(2\pi \carrier_2 \waveIdx) + Q_2(\waveIdx) \sin(2\pi \carrier_2 \waveIdx) \right] \nonumber \\
    & \propto \left[ I_1(\waveIdx) I_2(\waveIdx) - Q_1(\waveIdx)Q_2(\waveIdx) \right] \cdot\cos(2\pi(\carrier_1+\carrier_2)\waveIdx) + \left[ I_1(\waveIdx)Q_2(\waveIdx) + Q_1(\waveIdx)I_2(\waveIdx) \right] \cdot\sin(2\pi(\carrier_1+\carrier_2)\waveIdx) \nonumber \\
    & \quad + \left[ I_1(\waveIdx) I_2(\waveIdx) + Q_1(\waveIdx)Q_2(\waveIdx) \right] \cdot\cos(2\pi(\carrier_1-\carrier_2)\waveIdx) + \left[ -I_1(\waveIdx)Q_2(\waveIdx) + Q_1(\waveIdx)I_2(\waveIdx) \right] \cdot\sin(2\pi(\carrier_1-\carrier_2)\waveIdx) \nonumber \\
    & = \RE{\wave_1(\waveIdx) \wave_2(\waveIdx)}\cdot\cos(2\pi(\carrier_1+\carrier_2)\waveIdx) -\IM{\wave_1(\waveIdx) \wave_2(\waveIdx)}\cdot\sin(2\pi(\carrier_1+\carrier_2)\waveIdx) \nonumber \nonumber \\
    & \quad +\RE{\wave_1(\waveIdx) \conj{\wave}_2(\waveIdx)}\cdot\cos(2\pi(\carrier_1-\carrier_2)\waveIdx) -\IM{\wave_1(\waveIdx) \conj{\wave}_2(\waveIdx)}\cdot\sin(2\pi(\carrier_1-\carrier_2)\waveIdx),
    \label{eq: frequency-mixer-multiplication}
\end{align}
where $\conj{\wave}_{2}(\waveIdx)$ denotes the conjugate of $\wave_{2}(\waveIdx)$. Note that the output waveform of the mixer $\radio_o(\waveIdx)$ at frequency $\carrier_o$ can also be written in the form of
\begin{align}
    \radio_o(\waveIdx) = I_o(\waveIdx) \cos(2\pi \carrier_o \waveIdx) + Q_o(\waveIdx) \sin(2\pi \carrier_o \waveIdx),
    \label{eq:down-conversion}
\end{align}
where $\wave_o(\waveIdx) = I_o(\waveIdx) - \iu Q_o(\waveIdx)$ denotes the corresponding baseband waveform.
When the mixer is used for frequency up-conversion, the two input ports are IF and LO, the output port is RF, and their carrier frequency satisfies $\carrier_o = \carrier_1 + \carrier_2$. As long as the carrier frequencies are carefully selected without frequency aliasing, the mismatched frequency component $\carrier_1-\carrier_2$ will be filtered out by the LPF. Plugging this into equation {\eqref{eq: frequency-mixer-multiplication}} yields
\begin{align}
    &\begin{cases}
        I_o(\waveIdx) = \LPF{\radio_o(\waveIdx) \cdot \cos(2\pi(\carrier_1+\carrier_2)\waveIdx)}
        \propto \RE{\wave_1(\waveIdx) \wave_2(\waveIdx)} \\
        Q_o(\waveIdx) = \LPF{\radio_o(\waveIdx) \cdot \sin(2\pi(\carrier_1+\carrier_2)\waveIdx)} 
        \propto -\IM{\wave_1(\waveIdx) \wave_2(\waveIdx)}
    \end{cases}
    \Rightarrow \wave_o(\waveIdx) = I_o(\waveIdx) - \iu Q_o(\waveIdx) \propto \wave_1(\waveIdx) \cdot \wave_2(\waveIdx).
    \label{eq: up-conversion-multiplication}
\end{align}
When the mixer is used for frequency down-conversion, the RF and LO ports become the input ports for $\radio_1(\waveIdx)$ and $\radio_2(\waveIdx)$, respectively, and the IF port becomes the output port for $\radio_o(\waveIdx)$.
In this case, $\carrier_o = \carrier_1 - \carrier_2$, and the frequency component $\carrier_1+\carrier_2$ is filtered out by the LPF.
Similarly, we can derive $\wave_o(\waveIdx)$ from equation {\eqref{eq: frequency-mixer-multiplication}} as
\begin{align}
    &\begin{cases}
        I_o(\waveIdx) = \LPF{\radio_o(\waveIdx) \cdot \cos(2\pi(\carrier_1-\carrier_2)\waveIdx)} \propto \RE{\wave_1(\waveIdx) \cdot \conj{\wave}_2(\waveIdx)} \\
        Q_o(\waveIdx) = \LPF{\radio_o(\waveIdx) \cdot \sin(2\pi(\carrier_1-\carrier_2)\waveIdx)} \propto -\IM{\wave_1(\waveIdx) \cdot \conj{\wave}_2(\waveIdx)}
    \end{cases}
    \Rightarrow \wave_o(\waveIdx) = I_o(\waveIdx) - \iu Q_o(\waveIdx) \propto \wave_1(\waveIdx) \cdot \conj{\wave}_2(\waveIdx).
    \label{eq: down-conversion-multiplication}
\end{align}
Compared to equation {\eqref{eq: up-conversion-multiplication}}, the only difference in the down-conversion case is the requirement for a conjugate operation on the waveform input to the LO port, $\wave_2(\waveIdx)$.
In practice, the LO signal supplied to the mixer may exhibit a carrier frequency offset (CFO), $\Delta\carrier$, compared to the desired frequency of the target signal for up-conversion and down-conversion, this effect can be modeled as
\begin{align}
    \begin{cases}
        \wave_o(\waveIdx) \propto \wave_1(\waveIdx) \cdot \wave_2(\waveIdx) \cdot \eu^{\iu \Delta\carrier \waveIdx}, ~\textrm{where}~\carrier_o=\carrier_1 + \carrier_2 + \Delta\carrier,~
        \textrm{for signal up-conversion}, \\
        \wave_o(\waveIdx) \propto \wave_1(\waveIdx) \cdot \conj{\wave}_2(\waveIdx) \cdot \eu^{\iu \Delta\carrier \waveIdx}, ~\textrm{where}~\carrier_o=\carrier_1 - \carrier_2 + \Delta\carrier,~
        \textrm{for signal down-conversion}.
    \end{cases}
    \label{eq: up-down-conversion-multiplication-cfo}
\end{align}
%%

%%%%%
%%%%%
\section{Orthogonal Frequency-Division Multiplexing (OFDM) System}
\label{sec:supplementary-theory-ofdm-system}

OFDM is a technique of modulating data symbols onto multiple overlapping but orthogonal subcarriers within a given bandwidth, which is widely used in modern communication systems, including Wi-Fi (e.g., IEEE 802.11n/ac/ax) and cellular (e.g., LTE/5G). 
Assuming that an OFDM system occupies a bandwidth of $[-\band/2, +\band/2]$ in the baseband. This bandwidth $\band$ is divided into $\fftSize$ overlapping but orthogonal subcarriers with a subcarrier spacing of $\freqSub = \band/\fftSize$. Without loss of generality, $\fftSize$ is assumed to be an even number. In the baseband, the $\specIdx$-th subcarrier is at frequency
\begin{align}
    \freq_{\specIdx} = \left( \specIdx - \frac{\fftSize}{2} \right) \cdot \freqSub = \frac{\specIdx-\fftSize/2}{\fftSize} \cdot \band, ~\forall \specIdx = 0, \dots, \fftSize-1.
\end{align}
Generally, the OFDM system is structured by OFDM symbols in the time domain.
Within one OFDM symbol, the time domain I/Q samples are converted to/from frequency domain data symbols using an $\fftSize$-point DFT, so there are $\fftSize$ I/Q samples per OFDM symbol in the time domain corresponding to the $\fftSize$ subcarriers in the frequency domain. Hereby, we denote the time domain I/Q samples of an OFDM symbol as $\sampVec = [\samp[\sampIdx]] \in \mathbb{C}^{\fftSize}$, and its frequency domain data symbols as $\specVec = [\spec[\specIdx]] \in \mathbb{C}^{\fftSize}$.
The data symbols can be derived from the I/Q samples via an $\fftSize$-point DFT, i.e.,
\begin{align}
    \specVec = \shiftMat^{\fftSizeHalf} \cdot \dft(\sampVec) = \sqrt{\fftSize} \cdot \shiftMat^{\fftSizeHalf} \cdot \dftMat \cdot \sampVec,
    \label{eq: ofdm-system-dft}
\end{align}
where $\shiftMat^{\fftSizeHalf}$ is the circular shift matrix that shifts the zero-frequency (DC) subcarrier symbol originally indexed at $\specIdx=0$ to the center of the spectrum at $\specIdx = \fftSize/2$.
As a result, we have
\begin{align}
    \spec[\specIdx] = \sum_{\sampIdx=0}^{\fftSize-1} \samp[\sampIdx] \cdot \eu^{-\iu 2\pi \frac{\specIdx - \fftSizeHalf}{\fftSize}\sampIdx}, ~\forall \specIdx = 0, \dots, \fftSize-1.
    \label{eq: ofdm-digital-time-to-freq}
\end{align}
Similarly, the time domain I/Q waveform can be derived from the circularly shifted frequency domain data symbols using an $\fftSize$-point IDFT given by
\begin{align}
    \sampVec = \idft (\shiftMat^{\fftSizeHalf} \cdot \specVec) = \frac{1}{\sqrt{\fftSize}} \cdot \idftMat \cdot \shiftMat^{\fftSizeHalf} \cdot \specVec,
    \label{eq: ofdm-system-idft}
\end{align}
where the recovered time domain I/Q samples $\sampVec = [\samp[\sampIdx]] \in \mathbb{C}^{\fftSize}$ is given by
\begin{align}
    \samp[\sampIdx] = \frac{1}{\fftSize} \sum_{\specIdx=0}^{\fftSize-1} \spec[\specIdx] \cdot \eu^{\iu 2\pi \frac{\specIdx - \fftSizeHalf}{\fftSize}\sampIdx},~\forall \sampIdx=0, \dots, \fftSize-1.
    \label{eq: ofdm-digital-freq-to-time}
\end{align}
Based on the Nyquist-Shannon sampling theorem, the minimum sampling required for a system employing I/Q modulation to (re)construct the signal without aliasing is $\sampRate = \band$. Under this sampling rate, an OFDM symbol has a duration of $\waveLen = \fftSize/\sampRate = \fftSize/\band$. 

On the TX side, we consider an ideal DAC reconstruction kernel for the OFDM system that substitutes $\sampIdx$ by $\sampRate \cdot \waveIdx$ in equation {\eqref{eq: ofdm-digital-freq-to-time}}. Correspondingly, the length-$\fftSize$ sequence of I/Q samples becomes a waveform that lasts for a time duration of $\fftSize/\sampRate$.
In this case, the transmitted waveform can be written as
\begin{align}
    \wave(\waveIdx) & = \DAC{\samp[\sampIdx]} = \sum_{\specIdx=0}^{\fftSize-1} \spec[\specIdx] \cdot \eu^{\iu2\pi\frac{\specIdx - \fftSizeHalf}{\waveLen}\waveIdx},~\forall \waveIdx \in [0, \waveLen).
    \label{eq: ofdm-analog-freq-to-time}
\end{align}
This process can also be described using the Fourier series $\fourier(\cdot)$, given by
\begin{align}
    \wave(\waveIdx) & = \fourier(\specVec) = \sum_{\specIdx=-\fftSize/2}^{\fftSize/2-1} \spec[\specIdx] \cdot \eu^{\iu2\pi\frac{\specIdx}{\waveLen}\waveIdx},~\forall \waveIdx \in [0, \waveLen).
\end{align}
This baseband waveform is then modulated to  carrier frequency $\carrier$ as $\radio(\waveIdx) = \RE{ \wave(\waveIdx) \cdot \eu^{\iu 2\pi \carrier \waveIdx}}$.

On the RX side, the received waveform $\widetilde{\radio}(\waveIdx)$ at carrier frequency $\carrier$ is I/Q demodulated to $\widetilde{\wave}(\waveIdx)$, which is then filtered by an LPF with a cutoff frequency of $\sampRate/2$, and then sampled by two ADCs at the sampling rate of $\sampRate$ to acquire the I/Q samples $\widetilde{\sampVec} = [\widetilde{\samp}[\sampIdx]]$, i.e.,
\begin{align}
    \widetilde{\samp}[\sampIdx] & = \ADC{\LPF{\widetilde{\wave}(\waveIdx)}},~\sampIdx = 0, 1, \dots, \fftSize-1.
    \label{eq: ofdm-adc-sampling}
\end{align}
The symbols $\widetilde{\specVec}$ can then be recovered by equation {\eqref{eq: ofdm-system-dft}}.
% Given a channel with flat frequency response, e.g., a wired channel, we have $\widetilde{\wave}(\waveIdx) \propto \wave(\waveIdx)$ (excluding noises), so we have $\widetilde{\sampVec} \propto \sampVec$ in the time domain and $\widetilde{\specVec} \propto \specVec$ in the frequency domain.

During wireless transmissions, multipath propagation causes delayed copies of the transmitted signal to arrive at the receiver, leading to potential inter-symbol interference (ISI). To mitigate this effect, a replica of the ending I/Q samples in $\sampVec$ is appended to the beginning of $\sampVec$ as the \emph{cyclic prefix}, ensuring that multipath delays do not cause overlap between consecutive OFDM symbols.
Assume that a cyclic prefix of $\fftSizeCP$ I/Q samples, extended OFDM symbol with cyclic prefix, $\sampVecCP \in \mathbb{C}^{\fftSize + \fftSizeCP}$, can be written as
\begin{align}
    \sampCP[\sampIdx] = 
    \begin{cases}
        \samp[\fftSize+\sampIdx-\fftSizeCP], & \text{if}~\sampIdx < \fftSizeCP, \\
        \samp[\sampIdx-\fftSizeCP], & \text{if}~\sampIdx \geq \fftSizeCP,
    \end{cases}
    \quad \forall \sampIdx = 0, 1, \dots, \fftSize+\fftSizeCP-1.
\end{align}
Consider a timing delay of $\Delta \sampIdx$ I/Q samples with $\Delta \sampIdx \leq \fftSizeCP$, the received $\widetilde{\sampVec} \in \mathbb{C}^{\fftSize}$ after removing the cyclic prefix is
\begin{align}
    \widetilde{\samp}[\sampIdx] \propto
    \begin{cases}
        \samp[\fftSize+\sampIdx-\fftSizeCP+\Delta\sampIdx], & \text{if}~\sampIdx < \fftSizeCP-\Delta\sampIdx, \\
        \samp[\sampIdx-\fftSizeCP+\Delta\sampIdx], & \text{if}~\sampIdx \geq \fftSizeCP-\Delta\sampIdx,
    \end{cases}
    \quad \forall \sampIdx = 0, 1, \dots, \fftSize-1.
\end{align}
This is equivalent to the original transmitted $\sampVec$ with a circular shift of $(\fftSizeCP-\Delta\sampIdx)$ I/Q samples.
According to the DFT shifting theorem, it holds that
\begin{align}
    \widetilde{\spec}[\specIdx] \propto \spec[\specIdx] \cdot \eu^{-\iu 2\pi \frac{(\fftSizeCP-\Delta\sampIdx)}{\fftSize} \specIdx},\ \forall \specIdx = 0, 1, \dots, \fftSize-1.
\end{align}
This means that the received $\widetilde{\specVec}$ is proportional to the desired $\specVec$ with a phase shift of $2\pi (\fftSizeCP-\Delta\sampIdx) \specIdx /\fftSize$.
Since the timing delay $\Delta\sampIdx$ is a constant over OFDM symbols, it can be estimated using a reference OFDM symbol and then used to calibrate the remaining OFDM symbols.

%%%%%
%%%%%
\section{Hybrid Convolution Theorem}
\label{sec:supplementary-theory-convolution-theorem}

The convolution theorem~\cite{oppenheim1999discrete} states that the multiplication of two time domain signals equals the convolution of their frequency domain spectrums.
Specifically, there are two representations of the convolution theorem in the analog and digital domains: 
(\emph{i}) in the analog domain, the multiplication of two continuous waveforms corresponds to the linear convolution of their spectrums, and
(\emph{ii}) in the digital domain, the element-wise multiplication of two discrete I/Q samples corresponds to the circular convolution.
We consider a hybrid convolution theorem of these two, which is built on the discrete I/Q samples while it corresponds to the linear convolution in the frequency domain.

Consider an OFDM system with an FFT size of $\fftSize$ and subcarrier spacing of $\freqSub$. Let $\specVec_{1} = [\spec_1[\specIdx]] \in \mathbb{C}^{\fftSize}$ and $\specVec_{2} = \spec_2[\specIdx] \in \mathbb{C}^{\fftSize}$ denote two frequency-domain OFDM symbols, whose time-domain waveforms are given by $\wave_1(\waveIdx)$ and $\wave_2(\waveIdx),~\waveIdx\in[0,\waveLen)$, where $\waveLen=1/\freqSub$.
According to equation {\eqref{eq: ofdm-analog-freq-to-time}},
\begin{align}
    \wave_1(\waveIdx) = \fourier(\specVec_1) = \sum_{\specIdx=0}^{\fftSize-1} \spec_1[\specIdx] \cdot \eu^{\iu2\pi\frac{\specIdx - \fftSizeHalf}{\waveLen}\waveIdx},\
    \wave_2(\waveIdx) = \fourier(\specVec_2) = \sum_{\specIdx=0}^{\fftSize-1} \spec_2[\specIdx] \cdot \eu^{\iu2\pi\frac{\specIdx - \fftSizeHalf}{\waveLen}\waveIdx},\
    \forall \waveIdx \in [0, \waveLen).
    \label{eq: convolution-theorem-proof-input}
\end{align}
Let `$\convolution$' denote the linear convolution operation that maps $\mathbb{C}^{\fftSize} \convolution \mathbb{C}^{\fftSize} \rightarrow \mathbb{C}^{2\fftSize-1}$.
Specifically, the linear convolution of two OFDM symbols, denoted by $\specVec_o = [\spec_o[\specIdx]] \in \mathbb{C}^{2\fftSize-1}$, is given by
\begin{align}
    \specVec_o = \specVec_1 \convolution \specVec_2,~\text{where}~
    \spec_o[\specIdx] = \sum_{\convIdx=\max\{0, \specIdx-\fftSize+1\}}^{\min\{\fftSize-1, \specIdx\}} \spec_{1}[\convIdx] \cdot \spec_{2}[\specIdx-\convIdx],\
    \forall \specIdx = 0, 1, \dots, 2\fftSize-2.
    \label{eq: convolution-definition}
\end{align}
Note that the output symbol $\specVec_o$ has an extended length of $(2\fftSize-1)$ while maintaining the same subcarrier spacing ($\freqSub$) and waveform time ($\waveLen = 1/\freqSub$). According to equation~\eqref{eq: ofdm-analog-freq-to-time}, its time-domain waveform, $\wave_o(\waveIdx)$, is given by
\begin{align}
    \wave_o(\waveIdx)
    = \fourier(\specVec_o) = \sum_{\specIdx=0}^{2\fftSize-2} \spec_o[\specIdx] \cdot \eu^{\iu2\pi\frac{\specIdx - (2\fftSize-1)/2}{\waveLen}\waveIdx} 
    = \sum_{\specIdx=0}^{2\fftSize-2} \left( \sum_{\convIdx} \spec_{1}[\convIdx] \cdot \spec_{2}[\specIdx-\convIdx] \right) \cdot \eu^{\iu2\pi\frac{\specIdx - (2\fftSize-1)/2}{\waveLen}\waveIdx},\ \forall \waveIdx \in [0, \waveLen).
    \label{eq: convolution-theorem-proof-output}
\end{align}
Notice that in equation~\eqref{eq: convolution-theorem-proof-input}, symbol $\spec_1[\convIdx]$ is located at frequency $\frac{\convIdx-\fftSize/2}{\waveLen}$, and symbol $\spec_2[\specIdx-\convIdx]$ is located at frequency $\frac{\specIdx-\convIdx-\fftSize/2}{\waveLen}$. The multiplication of these two terms results in a symbol located at frequency $\frac{\specIdx-\fftSize}{\waveLen}$, which has a frequency shift of $\freqSub/2$ compared to the symbol $\spec_o[\specIdx]$ located at frequency $\frac{\specIdx - (2\fftSize-1)/2}{\waveLen}$ in equation~\eqref{eq: convolution-theorem-proof-output}.
Therefore, the following relationship between the time-domain waveforms holds,
\begin{align}
    \wave_o(\waveIdx)
    = \wave_1(\waveIdx) \cdot \wave_2(\waveIdx) \cdot \eu^{\iu 2\pi \frac{\freqSub}{2} \waveIdx}
    = \wave_1(\waveIdx) \cdot \wave_2(\waveIdx) \cdot \eu^{\iu \pi \freqSub \waveIdx}.
\end{align}
In summary, the hybrid convolution theorem can be written as
\begin{align}
    \wave_{1}(\waveIdx) \cdot \wave_{2}(\waveIdx) \cdot \eu^{\iu \pi \freqSub \cdot \waveIdx} = \fourier\big( (\specVec_{1} \convolution \specVec_{2}) \big),\
    \text{where}~\wave_1(\waveIdx) = \fourier(\specVec_1)~\textrm{and}~\wave_2(\waveIdx) = \fourier(\specVec_2).
    \label{eq: convolution-theorem-hybrid}
\end{align}

The bandwidth of $\specVec_1$ or $\specVec_2$ is given by $\band = \fftSize \cdot \freqSub$; as for $\specVec_o$, the subcarrier spacing remains the same while the FFT size becomes $(2\fftSize-1)$, so $\specVec_o$ occupies a bandwidth of $(2\fftSize-1) \cdot \freqSub$. This means that to capture the full spectral information of $\wave_o(\waveIdx)$, it is required that the ADC after I/Q demodulation operates at a minimum sampling rate of 
\begin{align}
    \sampRate' = (2\fftSize-1)\cdot \freqSub.
    \label{eq: convolution-theorem-new-sample-rate}
\end{align}

Recall from Supplementary Section~\ref{sec:supplementary-theory-frequency-mixer} that a computing mixer performs analog multiplication of two signals, similar to the form of the convolution theorem in equation~\eqref{eq: convolution-theorem-hybrid}. Hence, we can configure the computing mixer to calculate the convolution between two discrete signals.
In the case of frequency up-conversion (equation~\eqref{eq: up-conversion-multiplication}), the carrier frequency of the output signal, $\carrier_o$, satisfies 
\begin{align}
    \carrier_o = \carrier_1 + \carrier_2 + \freqSub/2,
    \label{eq: up-conversion-frequency-new}
\end{align}
which cancels the frequency shift term $\eu^{\iu\pi\freqSub\cdot\waveIdx}$.
In the case of frequency down-conversion (equation~\eqref{eq: down-conversion-multiplication}), $\carrier_o$ satisfies
\begin{align}
    \carrier_o = \carrier_1 - \carrier_2 + \freqSub/2,
    \label{eq: down-conversion-frequency-new}
\end{align}
and an extra flipping on $\specVec_2$ is needed to incorporate the conjugated waveform $\conj{\wave}_2(\waveIdx)$ in equation~\eqref{eq: down-conversion-multiplication}, given by
\begin{align}
    \spec'_2[\specIdx] = \spec_2[\fftSize-1-\specIdx],~\forall \specIdx=0, 1, \dots, \fftSize-1.
    \label{eq: down-conversion-modification}
\end{align}

%%%%%
%%%%%
\section{In-Physics MVM Computation Based on Frequency-Encoded OFDM Symbols}
\label{sec:supplementary-theory-vanilla-maft}

In this section, we present a subcarrier mapping algorithm that converts the linear convolution to the in-physics MVM computation. Furthermore, the OFDM system, as discussed in Supplementary Section~\ref{sec:supplementary-theory-ofdm-system}, is employed to efficiently convert the time domain I/Q samples from/to the frequency domain subcarrier symbols by IFFT/FFT.
Without loss of generality, we consider the case where the computing mixer performs signal up-conversion.

\subsection{Subcarrier Mapping Algorithm}

Consider the MVM between a complex-valued matrix $\weightMat = [\weightElem_{\outputIdx, \inputIdx}] \in \mathbb{C}^{\outputSize \times \inputSize}$ and a complex-valued vector $\inputVec = [\inputElem_{\inputIdx}] \in \mathbb{C}^{\inputSize}$, and their MVM results is given by $\outputVec = \weightMat \cdot \inputVec = [\outputElem_{\outputIdx}] \in \mathbb{C}^{\outputSize}$.
Consider an OFDM system with a subcarrier spacing of $\freqSub$ and an FFT size of $\fftSize = \inputSize \outputSize$, occupying a signal bandwidth of $\band = \fftSize \cdot \freqSub = \inputSize \outputSize \cdot \freqSub$. We encode $\weightMat$ and $\inputVec$ into OFDM symbols $\specWeightVec = [\specWeight[\specIdx]] \in \mathbb{C}^{\fftSize}$ and $\specInputVec = [\specInput[\specIdx]] \in \mathbb{C}^{\fftSize}$, respectively. This encoding process essentially maps elements of $\weightMat$ and $\inputVec$ onto different subcarriers of their respective OFDM symbol.
The encoding of $\weightMat$ into $\specWeightVec$ is given by
\begin{align}
    \specWeight[\specIdx] = \weightElem_{\outputIdx, \inputIdx},\ \forall \inputIdx = 0, \dots, \inputSize-1,\ \forall \outputIdx = 0, \dots, \outputSize-1,~\textrm{and}~\forall \specIdx = \inputSize\outputSize-\outputIdx-\inputIdx\outputSize-1,
    \label{eq: original-mvm-weight-to-spec-weight}
\end{align}
and the encoding of $\inputVec$ into $\specInputVec$ is given by
\begin{align}
    \specInput[\specIdx] &= 
    \begin{cases}
        \inputElem_{\inputIdx}, & \text{if}~\specIdx = \inputIdx \cdot \outputSize,~\forall \inputIdx = 0, \dots, \inputSize-1, \\
        0, & \text{otherwise}.
    \end{cases}
    \label{eq: original-mvm-input-to-spec-input}
\end{align}
According to the OFDM system, the I/Q waveforms corresponding to $\specWeightVec$ and $\specInputVec$ can be derived by equations {\eqref{eq: ofdm-system-idft}}--{{\eqref{eq: ofdm-digital-freq-to-time}}.
Specifically, let $\sampWeightVec = [\sampWeight[\sampIdx]] \in \mathbb{C}^{\fftSize}$ denote the I/Q waveform corresponding to $\weightMat$, which is obtained by an $\fftSize$-point IDFT given by
\begin{align}
    \sampWeight[\sampIdx] = \frac{1}{\fftSize} \sum_{\specIdx=0}^{\fftSize-1} \specWeight[\specIdx] \cdot \eu^{\iu 2\pi \frac{\specIdx - \fftSize/2}{\fftSize}\sampIdx} 
    = \frac{1}{\inputSize\outputSize} \sum_{\inputIdx=0}^{\inputSize} \sum_{\outputIdx=0}^{\outputSize} \weightElem_{\outputIdx, \inputIdx} \cdot \eu^{-\iu 2\pi \frac{1+\outputIdx+\inputIdx\outputSize + \inputSize\outputSize/2}{\inputSize\outputSize}\sampIdx},\ \forall \sampIdx = 0, 1, \dots, \fftSize-1.
    \label{eq: original-mvm-spec-to-samp-weight}
\end{align}
Similarly, the I/Q waveform $\sampInputVec = [\sampInput[\sampIdx]] \in \mathbb{C}^{\fftSize}$ corresponding to $\inputVec$ is given by
\begin{align}
    \sampInput[\sampIdx] = \frac{1}{\fftSize} \sum_{\specIdx=0}^{\fftSize-1} \specInput[\specIdx] \cdot \eu^{\iu 2\pi \frac{\specIdx - \fftSize/2}{\fftSize}\sampIdx} 
    = \frac{1}{\inputSize\outputSize} \sum_{\inputIdx=0}^{\inputSize} \inputElem_{\inputIdx} \cdot \eu^{-\iu 2\pi \frac{\inputIdx + \inputSize/2}{\inputSize}\sampIdx},\ \forall \sampIdx = 0, 1, \dots, \fftSize-1.
    \label{eq: original-mvm-spec-to-samp-input}
\end{align}
We assume ideal DACs following equation {\eqref{eq: ofdm-analog-freq-to-time}} at the sampling rate of $\sampRate = \fftSize \cdot \freqSub$, waveforms carrying $\weightMat$ and $\inputVec$ have the same duration $\waveLen$ given by
\begin{align}
    \waveLen = \frac{1}{\freqSub} = \frac{\fftSize}{\band} = \frac{\inputSize\outputSize}{\band}.
\end{align}
Specifically, the waveform $\waveWeight(\waveIdx)$ carrying $\weightMat$ is
\begin{align}
    \waveWeight(\waveIdx) = \DAC{\sampWeightVec} = \fourier(\specWeightVec)
    = \frac{1}{\inputSize\outputSize} \sum_{\inputIdx=0}^{\inputSize} \sum_{\outputIdx=0}^{\outputSize} \weightElem_{\outputIdx, \inputIdx} \cdot \eu^{-\iu 2\pi \frac{1+\outputIdx+\inputIdx\outputSize + \inputSize\outputSize/2}{\waveLen}\waveIdx},\ 
    \forall \waveIdx \in [0, \waveLen),
    \label{eq: original-mvm-samp-to-wave-weight}
\end{align}
and the waveform $\waveInput(\waveIdx)$ carrying $\inputVec$ is
\begin{align}
    \waveInput(\waveIdx) = \DAC{\sampInputVec} = \fourier(\specInputVec) 
    = \frac{1}{\inputSize\outputSize} \sum_{\inputIdx=0}^{\inputSize} \inputElem_{\inputIdx} \cdot \eu^{-\iu 2\pi \frac{\inputIdx + \inputSize/2}{\waveLen}\outputSize\waveIdx},\
    \forall \waveIdx \in [0, \waveLen).
    \label{eq: original-mvm-samp-to-wave-input}
\end{align}
Assume that waveforms carrying $\weightMat$, $\inputVec$, and $\outputVec$ are I/Q modulated to carrier frequencies of $\carrierWeight$, $\carrierInput$, $\carrierOutput$, respectively. For analog computing using the computing mixer for frequency up-conversion {\eqref{eq: up-conversion-frequency-new}}, it holds that
\begin{align}
    \carrierOutput = \carrierInput + \carrierWeight + \freqSub/2
    \Rightarrow
    \waveOutput(\waveIdx) = \waveInput(\waveIdx) \cdot \waveWeight(\waveIdx) \cdot \eu^{\iu\pi\freqSub \cdot \waveIdx},\ \forall \waveIdx \in [0, \waveLen).
    \label{eq: original-mvm-up-conversion-frequency}
\end{align}
The output waveform $\waveOutput(\waveIdx)$ spans a frequency over $(2\fftSize-1)\cdot \freqSub$.
Therefore, the time-domain I/Q samples $\sampOutputVec = [\sampOutput[\sampIdx]] \in \mathbb{C}^{2\fftSize-1}$ can be captured by a pair of ADCs operating at the sampling rate of $(2\fftSize-1)\cdot \freqSub$, or a per-sample duration of $\waveLen/(2\fftSize-1)$,
\begin{align}
    \sampOutput[\sampIdx] = \ADC{\waveOutput(\waveIdx)} = \waveOutput \left(\frac{\sampIdx}{2\fftSize-1} \cdot \waveLen \right),\ \forall \sampIdx=0, 1, \dots, 2\fftSize-2.
\end{align}
Finally, the frequency domain subcarrier symbols $\sampOutputVec = [\sampOutput[\specIdx]] \in \mathbb{C}^{2\fftSize-1}$ can be acquired by a $(2\fftSize-1)$-point DFT as
\begin{align}
    \specOutput[\specIdx] = \sum_{\sampIdx=0}^{2\fftSize-2} \sampOutput[\sampIdx] \cdot \eu^{-\iu 2\pi \frac{\specIdx}{2\fftSize-1}\sampIdx}, ~\forall \specIdx = 0, 1, \dots, 2\fftSize-1.
    \label{eq: original-mvm-samp-to-spec-output}
\end{align}

According to the hybrid convolution theorem {\eqref{eq: convolution-theorem-hybrid}} (see Supplementary Section~\ref{sec:supplementary-theory-convolution-theorem}), the output symbols $\specOutputVec = [\specOutput[\specIdx]] \in \mathbb{C}^{2\fftSize-1}$ as the convolution between $\specWeightVec$ and $\specInputVec$, with ``extended'' frequency components is given by
\begin{align}
    \specOutputVec = \specWeightVec \convolution \specInputVec,\ 
    \textrm{where}~
    \specOutput[\specIdx] = \sum_{\convIdx=\max\{0, \specIdx-\fftSize+1\}}^{\min\{\fftSize-1, \specIdx\}} \specWeight[\convIdx] \cdot \specInput[\specIdx-\convIdx],\
    \forall \specIdx = 0, 1, \dots, 2\fftSize-2.
\end{align}
Note that the output symbols carried by the middle $\outputSize$ subcarriers indexed at $\specOutput[\inputSize\outputSize-\outputSize, \dots, \inputSize\outputSize-1]$ satisfy
\begin{align}
    \specOutput[\inputSize\outputSize-1-\outputIdx]
    & = \sum_{\convIdx=0}^{\inputSize\outputSize-\outputIdx} \specInput[\convIdx] \cdot \specWeight[\inputSize\outputSize-1-\outputIdx - \convIdx]
    = \sum_{\inputIdx=0}^{\inputSize-1} \specInput[\inputIdx\outputSize] \cdot \specWeight[\inputSize\outputSize-1-\outputIdx - \inputIdx\outputSize] \nonumber \\
    & = \sum_{\inputIdx=0}^{\inputSize-1} \inputElem_{\inputIdx} \cdot \weightElem_{\outputIdx, \inputIdx} = \outputElem_{\outputIdx},\
    \forall \outputIdx = 0, 1, \dots, \outputSize-1.
\end{align}
This means that the desired output vector $\outputVec = \weightMat \cdot \inputVec$ is embedded in the spectrum of $\specOutputVec$.
As a result, the output waveform $\waveOutput(\waveIdx)$ can then be I/Q demodulated and sampled using an ADC at a sampling rate of $\sampRate' = (2\fftSize-1) \freqSub$ to acquire I/Q samples $\sampOutputVec = [\sampOutput[\sampIdx]] \in \mathbb{C}^{2\fftSize-1}$ with no frequency aliasing. Finally, the output symbol $\specOutputVec = [\specOutput[\specIdx]] \in \mathbb{C}^{2\fftSize-1}$ can be obtained by a $(2\fftSize-1)$-point DFT,
\begin{align}
    \specOutput[\specIdx] &= \sum_{\sampIdx=0}^{2\fftSize-1} \sampOutput[\sampIdx] \cdot \eu^{-\iu 2\pi \frac{\specIdx - \fftSize'/2}{\fftSize'}\sampIdx},\ \forall \specIdx = 0, 1, \dots, 2\fftSize-2,
    \label{eq: original-mvm-samp-to-spec}
\end{align}
from which the MVM result, $\outputVec$, can be extracted as
\begin{align}
    \outputElem_{\outputIdx} = \specOutput[\inputSize\outputSize-1-\outputIdx],\ \forall \outputIdx = 0, 1, \dots, \outputSize-1.
    \label{eq: original-mvm-spec-to-output}
\end{align}
A similar analysis also holds in the case where the computing mixer performs frequency down-conversion based on equation {\eqref{eq: down-conversion-modification}} with $\carrierOutput = \carrierInput - \carrierWeight + \freqSub/2$.

\subsection{Energy Efficiency Analysis}
We analyze the \emph{energy efficiency} of this ``vanilla'' in-physics MVM computation on the OFDM system, i.e., energy consumed per MAC operation ($\energyMAC$), and the \emph{computation efficiency}, i.e., the number of MAC operations per second per Watt ($\energyMAC^{-1}$). Specifically, there are three energy consumption components: (\emph{i}) $\energyMVMTx$ for the waveform generation of $\waveInput(\waveIdx)$ and I/Q (de)modulation, (\emph{ii}) $\energyMVMADC$ for the I/Q sampling of waveform $\waveOutput(\waveIdx)$ using two ADCs after I/Q demodulation, and (\emph{iii}) $\energyMVMDec$ for the digital computing based encoding (prior to waveform generation) and decoding (after waveform reception).
Note that we only include the energy on the client while excluding that by the central radio broadcasting $\waveWeight(\waveIdx)$.

We first derive $\energyMVMTx$ as follows. Let $\powerTx$ denote the transmit power of $\waveInput(\waveIdx)$ with a radio hardware efficiency of $\efficiencyTX$. The total energy required for the client radio to generate the waveform carrying the inference request, $\inputVec$, is given by
\begin{align}
    \energyMVMTx = (\efficiencyTX)^{-1} \powerTx \cdot \waveLen = (\efficiencyTX)^{-1} \powerTx \cdot \frac{\inputSize\outputSize}{\band}.
    \label{eq: original-mvm-energy-tx-power}
\end{align}
Let $\efficiencyMixer$ denote the efficiency of the computing mixer. The received signal power at the RX after the in-physics computing process carried out by the computing mixer is given by
\begin{align}
    \powerRx = \efficiencyMixer \cdot \powerTx.
    \label{eq: original-mvm-power-signal}
\end{align}
At radio frequency, the thermal noise power spectrum density is given by $k T_{0} = -{174}\thinspace\textrm{dBm/Hz}$, where $k=1.38 \times {10}^{-23}\thinspace\textrm{J/K}$ is the Boltzmann constant and $T_{0} = {300}\thinspace\textrm{K}$ is the room temperature.
After the mixing, the bandwidth of $\specOutputVec$ is $(2\fftSize-1) \cdot \freqSub$, so the total noise power over this bandwidth, $P_{\textrm{noise}}$, is given by
\begin{align}
    \powerNoise
    = (\efficiencyRX)^{-1} \cdot \boltzmann \temperature \cdot (2\fftSize-1) \cdot \freqSub
    = (\efficiencyRX)^{-1} \cdot \boltzmann \temperature \cdot (2\inputSize\outputSize-1) \cdot \freqSub \approx (\efficiencyRX)^{-1} \cdot \boltzmann \temperature \cdot 2 \band,
    \label{eq: original-mvm-power-noise}
\end{align}
where $(\efficiencyRX)^{-1}$ denotes the noise figure of the RX.
Combining equations {\eqref{eq: original-mvm-power-signal}} and {\eqref{eq: original-mvm-power-noise}} yields
\begin{align}
    \textsf{SNR}
    = \frac{\powerRx}{\powerNoise}
    \approx \frac{\efficiencyMixer \cdot \powerTx}{(\efficiencyRX)^{-1} \cdot \boltzmann \temperature \cdot 2 \band}
    \Rightarrow \powerTx = 2 (\efficiencyMixer \cdot \efficiencyRX)^{-1} \cdot \textsf{SNR} \cdot \boltzmann \temperature \cdot \band.
\end{align}
Plugging in $\powerTx$ from equation~\eqref{eq: original-mvm-energy-tx-power} and denote $\efficiencyTRX = \efficiencyTX \cdot \efficiencyMixer \cdot \efficiencyRX$ as the overall hardware efficiency, we have
\begin{align}
    \energyMVMTx = 2\inputSize\outputSize \cdot \efficiencyTRX^{-1} \cdot \textsf{SNR} \cdot \boltzmann \temperature.
    \label{eq: overhead-analysis-energy-tx-snr}
\end{align}

$\energyMVMADC$ represents the energy consumption by the ADC, which is proportional to the number of captured I/Q samples. Given a per-sample energy consumption of $e_{\textrm{adc}}$ (e.g., $e_{\textrm{adc}} = 1\thinspace\textrm{pJ/sample}$~\cite{adc_survey}), the capturing of $\sampOutputVec$ consisting of $(2\fftSize-1)$ complex-valued I/Q samples incurs a total energy consumption of
\begin{align}
    \energyMVMADC = (2\inputSize\outputSize-1) \cdot 2 e_{\textrm{adc}} \approx 4 \inputSize\outputSize \cdot e_{\textrm{adc}}.
\end{align}
Finally, the digital computing energy $\energyMVMDec$ is proportional to the number of real-valued MACs, where we denote the energy consumption per real-valued MAC in the state-of-the-art ASICs as $e_{\textrm{dig}}$.
This term includes the energy consumption associated with an $\fftSize$-point IFFT (encoding of $\specInputVec$ to $\sampInputVec$, equation {\eqref{eq: original-mvm-spec-to-samp-input}}) and a $(2\fftSize-1)$-point FFT (decoding of $\sampOutputVec$ to $\specOutputVec$, equation {\eqref{eq: original-mvm-samp-to-spec-output}}), i.e.,
\begin{align}
    \energyMVMDec
    & = \left[ 2 \fftSize \cdot \log_2 \fftSize + 2(2\fftSize-1) \cdot \log_2 (2\fftSize-1) \right] \cdot e_{\textrm{dig}} \nonumber \\
    &= \left[ 2\inputSize\outputSize \cdot \log_2 (\inputSize\outputSize) + 2(2\inputSize\outputSize-1) \cdot \log_2(2\inputSize\outputSize-1) \right] \cdot e_{\textrm{dig}} \nonumber \\
    & \approx 6 \inputSize\outputSize \cdot \log_2 (\inputSize\outputSize) \cdot e_{\textrm{dig}}.
    \label{eq: overhead-analysis-energy-vanilla-per-mvm-digital}
\end{align}
Putting equations {\eqref{eq: overhead-analysis-energy-tx-snr}}--{\eqref{eq: overhead-analysis-energy-vanilla-per-mvm-digital}} together, the total energy consumption $\energyMVM$ is given by
\begin{align}
    \energyMVM
    = \energyMVMTx + \energyMVMADC + \energyMVMDec
    = \underbrace{2\inputSize\outputSize \cdot \efficiencyTRX^{-1} \cdot \textsf{SNR} \cdot \boltzmann \temperature}_{\energyMVMTx} + \underbrace{4 \inputSize\outputSize \cdot e_{\textrm{adc}}}_{\energyMVMADC} + \underbrace{6 \inputSize\outputSize \cdot \log_2 (\inputSize\outputSize) \cdot e_{\textrm{dig}}}_{\energyMVMDec}.
    \label{eq: overhead-analysis-energy-vanilla-per-mvm}
\end{align}
The corresponding energy efficiency, measured by energy per MAC, $\energyMAC$, is 
\begin{align}
    \energyMAC
    = \frac{\energyMVM}{4\inputSize\outputSize}
    = \energyMACTx + \energyMACADC + \energyMACDec
    = \underbrace{\frac{1}{2} \cdot \efficiencyTRX^{-1} \cdot \textsf{SNR} \cdot \boltzmann \temperature}_{\energyMACTx} + \underbrace{e_{\textrm{adc}}}_{\energyMACADC} + \underbrace{\frac{3}{2} \cdot \log_2 (\inputSize\outputSize) \cdot e_{\textrm{dig}}}_{\energyMACDec}.
    \label{eq: overhead-analysis-energy-vanilla-per-mac}
\end{align}
It can be seen that $\energyMAC$ is dominated by the term $\energyMACDec$ and scales as $\complexity(\log (\inputSize\outputSize))$ as the problem size grows. 

%% figure begins
\begin{table}[!t]
    \caption{Comparison of energy consumption and energy efficiency complexity analysis between the vanilla in-physics matrix-vector multiplication (MVM) computation and the three {\name} schemes.}
    \label{fig:supp-energy-table}
    \centering
    \begin{tabular}{c}
    \includegraphics[width=1.0\columnwidth]{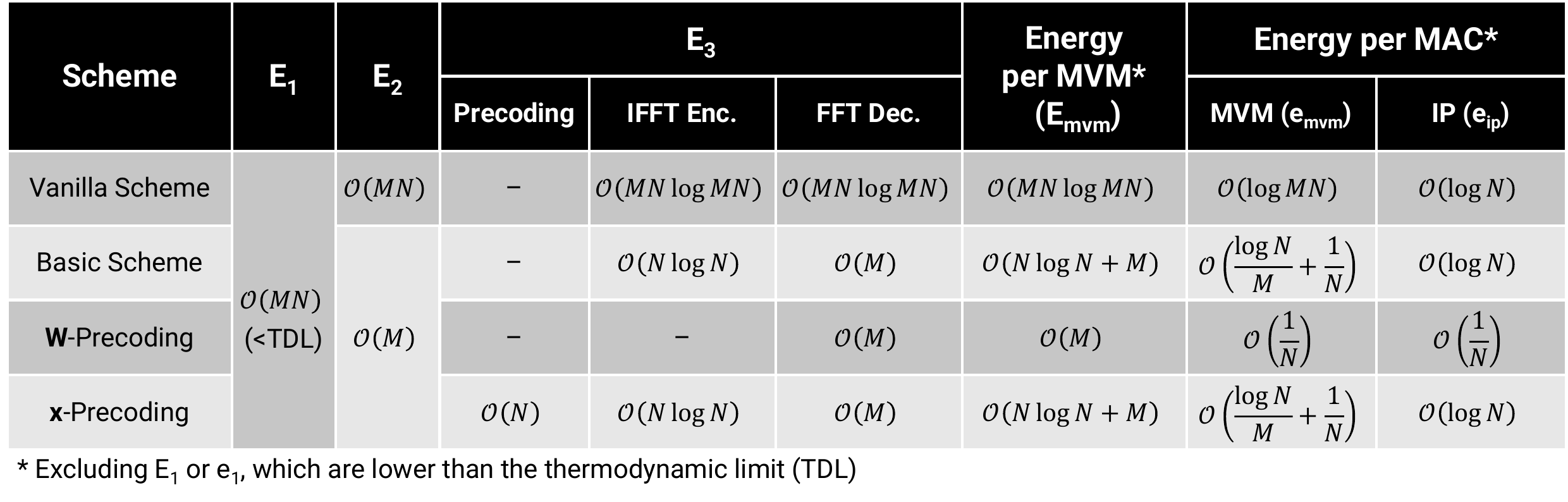}
    \end{tabular}
\end{table}
%% figure ends 

Therefore, {\name} incorporates several optimization strategies to significantly reduce the energy consumption for in-physics MVM computation.
As shown in \autoreftab{fig:supp-energy-table}, there are three schemes for {\name}:
(\emph{i}) the basic scheme without wireless channel precoding, designed for the wired case or when the CSI is not available;
(\emph{ii}) the $\weightMat$-precoding scheme, which precodes the $\weightMat$ on the central radio without incurring additional energy consumption on the clients while further reducing the energy consumption by time-encoding $\inputVec$;
(\emph{iii}) the $\inputVec$-precoding scheme, which precodes $\inputVec$ on the client that supports individual CSI for each client for higher computing accuracy.
The $\weightMat$-precoding scheme is evaluated in Section~\ref{sec: main-result}, and the detailed performance of the other two schemes can be found in Supplementary Section~\ref{sec:supplementary-experiment-ml-calibration}.

%%%%%
%%%%%
\section{{\namebf}'s Basic Scheme}
\label{sec:supplementary-theory-basic-scheme}

We first present a basic scheme of {\name} for a wired/cabled channel that significantly optimizes the energy efficiency of the in-physics MVM described in Supplementary Section~\ref{sec:supplementary-theory-vanilla-maft} via three techniques:
(\emph{i}) encoding FFT size reduction,
(\emph{ii}) ADC sampling rate reduction, and
(\emph{iii}) MVM decomposition.
We also introduce zero-subcarrier padding and cyclic prefix to overcome two practical issues stemming from the three techniques.

\subsection{Encoding FFT Size Reduction}
The first technique reduces the FFT size from $\inputSize\outputSize$ to $\inputSize$ to save the number of real-valued MACs required for encoding $\specInputVec$ into $\sampInputVec$.
Note from equation {\eqref{eq: original-mvm-spec-to-samp-input}} that the generated waveform $\sampInputVec$ is independent of the output size, $\outputSize$, and is with a period of $\inputSize$ samples, i.e.,
\begin{align}
    \sampInput[\sampIdx + \inputSize]
    & = \frac{1}{\inputSize\outputSize} \sum_{\inputIdx=0}^{\inputSize-1} \inputElem_{\inputIdx} \cdot \eu^{-\iu 2\pi \frac{\inputIdx + \inputSize/2}{\inputSize}(\sampIdx+\inputSize)} \nonumber \\
    & = \frac{1}{\inputSize\outputSize} \sum_{\inputIdx=0}^{\inputSize-1} \inputElem_{\inputIdx} \cdot \eu^{-\iu 2\pi \frac{\inputIdx + \inputSize/2}{\inputSize}\sampIdx} = \sampInput[\sampIdx],\
    \forall \sampIdx = 0, 1, \dots, \inputSize\outputSize-\inputSize-1.
\end{align}
Therefore, we only need to generate the first $\inputSize$ I/Q samples in $\sampInputVec$, which can then be repeated for $\outputSize$ times to obtain $\sampInputVec$. The generation of the first $\inputSize$ I/Q samples has a similar form as an $\inputSize$-point IDFT as
\begin{align}
    \sampInputVec &= \frac{1}{\outputSize\sqrt{\inputSize}} \cdot \underbrace{\Big[ \idftMat \cdot \shiftMat^{\inputSize/2} \cdot \inputVec , \dots, \idftMat \cdot \shiftMat^{\inputSize/2} \cdot \inputVec \Big]}_{\textrm{repeated}~ \outputSize~\textrm{times}},
    \label{eq: basic-scheme-input-to-samp}
\end{align}
but requires only an $\inputSize$-point IFFT involving $2 \inputSize \log_2 \inputSize$ real-valued MACs.

\subsection{ADC Sampling Rate Reduction}
An RX operating at the sampling rate of $(2\inputSize\outputSize-1) \cdot \freqSub$ is required to capture a total number of $(2\inputSize\outputSize-1)$ I/Q samples of $\sampOutputVec$ in order to recover the full spectral information of $\specOutputVec$. This process incurs significant energy consumption on the waveform reception term $\energyMACADC$ on ADC, and the digital computing term $\energyMACDec$ for FFT.
Fortunately, in equation {\eqref{eq: original-mvm-spec-to-output}}, we notice that the output vector $\outputVec$ can be demodulated from a set of consecutive subcarriers located in the middle of the spectrum of $\specOutputVec$, indexed from $(\inputSize\outputSize-\outputSize)$ to $(\inputSize\outputSize-1)$. Let $\specOutputVecLow = [\specOutputLow[\specIdx]] \in \mathbb{C}^{\outputSize}$, where $\specOutputLow[\specIdx] = \specOutput[\inputSize\outputSize-\outputSize+\specIdx], \forall \specIdx = 0, 1, \dots, \outputSize-1$, equation {\eqref{eq: original-mvm-spec-to-output}} can be written as
\begin{align}
    \outputElem_{\outputIdx} = \specOutputLow[\outputSize-1-\outputIdx],\ \forall \outputIdx = 0, 1, \dots, \outputSize-1.
    \label{eq: basic-scheme-samp-to-output}
\end{align}
These $\outputSize$ subcarriers in $\specOutputVecLow$ only occupy a narrow bandwidth of $\bandLow = \outputSize \cdot \freqSub$, which is a fraction $1/\inputSize$ of the original signal bandwidth of $\weightMat$ and $\inputVec$.
Therefore, one can employ an LPF with a cutoff frequency of $\bandLow/2$ and an ADC with a sampling rate of as small as $\sampRateLow = \bandLow$ to capture the waveform $\waveOutput(\waveIdx)$, which yields
\begin{align}
    \sampOutputVecLow = [\sampOutputLow[\sampIdx]] = \ADC{\LPF{\waveOutput(\waveIdx)}},\ \forall \waveIdx \in [0, \waveLen),\ \sampIdx = 0, 1, \dots, \outputSize-1.
    \label{eq: basic-scheme-wave-to-samp}
\end{align}
In this way, only $\outputSize$ I/Q samples are captured, and the decoding of $\outputVec$ from $\specOutputVecLow$ can be done using an $\outputSize$-point FFT following equation {\eqref{eq: original-mvm-samp-to-spec-output}} as
\begin{align}
    \specOutputVecLow = \sqrt{\outputSize} \cdot \shiftMat^{\outputSize/2} \cdot \dftMat \cdot \sampOutputVecLow,\ \textrm{where}~
    \specOutputLow[\specIdx] = \sum_{\sampIdx=0}^{\outputSize-1} \sampOutput[\sampIdx] \cdot \eu^{-\iu 2\pi \frac{\specIdx - \outputSize/2}{\outputSize}\sampIdx},\
    \forall \specIdx = 0, 1, \dots, \outputSize-1.
    \label{eq: basic-scheme-samp-to-spec}
\end{align}
In this way, only $\outputSize$ I/Q samples need to be captured by the ADC on the RX side, and the subsequent decoding of $\outputVec$ involves $2\outputSize\log_2 \outputSize$ real-valued MACs.

\subsection{MVM Decomposition}

%% figure begins
\begin{figure*}[!t]
    \centering
    \includegraphics[width=1.0\textwidth]{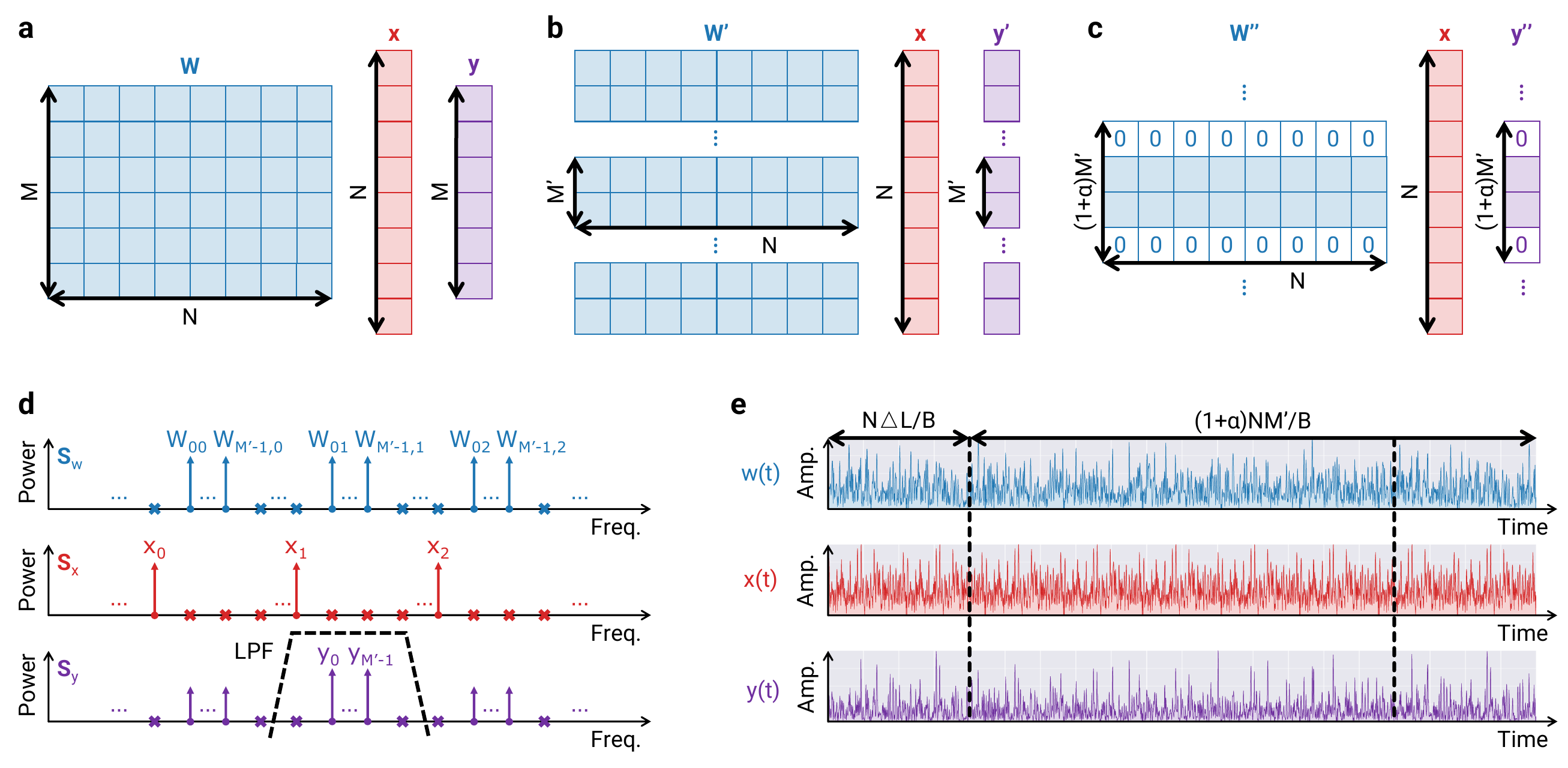}
    \caption{
    \textbf{The step-by-step waveform generation of {\namebf}'s MVM computation.}
    \textbf{a}, The original matrix-vector multiplication (MVM), $\outputVec = \weightMat \cdot \inputVec$, with input size $\inputSize$ and output size $\outputSize$. 
    \textbf{b}, Decomposition of the original MVM into small MVMs along the output dimension using a reduced output size, $\outputSizeNew$, which lowers the complexity of DFT-based decoding.
    \textbf{c}, Zero-padding is applied to $\weightMat$ and $\outputVec$, introducing all-zero rows that correspond to zero-subcarriers in the signal spectrum $\specWeightVec$ and $\specOutputVecLow$.
    \textbf{d}, The inserted zero-subcarriers in $\specOutputVecLow$ effectively mitigate the roll-off effect introduced by the low-pass filter (LPF).
    \textbf{e}, In the time domain, a cyclic prefix is added prior to the signal to mitigate potential synchronization errors.}
    \label{fig:supp-mvm-splitting}
\end{figure*}
%% figure ends 

Furthermore, we can decompose the large MVM $\outputVec = \weightMat \cdot \inputVec$ in the output dimension $\outputSize$ into $\outputSize/\outputSizeNew$ smaller MVMs, $\outputVecNew = \weightMatNew \cdot \inputVec$, as shown in 
\autorefsubfig{fig:supp-mvm-splitting}{a--b}, with $\weightMatNew \in \mathbb{C}^{\outputSizeNew \times \inputSize}$ and $\outputVecNew \in \mathbb{C}^{\outputSizeNew}$.
Since each smaller MVM requires only an $\outputSizeNew$-point FFT for decoding that involves $2\outputSizeNew \log_{2} \outputSizeNew$ MACs, the total number of MACs required for all $\outputSize/\outputSizeNew$ MVMs reduces to $2\outputSize \log_2 \outputSizeNew$.
Note that the waveform time for a single decomposed MVM is $\outputSizeNew\inputSize/\band$, and the total computation time for the original MVM remains $\frac{\outputSize}{\outputSizeNew} \cdot \frac{\outputSizeNew\inputSize}{\band} = \frac{\outputSize\inputSize}{\band}$, which matches the waveform duration $\waveLen$ of the original MVM computation without decomposition. Therefore, this MVM decomposition incurs no additional overhead in waveform time, and ensures that the energy consumption and computation throughput remain unchanged.
This MVM decomposition also proportionally reduces the number of subcarriers in both the TX channel of $\specWeightVec$ and $\specInputVec$, i.e., $\inputSize\outputSizeNew$ subcarriers, and the RX channel of $\specOutputVecLow$, i.e., $\outputSizeNew$ subcarriers. As a result, the downsampling ratio from TX to RX remains $\inputSize$, and given the same available bandwidth $\band$, the TX and RX sampling rates remain unchanged.

\subsection{Zero-Subcarrier Padding}

To mitigate the frequency aliasing effects, the ADC operating at a low sampling rate of $\outputSizeNew\freqSub$ relies on an ideal LPF to filter out frequency components outside of $[-\outputSizeNew\freqSub/2, +\outputSizeNew\freqSub/2]$.
An ideal LPF has a brick-wall shape, corresponding to a flat frequency response across the passband, and ``zero" frequency response elsewhere. However, such a brick-wall LPF is not practical due to its non-casualty in the time domain; rather, a practical LPF before ADC usually exhibits non-negligible roll-off effects around its cutoff frequency at $\outputSizeNew\freqSub/2$, i.e., there exists a transient frequency range around $\outputSizeNew\freqSub/2$ where the frequency response gradually drops to zero (see Supplementary Section~\ref{sec:supplementary-experiment-setup} for the detailed measurements).

To overcome the LPF's roll-off effect, zero subcarriers that carry no symbols are padded to the LPF's transient frequency. Consider a padded weight matrix $\weightMatPad \in \mathbb{C}^{(\outputSizeNew+2\outputSizePad)\times \inputSize}$, where $\outputSizePad$ rows of 0's are padded to the top and bottom of the decomposed weight matrix $\weightMatNew$, as shown in \autorefsubfig{fig:supp-mvm-splitting}{c}. As a result, the output vector becomes $\outputVecPad \in \mathbb{C}^{\outputSizeNew+2\outputSizePad}$ with $\outputSizePad$ 0's padded to the beginning and end of the vector. This process can be written as
\begin{align}
    \weightElemPad_{\outputIdx, \inputIdx} = 
    \begin{cases}
        \weightElemNew_{(\outputIdx-\outputSizePad), \inputIdx}, & \textrm{if}~\outputSizePad \leq \outputIdx < \outputSizeNew + \outputSizePad, \\
        0, & \textrm{otherwise},
    \end{cases}
    ~\textrm{and}~
    \outputElemPad_{\outputIdx} = 
    \begin{cases}
        \outputElemNew_{\outputIdx-\outputSizePad}, & \textrm{if}~\outputSizePad \leq \outputIdx < \outputSizeNew + \outputSizePad, \\
        0, & \textrm{otherwise}.
    \end{cases}
\end{align}
Based on equation {\eqref{eq: basic-scheme-samp-to-output}}, the received output signal spectrum after LPF and with padded zero subcarriers $\specOutputVecLow$ satisfies
\begin{align}
    \specOutputLow[\outputIdx] = 
    \begin{cases}
        \outputElemNew_{\outputSizeNew-1-\outputIdx+\outputSizePad}, & \textrm{if}~\outputSizePad \leq \outputIdx < \outputSizeNew + \outputSizePad, \\
        0, & \textrm{otherwise},
    \end{cases}
\end{align}
which implies that the output spectrum $\specOutputVecLow$ has $\outputSizePad$ zero subcarriers on both edges, i.e., the transient frequency range of a practical LPF, as shown in \autorefsubfig{fig:supp-mvm-splitting}{d}.
Let $\overheadPad = 2\outputSizePad/\outputSizeNew$ denote the overhead coefficient associated with zero subcarrier padding, the number of subcarriers per decomposed MVM is $(1+\overheadPad)\inputSize\outputSizeNew$ for $\specInputVec$ and $\specWeightVec$, and $(1+\overheadPad)\outputSizeNew$ for $\specOutputVecLow$.
In general, a larger value of $\overheadPad$ is required if the LPF exhibits a larger transient frequency range, leading to a larger overhead on the waveform duration that is inversely proportional to the subcarrier spacing, $\freqSub$.

\subsection{Cyclic Prefix}

So far, we assume that the client's DACs on the TX side and ADCs on the RX side are perfectly synchronized when generating and capturing $\sampInputVec$ and $\sampOutputVecLow$. The assumption does not hold in practice, especially when the RX employed an ADC operating at a reduced sampling rate with a downsampling ratio of $\inputSize$.
Inspired by OFDM-based wireless communication systems, we introduce a cyclic prefix to mitigate the potential delay and timing offset between the DAC and ADC, as shown in \autorefsubfig{fig:supp-mvm-splitting}{e}.
Different from conventional OFDM-based communication systems, the downsampling ratio of $\inputSize$ means that every $\inputSize$ I/Q sample in $\sampInputVec$ and $\sampWeightVec$ corresponds to a single I/Q sample in $\sampOutputVecLow$.
To ensure an integer number of I/Q samples for the cyclic prefix removal on $\sampOutputVecLow$, the cyclic prefix length on $\sampInputVec$ and $\sampWeightVec$ must be a multiple of $\inputSize$.
Hereby, we consider a cyclic prefix length of $\fftSizeCP \in \mathbb{N}$ for $\sampOutputVecLow$, and thus the cyclic prefix length is $\inputSize \cdot \fftSizeCP$ on $\sampInputVec$ and $\sampWeightVec$.
Then, the cyclic prefix attachment from $\sampInputVec$ to $\sampInputVecCP \in \mathbb{C}^{(1+\overheadPad)\inputSize\outputSizeNew+\inputSize\fftSizeCP}$ can be written as
\begin{align}
    \sampInputCP[\sampIdx]
    \begin{cases}
        \sampInput[(1+\overheadPad)\inputSize\outputSizeNew+\sampIdx-\inputSize\fftSizeCP],\ \textrm{if}~\sampIdx < \inputSize\fftSizeCP, \\
        \sampInput[\sampIdx-\inputSize\fftSizeCP],\ \textrm{if}~\sampIdx \geq \inputSize\fftSizeCP,
    \end{cases}\
    \forall \sampIdx = 0, 1, \dots, (1+\overheadPad)\inputSize\outputSizeNew+\inputSize\fftSizeCP-1.
\end{align}
Similarly, the cyclic prefix attachment from $\sampWeightVec$ to $\sampWeightVecCP \in \mathbb{C}^{(1+\overheadPad)\inputSize\outputSizeNew+\inputSize\fftSizeCP}$ can be written as
\begin{align}
    \sampWeightCP[\sampIdx] = 
    \begin{cases}
        \sampWeight[(1+\overheadPad)\inputSize\outputSizeNew+\sampIdx-\inputSize\fftSizeCP],\ \textrm{if}~\sampIdx < \inputSize\fftSizeCP, \\
        \sampWeight[\sampIdx-\inputSize\fftSizeCP],\ \textrm{if}~\sampIdx \geq \inputSize\fftSizeCP,
    \end{cases}\
    \forall \sampIdx = 0, 1, \dots, (1+\overheadPad)\inputSize\outputSizeNew+\inputSize\fftSizeCP-1.
\end{align}
On the RX side, only the first $(1+\overheadPad)\outputSizeNew$ I/Q samples need to be captured.
Similarly, we define an overhead coefficient $\overheadCP=\fftSizeCP/(\outputSizeNew+2\outputSizePad)$ such that the length of $\sampInputVec$ and $\sampWeightVec$ per decomposed MVM is $(1+\overheadPad)(1+\overheadCP)\inputSize\outputSizeNew$, and the length of $\sampOutputVecLow$ is $(1+\overheadPad)(1+\overheadCP)\outputSizeNew$.

\subsection{Energy Efficiency Analysis}
Similar to the energy efficiency analysis in Supplementary Section~\ref{sec:supplementary-theory-vanilla-maft}, the energy consumption for the {\name} basic scheme contains three components:
(\emph{i}) $\energyMVMTx$ for the waveform generation of $\waveInput(\waveIdx)$ and I/Q (de)modulation, (\emph{ii}) $\energyMVMADC$ for the I/Q sampling of waveform $\waveOutput(\waveIdx)$ using two ADCs operating at reduced sampling rate after I/Q demodulation, and (\emph{iii}) $\energyMVMDec$ for the digital computing based encoding and decoding.

%% figure begins
\begin{figure*}[!t]
    \centering
    \includegraphics[width=0.9\textwidth]{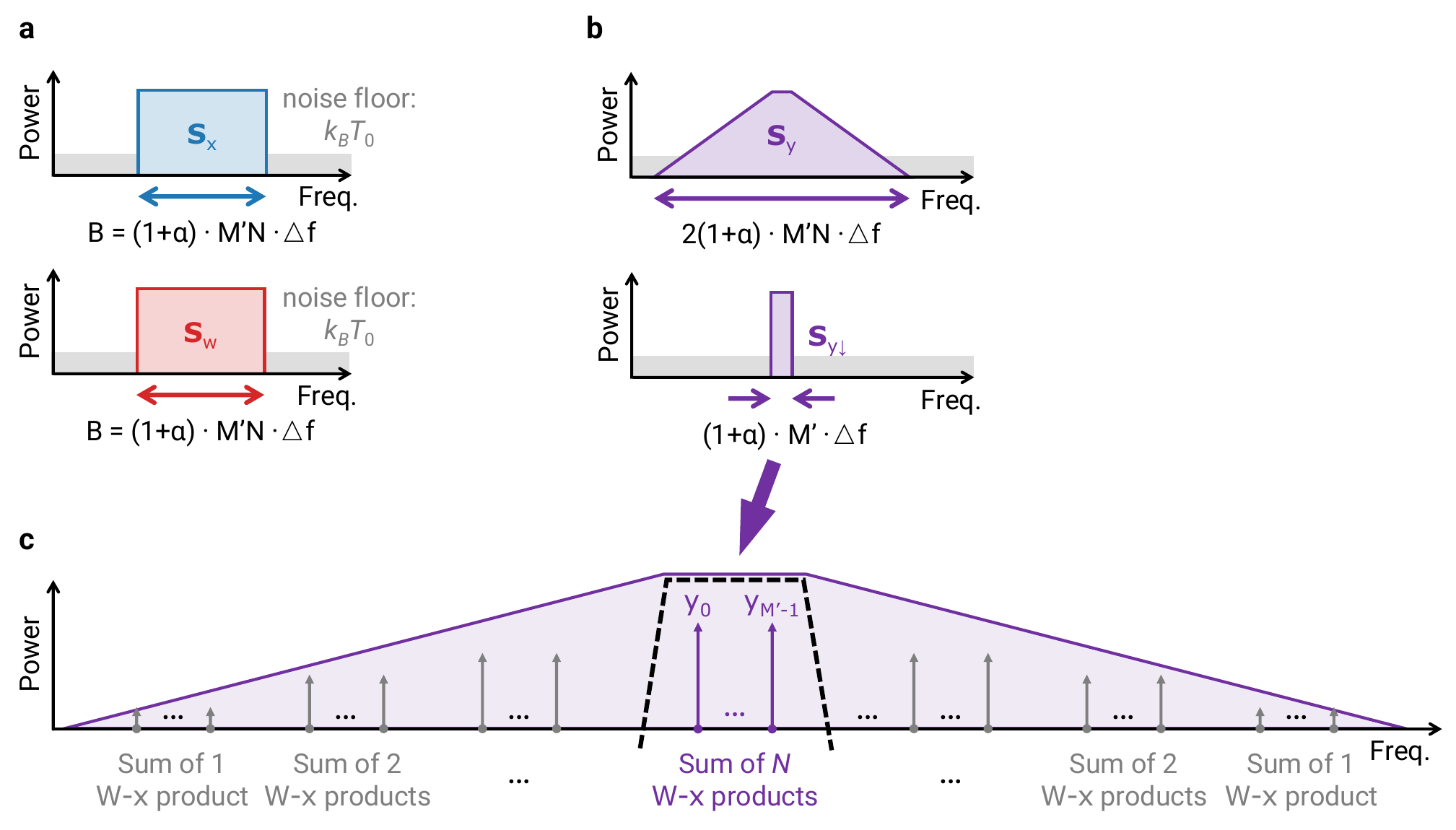}
    \caption{
    \textbf{The frequency-domain power spectral density comparison of $\specInputVec$, $\specWeightVec$, $\specOutputVec$ and $\specOutputVecLow$.}
    \textbf{a}, The spectrum of $\specInputVec$ and $\specWeightVec$ occupies a signal bandwidth of $\inputSize\outputSizeNew \cdot \freqSub$.
    \textbf{b}, The spectrum of the computing mixer's output signal, $\specOutputVec$, occupies a signal bandwidth of $(2\inputSize\outputSizeNew-1) \cdot \freqSub$, while the region of interest carrying information about $\specOutputVecLow$ is confined to a smaller bandwidth of $\outputSizeNew \cdot \freqSub$.
    \textbf{c}, A zoomed-in view of the output signal spectrum, $\specOutputVec$, which contains $\specOutputVecLow$ in the center of the bandwidth.}
    \label{fig:supp-power-density}
\end{figure*}
%% figure ends 

First, the new waveform duration for each decomposed MVM including the overhead is $(1+\overheadPad)(1+\overheadCP)\inputSize\outputSizeNew/\band$. Given the transmit power of $\powerTx$, the total energy required for the client radio to generate the waveforms carrying the total number of $\outputSize/\outputSizeNew$ inference requests is given by
\begin{align}
    \energyMVMTx = \frac{\outputSize}{\outputSizeNew} \cdot (\efficiencyTX)^{-1} \powerTx \cdot \frac{(1+\overheadPad)(1+\overheadCP) \cdot \inputSize\outputSizeNew}{\band} = (\efficiencyTX)^{-1} \powerTx \cdot \frac{(1+\overheadPad) (1+\overheadCP) \cdot \inputSize\outputSize}{\band}.
    \label{eq: basic-scheme-energy-tx-power}
\end{align}
Different from Supplementary Section~\ref{sec:supplementary-theory-vanilla-maft}, the SNR in the {\name} basic scheme is measured within the captured narrowband of $[-(1+\overheadPad)\outputSizeNew\freqSub/2, +(1+\overheadPad)\outputSizeNew\freqSub/2]$.
As illustrated in \autorefsubfig{fig:supp-power-density}{a--b}, the original $\specOutputVec$ spans over a bandwidth of $2(1+\overheadPad) \inputSize\outputSizeNew \cdot \freqSub$ with the total power of $\efficiencyMixer \cdot \powerTx$, while the bandwidth of interest is only the portion of $(1+\overheadPad) \outputSizeNew \cdot \freqSub$.
Note that the power is not evenly distributed over the subcarriers in $\specOutputVec$; each subcarrier $\specOutput[\specIdx]$ is the sum of multiple products of $\inputElem$-$\weightElem$ pairs, according to the convolution theorem.
Hereby, we assume all the products of the $\inputElem$-$\weightElem$ pairs are independent and identically distributed (i.i.d.). After the linear convolution based on equation {\eqref{eq: convolution-definition}}, the first $\outputSizeNew$ elements on the output $\specOutputVec$ (excluding the $\overheadPad\outputSizeNew$ padded zero subcarriers) only have one such product, the second $\outputSizeNew$ elements are the sum of two of such products, and so on, until the middle $\outputSizeNew$ elements, which is the sum of $\inputSize$ of such products as the captured $\specOutputVecLow$ or $\outputVec$. The second half of the subcarriers on $\specOutputVec$ follows the same trend but with a reversed symmetry compared to the first half of the subcarriers. According to the law of large numbers and the central limit theorem together with $\inputSize \gg 1$, elements of $\specOutputVec$ follow Gaussian distributions, whose variance (or power) is proportional to the number of products to be summed. Therefore, the power density spectrum of $\specOutputVec$ has a ``triangle'' shape, as shown in \autorefsubfig{fig:supp-power-density}{c}. Moreover, the total power of the middle $\outputSizeNew$ subcarriers of interest, i.e., $\specOutputVecLow$, is $1/\inputSize$ of the power of $\specOutputVec$. Therefore, the received power $\powerRx$ within the narrowband can be approximated by
\begin{align}
    \powerRx \approx \frac{1}{\inputSize} \cdot \efficiencyMixer \cdot \powerTx.
    \label{eq: basic-scheme-power-signal}
\end{align}
On the other hand, the noise power only spans the narrow band of $(1+\overheadPad)\outputSizeNew\freqSub$, so the noise power, $\powerNoise$, is given by
\begin{align}
    \powerNoise = (\efficiencyRX)^{-1} \cdot \boltzmann \temperature \cdot (1+\overheadPad) \cdot \outputSizeNew \cdot \freqSub.
    \label{eq: basic-scheme-power-noise}
\end{align}
Combining equations {\eqref{eq: basic-scheme-power-signal}} and {\eqref{eq: basic-scheme-power-noise}}, the relationship between $\powerTx$ and $\textsf{SNR}$ is given by
\begin{align}
    \textsf{SNR}
    = \frac{\powerRx}{\powerNoise} = \frac{{\inputSize}^{-1} \cdot \efficiencyMixer \cdot \powerTx}{(\efficiencyRX)^{-1} \cdot \boltzmann \temperature \cdot (1+\overheadPad) \cdot \outputSizeNew \cdot \freqSub} = \frac{\efficiencyMixer \cdot \efficiencyRX \cdot \powerTx}{\boltzmann \temperature \cdot \band} \Rightarrow \powerTx = (\efficiencyMixer \cdot \efficiencyRX)^{-1} \cdot \textsf{SNR} \cdot \boltzmann \temperature \cdot \band.
    \label{eq: basic-scheme-energy-tx-snr}
\end{align}
Plugging $\powerTx$ in equation {\eqref{eq: basic-scheme-energy-tx-power}} yields
\begin{align}
    \energyMVMTx = (1+\overheadPad) (1+\overheadCP) \cdot \inputSize\outputSize \cdot \efficiencyTRX^{-1} \cdot \textsf{SNR} \cdot \boltzmann \temperature.
    \label{eq: basic-scheme-energy-per-mvm-tx}
\end{align}

For each decomposed MVM, only $(1+\overheadPad)\outputSizeNew$ I/Q samples in $\sampOutputVecLow$ need to be captured by a pair of ADCs operating at a low sampling rate.
Therefore, the total energy consumption $\energyMVMADC$ across all $\outputSize/\outputSizeNew$ decomposed MVMs is given by
\begin{align}
    \energyMVMADC = \frac{\outputSize}{\outputSizeNew} \cdot (1+\overheadPad)\outputSizeNew \cdot 2 e_{\textrm{adc}} = 2(1+\overheadPad) \cdot \outputSize \cdot e_{\textrm{adc}}.
    \label{eq: basic-scheme-energy-per-mvm-adc}
\end{align}

For the encoding energy part of $\energyMVMDec$, it reduces to an $\inputSize$-point IFFT for encoding following equation~{\eqref{eq: basic-scheme-input-to-samp}}, including $2\inputSize \log_2(\inputSize)$ MACs, which needs to be performed only once and can be reused for all the decomposed MVMs. In addition, the decoding energy part of $\energyMVMDec$ in equation {\eqref{eq: basic-scheme-samp-to-spec}} consumes $(1+\overheadPad)\outputSizeNew$-point FFT per decomposed MVM. Therefore, $\energyMVMDec$ including both the encoding and decoding energy is given by
\begin{align}
    \energyMVMDec = \left( 2\inputSize \cdot \log_2 \inputSize + \frac{\outputSize}{\outputSizeNew} \cdot 2 (1+\overheadPad)\outputSizeNew \log_2 ((1+\overheadPad)\outputSizeNew) \right) \cdot e_{\textrm{dig}} = \left( 2\inputSize \cdot \log_2 \inputSize + 2(1+\overheadPad) \cdot \outputSize \cdot \log_2 ((1+\overheadPad)\outputSizeNew) \right) \cdot e_{\textrm{dig}}.
    \label{eq: basic-scheme-energy-per-mvm-digital}
\end{align}

Putting equations {\eqref{eq: basic-scheme-energy-per-mvm-tx}}, {\eqref{eq: basic-scheme-energy-per-mvm-adc}} and {\eqref{eq: basic-scheme-energy-per-mvm-digital}} together, the total energy consumption $\energyMVM$ is given by
\begin{align}
    \energyMVM = \energyMVMTx + \energyMVMADC + \energyMVMDec
    & = \underbrace{(1+\overheadPad) (1+\overheadCP) \cdot \inputSize\outputSize \cdot \efficiencyTRX^{-1} \cdot \textsf{SNR} \cdot \boltzmann \temperature}_{\energyMVMTx} + \underbrace{2(1+\overheadPad) \cdot \outputSize \cdot e_{\textrm{adc}}}_{\energyMVMADC} \nonumber \\
    & \quad + \underbrace{\left( 2\inputSize \cdot \log_2 \inputSize + 2(1+\overheadPad) \cdot \outputSize \cdot \log_2 ((1+\overheadPad)\outputSizeNew) \right) \cdot e_{\textrm{dig}}}_{\energyMVMDec}.
    \label{eq: basic-scheme-energy-per-mvm}
\end{align}
The corresponding energy efficiency, measured by energy per MAC, $\energyMAC$, is
\begin{align}
    \energyMAC = \frac{\energyMVM}{4\inputSize\outputSize} = \energyMACTx + \energyMACADC + \energyMACDec
    & = \underbrace{\frac{(1+\overheadPad) (1+\overheadCP)}{4} \cdot \efficiencyTRX^{-1} \cdot \textsf{SNR} \cdot \boltzmann \temperature}_{\energyMACTx} + \underbrace{\frac{1+\overheadPad}{2\inputSize} \cdot e_{\textrm{adc}}}_{\energyMACADC} \nonumber \\
    &\quad + \underbrace{\left(\frac{1}{2\outputSize} \cdot \log_2 \inputSize + \frac{1+\overheadPad}{2\inputSize} \cdot \log_2 ((1+\overheadPad)\outputSizeNew) \right) \cdot e_{\textrm{dig}}}_{\energyMACDec}.
    \label{eq: basic-scheme-energy-per-mac}
\end{align}

From equation {\eqref{eq: basic-scheme-energy-per-mac}}, it can be seen that the energy efficiency term $\energyMACTx$ is independent of the MVM dimension, and the terms $\energyMACADC$ and $\energyMACDec$ respectively scale as $\complexity(\frac{1}{\inputSize})$ and $\complexity(\frac{\log\inputSize}{\outputSize}+\frac{1}{\inputSize})$.
As a result, as the MVM dimensions increase, i.e.,  $\inputSize \rightarrow \infty$ and $\outputSize \rightarrow \infty$, $\energyMAC$ approaches $\energyMACTx$, which is bottlenecked by the thermal noise and hardware limit, thus achieving significantly enhanced energy efficiency compared to digital computing.
Moreover, under an ideal hardware with $\efficiencyTRX=1$ and $\overheadPad=\overheadCP=0$, we can derive {\name}'s thermal dynamic limit (TDL) as
\begin{align}
    \energyMACTDL :=
    \lim_{\inputSize \to +\infty, \outputSize \to +\infty} \energyMAC = \textsf{SNR} \cdot k T_{0}/4.
    \label{eq: basic-scheme-energy-per-mac-tdl}
\end{align}
This TDL can even exceed the energy efficiency for $b$-bit computation at the Landauer Limit, given by $e_{\textrm{Landauer}} = b^{2}\cdot\ln{2} \cdot k T_{0}$, when $\textsf{SNR}/4 < b^{2} \ln{2}$, or $\textsf{SNR} < 2.77 b^{2}$.

\subsection{MVM Decomposition into IPs}
According to equation {\eqref{eq: basic-scheme-energy-per-mac}}, while a decomposed MVM with a smaller value of $\outputSizeNew$ reduces the energy consumption term $\energyMACDec$, it usually requires a large overhead, i.e., larger values of $\overheadPad$ and/or $\overheadCP$.
For the extreme MVM decomposition case with $\outputSizeNew=1$, the original MVM is decomposed into $\outputSize$ IPs. We denote the total energy consumption and energy efficiency under this extreme decomposition as $\energyMVMIPMult$ and $\energyMACIPMult$, respectively.
Since the number of padded zero subcarriers must be a positive integer, we set the minimum $\outputSizePad=1$ that yields a frequency-domain overhead coefficient of $\overheadPad=2$.
In this case, the decoding process in equation {\eqref{eq: basic-scheme-samp-to-spec}} becomes a three-point FFT ($(1+\overheadPad)\outputSizeNew=3$), where the IP result is carried by the middle subcarrier, $\outputElem_{0} = \specOutputLow[1]$. This means the energy consumption term $\energyMVMDec$ given by equation {\eqref{eq: basic-scheme-energy-per-mvm-digital}} does not hold. Instead, the decoding process from $\sampOutputVecLow \in \mathbb{C}^{3}$ to $\specOutputLow[1]$ can be rewritten as
\begin{align}
    \outputElem_{0} = \specOutputLow[1] = \sum_{\sampIdx=0}^{2} \sampOutput[\sampIdx] \cdot \eu^{-\iu 2\pi \frac{1 - 3/2}{3}\sampIdx} = \sampOutput[0] + \sampOutput[1] \cdot \eu^{\iu \frac{\pi}{3}} + \sampOutput[2] \cdot \eu^{\iu \frac{2\pi}{3}},
    \label{eq: basic-scheme-three-point-fft}
\end{align}
which contains two complex-valued MACs or, equivalently, eight real-valued MACs.
In addition, the synchronization algorithm in Supplementary Section~\ref{sec:supplementary-experiment-setup} ensures the sub-sample-level of timing synchronization between the TX and RX sides, and the cyclic prefix with $\fftSizeCP=1$ is sufficient, corresponding to $\overheadCP=\fftSizeCP/(\outputSizeNew+2\outputSizePad)=1/3$.
By plugging $\overheadPad=2$ and $\overheadCP=1/3$ for other two terms $\energyMVMTx$ and $\energyMVMADC$, the energy required for the MVM computation, $\energyMVMIPMult$, is given by
\begin{align}
    \energyMVMIPMult = \underbrace{4 \inputSize\outputSize \cdot \efficiencyTRX^{-1} \cdot \textsf{SNR} \cdot \boltzmann \temperature}_{\energyMVMTx} + \underbrace{6\outputSize \cdot e_{\textrm{adc}}}_{\energyMVMADC} + \underbrace{(2\inputSize \cdot \log_2 \inputSize + 8\outputSize) \cdot e_{\textrm{dig}}}_{\energyMVMDec},
    \label{eq: basic-scheme-energy-mvm-multiple}
\end{align}
which corresponds to an energy efficiency of
\begin{align}
    \energyMACIPMult = \frac{\energyMVMIPMult}{4\inputSize\outputSize} = \underbrace{\efficiencyTRX^{-1} \cdot \textsf{SNR} \cdot \boltzmann \temperature}_{\energyMACTx} + \underbrace{\frac{3}{2\inputSize} \cdot e_{\textrm{adc}}}_{\energyMACADC} + \underbrace{\left( \frac{1}{2\outputSize} \cdot \log_2 \inputSize + \frac{2}{\inputSize} \right) \cdot e_{\textrm{dig}}}_{\energyMACDec}.
    \label{eq: basic-scheme-energy-mac-multiple}
\end{align}

In the special case with $\outputSize=1$, the MVM is reduced to a standalone IP computation task given by $c = \innerproduct{\mathbf{a}}{\mathbf{b}} = \sum_{\inputIdx=1}^{\inputSize} a_{\inputIdx} \cdot \overline{b_{\inputIdx}}$. The total energy consumption for a standalone IP, denoted by $\energyMVMIPSing$, can be obtained by plugging $\outputSize=1$ into equation {\eqref{eq: basic-scheme-energy-mvm-multiple}}, i.e.,
\begin{align}
    \energyMVMIPSing = \underbrace{4 \inputSize \cdot \efficiencyTRX^{-1} \cdot \textsf{SNR} \cdot \boltzmann \temperature}_{\energyMVMTx} + \underbrace{6 \cdot e_{\textrm{adc}}}_{\energyMVMADC} + \underbrace{(2\inputSize \cdot \log_2 \inputSize + 8) \cdot e_{\textrm{dig}}}_{\energyMVMDec}.
    \label{eq: basic-scheme-energy-mvm-single}
\end{align}
The corresponding energy efficiency, denoted by $\energyMACIPSing$, can then be derived by plugging $\outputSize=1$ into equation {\eqref{eq: basic-scheme-energy-mac-multiple}}, i.e.,
\begin{align}
    \energyMACIPSing = \frac{\energyMVMIPSing}{4\inputSize} = \underbrace{\efficiencyTRX^{-1} \cdot \textsf{SNR} \cdot \boltzmann \temperature}_{\energyMACTx} + \underbrace{\frac{3}{2\inputSize} \cdot e_{\textrm{adc}}}_{\energyMACADC} + \underbrace{\left( \frac{1}{2} \cdot \log_2 \inputSize + \frac{2}{\inputSize} \right) \cdot e_{\textrm{dig}}}_{\energyMACDec}.
    \label{eq: basic-scheme-energy-mac-single}
\end{align}
Note that since $\left( \frac{1}{2} \cdot \log_2 \inputSize + \frac{2}{\inputSize} \right) \cdot e_{\textrm{dig}} > e_{\textrm{dig}}$, $\energyMACIPSing$ is always higher than $e_{\textrm{dig}}$. This is because the IFFT-based encoding requires $2\inputSize \log_2 \inputSize$ MACs, which is higher energy consumption than directly computing the IP itself and cannot be amortized compared to the MVM task as $\outputSize$ scales. 
Fortunately, this limitation can be resolved in the $\weightMat$-precoding scheme, described in Supplementary Section~\ref{sec:supplementary-theory-weight-precoding-scheme}.

%%%%%
%%%%%
\section{{\namebf}'s $\weightMat$-Precoding Scheme: Wireless Channel Calibration at the Central Radio}
\label{sec:supplementary-theory-weight-precoding-scheme}

%% figure begins
\begin{figure*}[!t]
    \centering
    \includegraphics[width=0.95\textwidth]{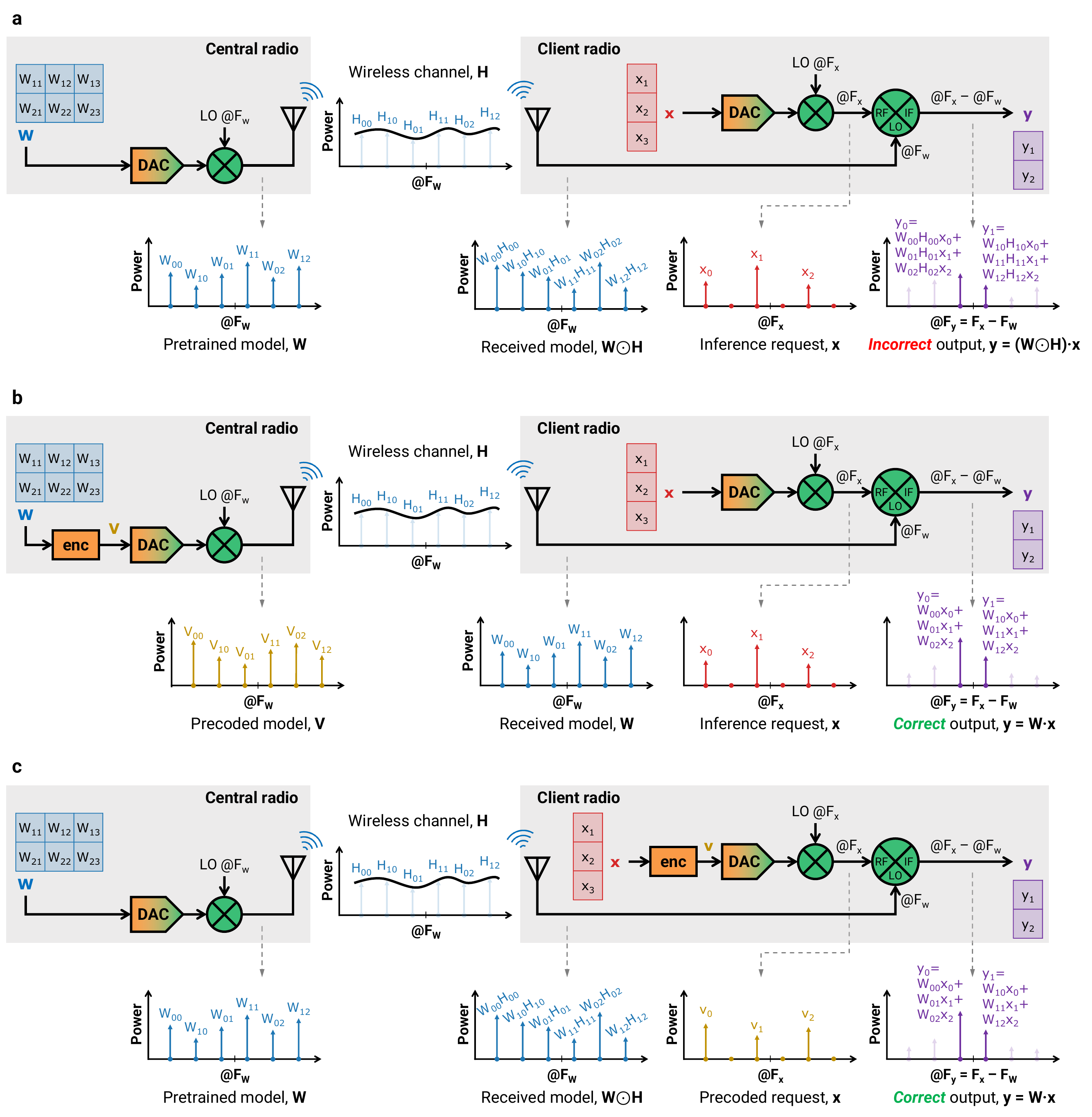}
    \caption{
    \textbf{The overview of the channel modeling and the two channel calibration schemes of {\namebf}.}
    \textbf{a}, The wireless channel introduces signal distortion, leading to incorrect matrix-vector multiplication (MVM) results when the channel state information (CSI), $\channelMat$, is uncalibrated.
    \textbf{b}, In the $\weightMat$-precoding scheme, model weights $\weightMat$ are precoded into $\precodeMat$ at the central radio to ensure that the correct model weights are received by the client after wireless transmission.
    \textbf{c}, In the $\inputVec$-precoding scheme, each inference request $\inputVec$ is precoded into $\precodeVec$ at the client to compensate for channel effects on the received signal.}
    \label{fig:supp-channel-calibration}
\end{figure*}
%% figure ends 

The basic scheme of {\name}, described in Supplementary Section~\ref{sec:supplementary-theory-basic-scheme}, assumes that the $\waveWeight(\waveIdx)$ carrying model weights $\weightMat$ is directly input into the computing mixer.
However, {\name} leverages the shared wireless medium to broadcast model weights, where signals carrying model weights experience timing delay, multi-path effect, and distortion as they propagate through the wireless channel from the central radio to each client, as illustrated in \autorefsubfig{fig:supp-channel-calibration}{a}. Therefore, channel state information (CSI) estimation and calibration are required to ensure that the signals received by the clients carry the desired model weights, $\weightMat$, after wireless transmission.

In this section, we consider channel calibration at the central radio, termed the ``$\weightMat$-precoding scheme'', as shown in \autorefsubfig{fig:supp-channel-calibration}{b}, which precodes the model weights $\weightMat$ into $\precodeMat = [\precodeMatElem_{\outputIdx, \inputIdx}] \in \mathbb{C}^{\outputSize \times \inputSize}$ before transmission to the clients. This approach does not incur additional computational or energy costs for the client. Moreover, multiple clients in close proximity sharing similar CSI can be served with the same precoded model weights.
The $\weightMat$-precoding scheme is derived from the basic scheme described in Supplementary Section~\ref{sec:supplementary-theory-basic-scheme}. For brevity, we use $\weightMat$ and $\outputVec$ to represent the decomposed weight matrix $\weightMat''$ and padded output $\outputVec''$ in Supplementary Section~\ref{sec:supplementary-theory-basic-scheme}, with $\outputSize$ referring $(1+\overheadPad)\outputSizeNew$.
By eliminating the IFFT-based encoding on $\inputVec$, the $\weightMat$-precoding scheme further improves the energy efficiency for in-physics MVM computation.

\subsection{Wireless Channel Modeling}
For the wireless channel from the central radio to the client, we define its wireless channel's frequency response $\channelFunc(\freq) \in \mathbb{C}$ as a complex-valued function of the frequency $\freq$.
As shown in \autorefsubfig{fig:supp-channel-calibration}{a}, for a general spectrum $\specVec_{\textrm{TX}}$, the frequency response on its $\specIdx$-th subcarrier at frequency $(\carrier + (\specIdx-\fftSize/2) \cdot \freqSub)$ can be acquired by $\channelFunc(\carrier + (\specIdx-\fftSize/2) \cdot \freqSub)$.
Define the channel state information (CSI) as a vector $\specChannelVec = [\specChannel[\specIdx]] \in \mathbb{C}^{\fftSize}$ of the same dimension, which has $\specChannel[\specIdx] = \channelFunc(\carrier + (\specIdx-\fftSize/2) \cdot \freqSub)$. Let `$\odot$' and `$\oslash$' denote the element-wise multiplication and division of two vectors with equal length, respectively. The impact of the wireless channel from $\specVec_{\textrm{TX}}$ to $\specVec_{\textrm{RX}}$ can be formulated as
\begin{align}
    \specVec_{\textrm{RX}} = \specVec_{\textrm{TX}} \odot \specChannelVec,\
    \textrm{where}~
    \spec_{\textrm{RX}}[\specIdx] = \spec_{\textrm{TX}}[\specIdx] \cdot \specChannel[\specIdx] = \spec_{\textrm{TX}}[\specIdx] \cdot \channelFunc\left(\carrier + \left(\specIdx-\frac{\fftSize}{2}\right) \cdot \freqSub\right),\ \forall \specIdx = 0, 1, \dots, \fftSize.
    \label{eq: weight-precoding-scheme-channel-definition}
\end{align}

\subsection{$\weightMat$-Precoding Scheme: Algorithm}
% \label{ssec:supplementary-theory-weight-precoding-scheme-algorithm}
Specifically for the MVM task, the FFT size is $\inputSize\outputSize$, so that we consider the CSI as $\specChannelVec = [\specChannel[\specIdx]] \in \mathbb{C}^{\inputSize\outputSize}$.
For this $\specChannelVec$, we define its corresponding CSI matrix, $\channelMat = [\channelMatElem_{\outputIdx, \inputIdx}] \in \mathbb{C}^{\outputSize \times \inputSize}$, given by
\begin{align}
    \channelMatElem_{\outputIdx, \inputIdx} = \specChannel[\outputSize\inputSize-1-\outputIdx-\inputIdx\outputSize],
    ~\forall \inputIdx = 0, 1, \dots, \inputSize-1,
    ~\textrm{and}~\outputIdx = 0, 1, \dots, \outputSize-1.
    \label{eq: weight-precoding-scheme-csi-reshape}
\end{align}
Based on equations {\eqref{eq: original-mvm-weight-to-spec-weight}}, {\eqref{eq: weight-precoding-scheme-channel-definition}}, and {\eqref{eq: weight-precoding-scheme-csi-reshape}}, the MVM including the wireless channel impact can be written as
\begin{align}
    \outputVec = \weightMat \cdot \inputVec = (\channelMat \odot \precodeMat) \cdot \inputVec,\
    \textrm{with}~ \outputElem_{\outputIdx} = \sum_{\inputIdx=0}^{\inputSize-1} \channelMatElem_{\outputIdx, \inputIdx} \cdot \precodeMatElem_{\outputIdx, \inputIdx} \cdot \inputElem_{\inputIdx},\ \forall \outputIdx &= 0, 1, \dots, \outputSize-1.
    \label{eq: weight-precoding-scheme-mvm-channel}
\end{align}
To estimate the channel $\channelMat$, we randomize a set of $\precodeMat^{(i)}$ and $\inputVec^{(i)}$; then we obtain the MVM results by (\emph{i}) simulating MVM with wireless channel effect following equation~\eqref{eq: weight-precoding-scheme-mvm-channel}, denoted as $\outputVec^{(i)}$, and (\emph{ii}) the analog computing by {\name} without channel calibration, denoted as $\outputVecEst^{(i)}$. Then, $\channelMat$ can be estimated using the minimum mean squared error (MMSE) method given by
\begin{align}
    \channelMat^{\star} &= \arg \min_{\channelMat} \sum_{i} \abs{\outputVec^{(i)} - \hat{\outputVec}^{(i)}}^2 = \arg \min_{\channelMat} \sum_{i} \abs{(\channelMat \odot \precodeMat^{(i)}) \cdot \inputVec^{(i)} - \hat{\outputVec}^{(i)}}^2.
    \label{eq: weight-precoding-scheme-mmse-optimization}
\end{align}
Due to the continuity nature of the channel response $\channelFunc(\freq)$, we only need to perform the MMSE optimization in equation {\eqref{eq: weight-precoding-scheme-mmse-optimization}} once with a large value of $\fftSize$ (e.g., $\fftSize=300$ for the {25}\thinspace{MHz} channel used in our experiments) for its $\channelMat^{\star}$. Then, the corresponding $\specChannelVec^{\star}$ can be acquired by reversely conducting equation {\eqref{eq: weight-precoding-scheme-csi-reshape}}, and the general $\channelFunc(\freq)$ can be estimated by nearest neighbor-based interpolation on amplitude and linear interpolation on phases, from which the $\specChannelVec$ of other $\fftSize$ and $\channelMat$ of other $\inputSize$ and $\outputSize$ can be inferred.
When there are multiple users, we optimize $\channelMat^{\star}$ for each user and average them over users to parameterize the overall $\channelMat^{\star}$.
With the optimized $\channelMat^{\star}$, and the precoded weight matrix $\precodeMat$ is given by
\begin{align}
    \precodeMat = \weightMat \oslash \channelMat^{\star},\
    \textrm{where}~\precodeMatElem_{\outputIdx, \inputIdx} = \frac{\weightElem_{\outputIdx, \inputIdx}}{\channelMatElem_{\outputIdx, \inputIdx}^{\star}},\ \forall \inputIdx = 0, 1, \dots, \inputSize-1, ~\textrm{and}~ \forall \outputIdx = 0, 1, \dots, \outputSize-1.
    \label{eq: weight-precoding-scheme-precoding-weight}
\end{align}
Note that the channel estimation process in equation~\eqref{eq: weight-precoding-scheme-mmse-optimization} can be preprocessed, whose cost can be averaged down over an unlimited number of inference requests. Also, the precoding process in equation~\eqref{eq: weight-precoding-scheme-precoding-weight} is performed on the central radio, which does not impact the energy consumption or the computation throughput of the client.

\subsection{$\weightMat$-Precoding Scheme: Time Encoding for $\inputVec$}
Note that in equation {\eqref{eq: basic-scheme-input-to-samp}}, the $\inputSize$-point IFFT applied to $\inputVec$ can be formulated as an MVM given by $(\idftMat \cdot \shiftMat^{\inputSize/2}) \cdot \inputVec$. Therefore, we can detach this IFFT process from $\inputVec$, and attach it as part of $\weightMat$.
Specifically, after the detachment, the new input vector $\inputVecNew = [\inputElemNew_{\inputIdx}] \in \mathbb{C}^{\inputSize}$ is given by
\begin{align}
    \inputVecNew = \matInv{(\idftMat \cdot \shiftMat^{\inputSize/2})} \cdot \inputVec,
    \label{eq: weight-precoding-scheme-new-MVM-input}
\end{align}
whose corresponding I/Q waveform $\sampInputVecNew = [\sampInputNew[\sampIdx]] \in \mathbb{C}^{\inputSize\outputSize}$ can be generated by
\begin{align}
    \sampInputVecNew
    = \frac{1}{\outputSize\sqrt{\inputSize}} \cdot \underbrace{\Big[ \idftMat \cdot \shiftMat^{\inputSize/2} \cdot \inputVecNew , \dots, \idftMat \cdot \shiftMat^{\inputSize/2} \cdot \inputVecNew \Big]}_{\textrm{repeated}~\outputSize~\textrm{times}}
    = \frac{1}{\outputSize\sqrt{\inputSize}} \cdot \underbrace{\Big[ \inputVec , \dots, \inputVec \Big]}_{\textrm{repeated}~\outputSize ~\text{times}}.
    \label{eq: weight-precoding-scheme-encoding-input-stage}
\end{align}
Therefore, $\sampInputVecNew$ can be generated by repeating the original $\inputVec$, i.e., via direct time encoding, and involved no MACs.
On the other hand, the new $\weightMatNew = [\weightElemNew_{\outputIdx, \inputIdx}] \in \mathbb{C}^{\outputSize \times \inputSize}$ combined with the IFFT operation is given by
\begin{align}
    \weightMatNew = \weightMat \cdot (\idftMat \cdot \shiftMat^{\inputSize/2}).
    \label{eq: weight-precoding-scheme-new-MVM-weight}
\end{align}
Plugging this new $\weightMatNew$ into the encoding process at the central radio, the corresponding I/Q waveform $\sampWeightVec = [\sampWeight[\sampIdx]] \in \mathbb{C}^{\inputSize\outputSize}$ is given by
\begin{align}
    & \sampWeightNew[\sampIdx] = \frac{1}{\inputSize\outputSize} \sum_{\inputIdx=0}^{\inputSize-1} \sum_{\outputIdx=0}^{\outputSize-1} \weightElemNew_{\outputIdx, \inputIdx} \cdot \eu^{-\iu 2\pi \frac{1+\outputIdx+\inputIdx\outputSize + \inputSize\outputSize/2}{\inputSize\outputSize}\sampIdx},\
    \textrm{where}~\weightElemNew_{\outputIdx, \inputIdx} = \frac{1}{\sqrt{\inputSize}}\sum_{\inputIdx'=0}^{\inputSize-1} \weightElem_{\outputIdx, \inputIdx'} \cdot \eu^{\iu 2\pi \frac{\inputIdx - \inputSize/2}{\inputSize}\inputIdx'},\ \forall \sampIdx = 0, 1, \dots, \fftSize-1.
\end{align}
Overall, the weight matrix $\weightMat$ is still frequency-encoded and, upon receiving the $\weightMat$-precoded model weights from the central radio, the client performs local MVM computation given by
\begin{align}
    \outputVec
    & = \precodeMat' \odot \channelMat \cdot \inputVecNew \nonumber \\
    & = (\weightMatNew \oslash \channelMat^{\star}) \odot \channelMat \cdot \inputVecNew
    = \left[\weightMat \cdot (\idftMat \cdot \shiftMat^{\inputSize/2}) \oslash \channelMat^{\star}\right] \odot \channelMat \cdot \inputVecNew\ \quad
    && \textrm{($\weightMat$-precoding at the central radio)} \nonumber \\
    & = \left[\weightMat \cdot (\idftMat \cdot \shiftMat^{\inputSize/2}) \oslash \channelMat^{\star}\right] \odot \channelMat \cdot \left[\matInv{(\idftMat \cdot \shiftMat^{\inputSize/2})} \cdot \inputVec\right]\ \quad && \textrm{(time encoding of $\inputVec$ at the client)} \nonumber \\
    & = \left[\weightMat \cdot (\idftMat \cdot \shiftMat^{\inputSize/2})\right] \oslash (\channelMat^{\star} \oslash \channelMat) \cdot \left[\matInv{(\idftMat \cdot \shiftMat^{\inputSize/2})} \cdot \inputVec\right] 
    \approx \weightMat \cdot \inputVec\ \quad && \textrm{(original MVM)}.
\end{align}
Note that the last approximation is due to the MMSE-based channel estimation, which corresponds to $\channelMat^{\star} \approx \channelMat$, or $\channelMat^{\star} \oslash \channelMat$ being approximately an all-ones matrix.

\subsection{$\weightMat$-Precoding Scheme: Energy Efficiency Analysis}
Compared to the basic scheme described in Supplementary Section~\ref{sec:supplementary-theory-basic-scheme}, the $\weightMat$-precoding scheme further improves the energy efficiency by applying ``IFFT-less'' time encoding on the client. In particular, under this scheme, the energy consumption term $\energyMVMDec$ becomes
\begin{align}
    \energyMVMDec = \frac{\outputSize}{\outputSizeNew} \cdot 2 (1+\overheadPad)\outputSizeNew \log_2 ((1+\overheadPad)\outputSizeNew) \cdot e_{\textrm{dig}} = 2(1+\overheadPad) \cdot \outputSize \cdot \log_2 ((1+\overheadPad)\outputSizeNew) \cdot e_{\textrm{dig}},
    \label{eq: weight-precoding-scheme-energy-per-mvm-digital}
\end{align}
where the encoding energy of $(2\inputSize \log_2 \inputSize \cdot e_{\textrm{dig}})$ in equation {\eqref{eq: basic-scheme-energy-per-mvm-digital}} is eliminated. As a result, the total energy consumption per MVM, $\energyMVM$, for the $\weightMat$-precoding scheme is given by
\begin{align}
    \energyMVM = \energyMVMTx + \energyMVMADC + \energyMVMDec
    & = \underbrace{(1+\overheadPad) (1+\overheadCP) \cdot \inputSize\outputSize \cdot \efficiencyTRX^{-1} \cdot \textsf{SNR} \cdot \boltzmann \temperature}_{\energyMVMTx} + \underbrace{2(1+\overheadPad) \cdot \outputSize \cdot e_{\textrm{adc}}}_{\energyMVMADC} \nonumber \\
    & \quad + \underbrace{ 2(1+\overheadPad) \cdot \outputSize \cdot \log_2 ((1+\overheadPad)\outputSizeNew) \cdot e_{\textrm{dig}}}_{\energyMVMDec}.
    \label{eq: weight-precoding-scheme-energy-per-mvm}
\end{align}
The corresponding energy efficiency, $\energyMAC$, is given by 
\begin{align}
    \energyMAC = \frac{\energyMVM}{4\inputSize\outputSize} = \energyMACTx + \energyMACADC + \energyMACDec &= \underbrace{\frac{(1+\overheadPad) (1+\overheadCP)}{4} \cdot \efficiencyTRX^{-1} \cdot \textsf{SNR} \cdot \boltzmann \temperature}_{\energyMACTx} + \underbrace{\frac{1+\overheadPad}{2\inputSize} \cdot e_{\textrm{adc}}}_{\energyMACADC} + \underbrace{ \frac{1+\overheadPad}{2\inputSize} \cdot \log_2 ((1+\overheadPad)\outputSizeNew) \cdot e_{\textrm{dig}}}_{\energyMACDec}.
    \label{eq: weight-precoding-scheme-energy-per-mac}
\end{align}
Note that equation {\eqref{eq: weight-precoding-scheme-energy-per-mac}} indicates that the energy efficiency of this $\weightMat$-precoding scheme is independent of the output size $\outputSize$, due to the elimination of the $\inputSize$-point IFFT for encoding.
The corresponding TDL under the $\weightMat$-precoding scheme is given by
\begin{align}
    \energyMACTDL := \lim_{\inputSize \to +\infty} \energyMAC = \textsf{SNR} \cdot k T_{0}/4.
    \label{eq: weight-precoding-scheme-energy-per-mac-tdl}
\end{align}
%%

%%%%%
%%%%%
\subsection{$\weightMat$-Precoding Scheme: MVM Decomposition into IPs}

Similarly, the energy efficiency $\energyMVMIPMult$ for the $\weightMat$-precoding scheme that decomposes the MVM into $\outputSize$ IPs can be rewritten from equation {\eqref{eq: basic-scheme-energy-mvm-multiple}} as
\begin{align}
    \energyMVMIPMult = \underbrace{4\inputSize\outputSize \cdot \efficiencyTRX^{-1} \cdot \textsf{SNR} \cdot \boltzmann \temperature}_{\energyMVMTx} + \underbrace{6\outputSize \cdot e_{\textrm{adc}}}_{\energyMVMADC} + \underbrace{ 8\outputSize \cdot e_{\textrm{dig}}}_{\energyMVMDec},
    \label{eq: weight-precoding-scheme-energy-per-mvm-multiple}
\end{align}
where the encoding energy consumption of $\left( \frac{1}{2} \cdot \log_2 \inputSize \right) \cdot e_{\textrm{dig}}$ is eliminated due to the time encoding of $\inputVec$.
Also, $\energyMACIPMult$ can be derived from equation {\eqref{eq: basic-scheme-energy-mac-multiple}}, which is
\begin{align}
    \energyMACIPMult = \frac{\energyMVMIPMult}{4\inputSize\outputSize} = \underbrace{\efficiencyTRX^{-1} \cdot \textsf{SNR} \cdot \boltzmann \temperature}_{\energyMACTx} + \underbrace{\frac{3}{2\inputSize} \cdot e_{\textrm{adc}}}_{\energyMACADC} + \underbrace{ \frac{2}{\inputSize} \cdot e_{\textrm{dig}}}_{\energyMACDec}.
    \label{eq: weight-precoding-scheme-energy-per-mac-multiple}
\end{align}
Here, notice that the energy efficiency for $\weightMat$-precoding scheme is independent of $\outputSize$. 
Therefore, for a standalone IP computation with $\outputSize=1$, it holds that
\begin{align}
    \energyMACIPSing = \energyMACIPMult = \underbrace{\efficiencyTRX^{-1} \cdot \textsf{SNR} \cdot \boltzmann \temperature}_{\energyMACTx} + \underbrace{\frac{3}{2\inputSize} \cdot e_{\textrm{adc}}}_{\energyMACADC} + \underbrace{ \frac{2}{\inputSize} \cdot e_{\textrm{dig}}}_{\energyMACDec}.
\end{align}
Compared to the basic scheme, this $\weightMat$-precoding scheme achieves further enhanced energy efficiency for standalone IP computations.

%%%%%
%%%%%
\section{{\namebf}'s $\inputVec$-Precoding Scheme: Wireless Channel Calibration at the Client}
\label{sec:supplementary-theory-input-precoding-scheme}

One alternative channel calibration scheme is performed for each client by adjusting the input $\inputVec$, which we term as the $\inputVec$-precoding scheme, as shown in \autorefsubfig{fig:supp-channel-calibration}{c}.
This scheme allows each client to estimate and apply its own CSI, especially when the users are away from each other and the CSIs are diverse among clients.
On the other hand, this scheme requires extra computation costs for clients, incurring higher energy consumption.
This scheme precodes $\inputVec$ into $\precodeVec = [\precodeVecElem_{\inputIdx}] \in \mathbb{C}^{\inputSize}$.

\subsection{$\inputVec$-Precoding Scheme: Algorithm}

For the same CSI $\specChannelVec$ as considered in Supplementary Section~\ref{sec:supplementary-theory-weight-precoding-scheme}, we define the equivalent channel vector $\channelVec = [\channelVecElem_{\inputIdx}] \in \mathbb{C}^{\inputSize}$.
Due to the smoothness of the channel response $\channelFunc(\freq)$ and thus $\specChannelVec$, we let $\channelVecElem_{\inputIdx}$ to approximate the channel responses as
\begin{align}
    \channelVecElem_{\inputIdx} \approx \specChannel[\outputSize\inputSize-1-\outputIdx-\inputIdx\outputSize] = \channelMatElem_{\outputIdx, \inputIdx}, ~\forall \inputIdx = 0, 1, \dots, \inputSize-1,\ \outputIdx = 0, 1, \dots, \outputSize-1.
\end{align}
This $\inputVec$-precoding scheme transmits the $\precodeVec$ in substitute of $\inputVec$, which compensates the channel impact on the received $\weightMat$ during the frequency mixing.
Then, we can rewrite equation~\eqref{eq: weight-precoding-scheme-mvm-channel} by this new equivalent channel vector $\channelVec$ and a precoded $\precodeVec$ as
\begin{align}
    \outputVec = \weightMat \cdot (\channelVec \odot \precodeVec) ~\textrm{with}~ \outputElem_{\outputIdx} = \sum_{\inputIdx=0}^{\inputSize-1} \weightElem_{\outputIdx, \inputIdx} \cdot \channelVecElem_{\inputIdx} \cdot \precodeVecElem_{\inputIdx}, ~\forall \outputIdx = 0, 1, \dots, \outputSize-1.
\end{align}
To estimate the equivalent channel vector $\channelVec$, we randomize a series of $\precodeVec^{(i)}$ and $\weightMat^{(i)}$, and generate their corresponding $\outputVec^{(i)}$ and $\outputVecEst^{(i)}$. Similarly, $\channelVec$ can be optimized by the MMSE method given by
\begin{align}
    \channelVec^{\star} = \arg \min_{\channelVec} \sum_{i} \abs{\outputVec^{(i)} - \hat{\outputVec}^{(i)}}^2 = \arg \min_{\channelVec} \sum_{i} \abs{\weightMat^{(i)} \cdot (\channelVec \odot \precodeVec^{(i)}) - \hat{\outputVec}^{(i)}}^2.
    \label{eq: channel-calibration-input-mse-optimization}
\end{align}
Given the optimized $\channelVec^{\star}$, the precoded input $\precodeVec$ is given by
\begin{align}
    \precodeVec = \inputVec \oslash \channelVec^{\star} ~\textrm{with}~ \precodeVecElem_{\inputIdx} = \frac{\inputElem_{\inputIdx}}{\channelVecElem_{\inputIdx}^{\star}},\ \forall \inputIdx = 0, 1, \dots, \inputSize.
    \label{eq: channel-calibration-input-precoding-input}
\end{align}
Essentially, the MVM computation of this $\inputVec$-precoding scheme can be summarized by
\begin{align}
    \outputVec
    & = \weightMat \odot \channelMat \cdot \precodeVec \nonumber \approx \weightMat \cdot (\channelVec \odot \precodeVec) \\
    & = \weightMat \cdot (\channelVec \odot \inputVec \oslash \channelVec^{\star}) \quad
    && \textrm{($\inputVec$-precoding at the client)} \nonumber \\
    & = \weightMat \cdot \left[ (\channelVec \oslash \channelVec^{\star}) \odot \inputVec \right]
    \approx \weightMat \cdot \inputVec\ \quad && \textrm{(original MVM)}.
\end{align}
Here, the last approximation is due to the MMSE-based channel estimation, which corresponds to $\channelVec^{\star} \approx \channelVec$, or $\channelVec^{\star} \oslash \channelVec$ being approximately an all-ones vector.

%%%%%
%%%%%
\subsection{$\inputVec$-Precoding Scheme: Energy Efficiency Analysis}

Compared to the basic scheme described in Supplementary Section~\ref{sec:supplementary-theory-basic-scheme}, this $\inputVec$-precoding scheme maintains the same energy consumption terms $\energyMVMTx$ and $\energyMVMADC$.
On the other hand, the precoding is performed in the frequency domain for $\inputVec$ leveraging the IFFT-based encoding. Moreover, the precoding process based on equation {\eqref{eq: channel-calibration-input-precoding-input}} requires additional $\inputSize$ complex-valued MACs (or $4\inputSize$ real-valued MACs). Hence, the energy consumption term $\energyMVMDec$ for the $\inputVec$-precoding scheme is given by
\begin{align}
    \energyMVMDec = \left(4\inputSize + 2\inputSize \cdot \log_2 \inputSize + 2(1+\overheadPad) \cdot \outputSize \cdot \log_2 ((1+\overheadPad)\outputSizeNew) \right) \cdot e_{\textrm{dig}}.
    \label{eq: input-precoding-scheme-energy-per-mvm-digital}
\end{align}
Then, the energy consumption per MVM under this $\inputVec$-precoding scheme is given by
\begin{align}
    \energyMVM = \energyMVMTx + \energyMVMADC + \energyMVMDec
    & = \underbrace{(1+\overheadPad) (1+\overheadCP) \cdot \inputSize\outputSize \cdot \efficiencyTRX^{-1} \cdot \textsf{SNR} \cdot \boltzmann \temperature}_{\energyMVMTx} + \underbrace{2(1+\overheadPad) \cdot \outputSize \cdot e_{\textrm{adc}}}_{\energyMVMADC} \nonumber \\
    & \quad + \underbrace{\left( 4\inputSize + 2\inputSize \log_2 \inputSize + 2(1+\overheadPad) \cdot \outputSize \cdot \log_2 ((1+\overheadPad)\outputSizeNew) \right) \cdot e_{\textrm{dig}}}_{\energyMVMDec}.
    \label{eq: input-precoding-scheme-energy-per-mvm}
\end{align}
The corresponding energy efficiency is given by
\begin{align}
    \energyMAC
    & = \frac{\energyMVM}{4\inputSize\outputSize} = \energyMACTx + \energyMACADC + \energyMACDec \nonumber \\
    & = \underbrace{\frac{(1+\overheadPad) (1+\overheadCP)}{4} \cdot \efficiencyTRX^{-1} \cdot \textsf{SNR} \cdot \boltzmann \temperature}_{\energyMACTx} + \underbrace{\frac{1+\overheadPad}{2\inputSize} \cdot e_{\textrm{adc}}}_{\energyMACADC} + \underbrace{ \left( \frac{1}{\outputSize} + \frac{\log_2 \inputSize}{2\outputSize} + \frac{1+\overheadPad}{2\inputSize} \cdot \log_2 ((1+\overheadPad)\outputSizeNew) \right) \cdot e_{\textrm{dig}}}_{\energyMACDec}.
    \label{eq: input-precoding-scheme-energy-per-mac}
\end{align}
Similar to the basic scheme, the TDL under this $\inputVec$-precoding scheme with $\inputSize \rightarrow \infty$ and $\outputSize \rightarrow \infty$ is given by
\begin{align}
    \energyMACTDL := \lim_{\inputSize \to +\infty, \outputSize \to +\infty} \energyMAC = \textsf{SNR} \cdot k T_{0}/4.
    \label{eq: input-precoding-scheme-energy-per-mac-tdl}
\end{align}

\subsection{$\inputVec$-Precoding Scheme: MVM Decomposition into IPs}

As for the IP-based MVM decomposition with $\outputSizeNew=1$, we have the additional precoding energy consumption of $4\inputSize \cdot e_{\textrm{dig}}$. Hence, equation {\eqref{eq: basic-scheme-energy-mvm-multiple}} can be rewritten as
\begin{align}
    \energyMVMIPMult = \underbrace{4 \inputSize\outputSize \cdot \efficiencyTRX^{-1} \cdot \textsf{SNR} \cdot \boltzmann \temperature}_{\energyMVMTx} + \underbrace{6\outputSize \cdot e_{\textrm{adc}}}_{\energyMVMADC} + \underbrace{ \left( 4\inputSize + 2\inputSize \log_2 \inputSize + 8\outputSize \right) \cdot e_{\textrm{dig}}}_{\energyMVMDec},
    \label{eq: input-precoding-scheme-energy-mvm-multiple}
\end{align}
and equation {\eqref{eq: basic-scheme-energy-mac-multiple}} can be rewritten as
\begin{align}
    \energyMACIPMult = \frac{\energyMVMIPMult}{4\inputSize\outputSize} = \underbrace{\efficiencyTRX^{-1} \cdot \textsf{SNR} \cdot \boltzmann \temperature}_{\energyMACTx} + \underbrace{\frac{3}{2\inputSize} \cdot e_{\textrm{adc}}}_{\energyMACADC} + \underbrace{ \left( \frac{1}{\outputSize} + \frac{\log_2 \inputSize}{2\outputSize} + \frac{2}{\inputSize} \right) \cdot e_{\textrm{dig}}}_{\energyMACDec}.
    \label{eq: input-precoding-scheme-energy-mac-multiple}
\end{align}
The energy efficiency of the standalone IP computation, $\energyMACIPSing$, can be obtained by plugging $\outputSize=1$ in equation {\eqref{eq: input-precoding-scheme-energy-mac-multiple}}, 
\begin{align}
    \energyMACIPSing = \underbrace{\efficiencyTRX^{-1} \cdot \textsf{SNR} \cdot \boltzmann \temperature}_{\energyMACTx} + \underbrace{\frac{3}{2\inputSize} \cdot e_{\textrm{adc}}}_{\energyMACADC} + \underbrace{ \left( 1 + \log_2 \inputSize + \frac{2}{\inputSize} \right) \cdot e_{\textrm{dig}}}_{\energyMACDec}.
    \label{eq: input-precoding-scheme-energy-mac-single}
\end{align}
Similar to equation {\eqref{eq: basic-scheme-energy-mac-single}} for the based scheme of {\name}, it can be seen that this $\inputVec$-precoding scheme is not energy efficient for standalone IP computation.

%%%%%
%%%%%
\section{Computation Throughput Analysis}
\label{sec:supplementary-theory-computation-analysis}

In this section, we analyze {\name}'s \emph{computation throughput}, i.e., the number of real-valued MACs per unit time.
Without loss of generality, we consider the MVM $\outputVec=\weightMat \cdot \inputVec$, where a single central radio wirelessly broadcasts model weights $\weightMat$ to $\userNum$ clients using an OFDM system with FFT size $\fftSize = \inputSize\outputSize$. Each client performs local inference on $\inputVec$ and generates $\outputVec$. In this case, a total number of $4\inputSize \outputSize$ real-valued MACs is involved per client, and a total number of $4\inputSize\outputSize \cdot \userNum$ real-valued MACs across all $\userNum$ clients.

To achieve an MVM involving $\userNum \cdot 4\inputSize\outputSize$ real-valued MACs, the latency is primarily determined by the waveform duration for transmitting $\waveWeight(\waveIdx)$ and $\waveInput(\waveIdx)$. The total waveform duration across all $\outputSize/\outputSizeNew$ decomposed MVMs is
\begin{align}
    \waveLen_{\textrm{mvm}} = \frac{\outputSize}{\outputSizeNew} \cdot \frac{(1+\overheadPad)(1+\overheadCP) \cdot \inputSize\outputSizeNew}{\band} = \frac{(1+\overheadPad)(1+\overheadCP) \cdot \inputSize\outputSize}{\band}.
    \label{eq: computation-throughput-waveform-time}
\end{align}
Thus, the computation throughput, denoted by $\throughput$, is given by
\begin{align}
    \throughput = \frac{\userNum \cdot 4\inputSize\outputSize}{\waveLen_{\textrm{mvm}}} = \frac{4 \userNum \cdot \band}{(1+\overheadPad) (1+\overheadCP)},
    \label{eq: computation-throughput-throughput}
\end{align}
which applies to all three schemes of {\name}, described in Supplementary Sections~\ref{sec:supplementary-theory-basic-scheme},~\ref{sec:supplementary-theory-weight-precoding-scheme}, and~\ref{sec:supplementary-theory-input-precoding-scheme}.
To conclude, the computation throughput of {\name} is proportional to the available bandwidth, $\band$, and number of users $\userNum$. With wirelessly broadcast model weights, the computation throughout will be determined by the available wireless bandwidth in the unlicensed (e.g., {25}\thinspace{MHz} in the {915}\thinspace{MHz} ISM band) or licensed bands.

%% file: tex/Supplementary_experiment.tex
\section*{\textbf{Supplementary Information: Experiment}}

\section{Experimental Setup}
\label{sec:supplementary-experiment-setup}

%% figure begins
\begin{figure*}[!h]
    \centering
    \includegraphics[width=0.9\textwidth]{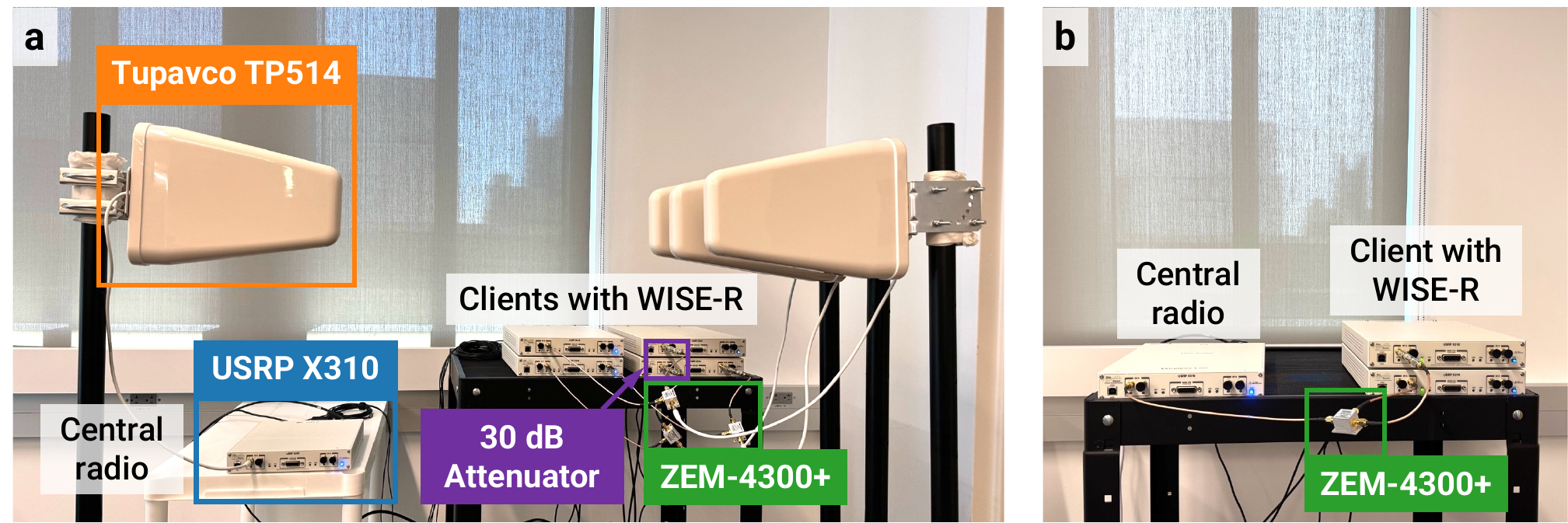}
    \caption{
    \textbf{The software-defined radio testbed for {\namebf}'s experimental implementation.}
    \textbf{a}, Experimental setup for {\name} with model weights broadcast over a wireless channel to three clients, each employing a passive ZEM-4300+ frequency mixer as the computing mixer for in-physics MVM computation.
    \textbf{b}, Experimental setup for {\name} with model weights transmitted over a wired channel to a single client.}
    \label{fig:supp-setup}
\end{figure*}
%% figure ends 

\autorefsubfig{fig:supp-setup}{a} shows the detailed experiment setup of {\name}, including the Tupavco TP514 Yagi directional antenna, Mini-Circuits ZEM-4300+~\cite{ZEM-4300+} as the computing frequency mixer, and USRP X310 software-defined radio (SDR) as the transceiver radio unit. One USRP X310 serves as the central radio that broadcasts $\waveWeight(\waveIdx)$ or $\wavePrecode(\waveIdx)$ (for the $\weightMat$-precoding scheme) over a wireless channel with a bandwidth of $\band={25}\thinspace\textrm{MHz}$ at the carrier frequency of $\carrierWeight={0.915}\thinspace\textrm{GHz}$. Our wireless experiments are conducted in the unlicensed industrial, scientific, and medical (ISM) band centered at {0.915}\thinspace{GHz}, which has a limited bandwidth of only {26}\thinspace{MHz} between {902--928}\thinspace{MHz}~\cite{ISM}.
For a client, one RX antenna receives the wirelessly broadcast model weights at the input to the LO port of the computing mixer; and one USRP X310 generates $\waveInput(\waveIdx)$ or $\wavePrecode(\waveIdx)$ (for the $\inputVec$-precoding scheme) at the carrier frequency of $\carrierInput={1.2}\thinspace\textrm{GHz}$, which is streamed into the computing mixer's RF port via a cable and a total of {30}\thinspace{dB} attenuator. The output signal of the computing mixer from the IF port, $\waveOutput(\waveIdx)$, at the carrier frequency of $\carrierOutput={0.285}\thinspace\textrm{MHz}$ is streamed to the RX channel of a USRP X310, which is configured with a low sampling rate of $\bandLow = \max\{ \frac{25}{\inputSize}, 0.20\}\thinspace\textrm{MHz}$.
The wireless link distance is $\approx${1}\thinspace{m}, which is limited by the USRP X310's TX power and the required LO input power to the computing mixer.

We also consider a wired setting of {\name} with a single client, as shown in \autorefsubfig{fig:supp-setup}{b}, with the same carrier frequency configuration of $\carrierWeight$, $\carrierInput$, and $\carrierOutput$. In this case, the TX channel of the central radio directly streams $\waveWeight(\waveIdx)$ to the computing mixer's LO port, where no precoding is used. A signal bandwidth of $\band={100}\thinspace\textrm{MHz}$ is employed in this setup with the wired channel, which is limited by the DAC sampling rate of the USRP X310.

\subsection{Tupavco TP514 Yagi Directional Antenna}
% \label{ssec:supplementary-experiment-setup-antenna}

In the wireless setup of {\name}, we use the Tupavco TP514 Yagi directional antenna as the TX/RX antenna to establish the wireless link between the central radio and each client. In general, a Yagi antenna consists of multiple parallel resonant antenna elements, which focus the transmitted/received RF signal power in a specific direction. The geometry of these antenna elements is determined by the target operating frequency. As a fully passive component, a Yagi antenna offers a higher gain in the intended direction where the RF signal is concentrated, while exhibiting lower gains in other directions when compared to an ideal isotropic antenna.
Specifically, the Tupavco TP514 Yagi antenna is designed and optimized for dual frequency bands:{0.80--0.96}\thinspace{GHz} and {1.7--2.5}\thinspace{GHz}. This frequency range includes the ISM band at {0.915}\thinspace{GHz} utilized in our experiments. The antenna provides an antenna gain of {9}\thinspace{dB} in the designated direction, and has horizontal and vertical beamwidths of {65}$^{\circ}$ and {55}$^{\circ}$, respectively. It is connected to the transceiver radio, a USRP X310 SDR, via an SMA cable.

\subsection{Computing Frequency Mixer, ZEM-4300+}

%% figure begins
\begin{figure*}[!t]
    \centering
    \includegraphics[width=0.9\columnwidth]{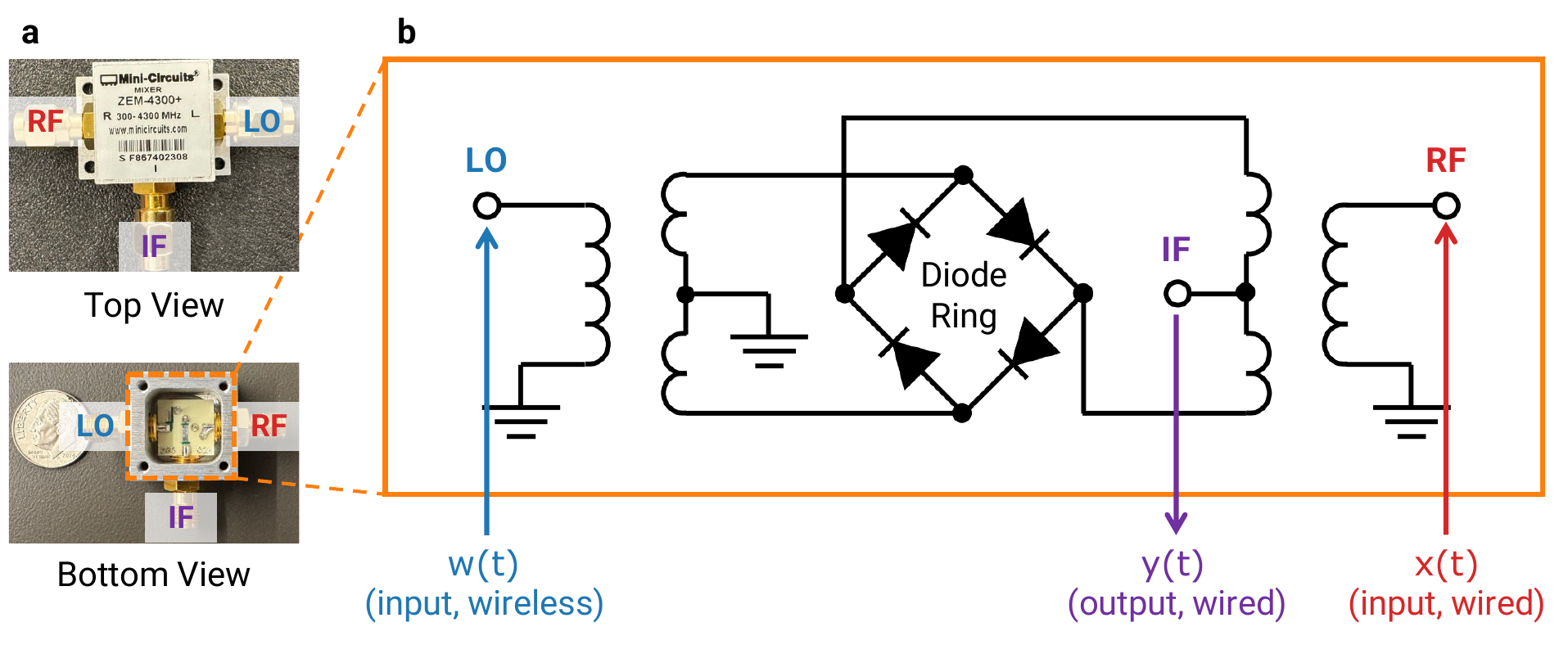}
    \caption{
    \textbf{The detailed structure of the computing mixer, Mino-Circuits ZEM-4300+~\cite{ZEM-4300+}.}
    \textbf{a}, The external and internal view of the employed frequency mixer, ZEM-4300+.
    \textbf{b}, The schematic of a double-balanced diode mixer composed of a four-diode-bridge, producing an output signal $\radio_{\textrm{IF}}(\waveIdx) \propto \textsf{sgn}(\radio_{\textrm{LO}}(\waveIdx)) \cdot \radio_{\textrm{RF}}(\waveIdx)$. This mixing process approximates the time-domain multiplication of two RF signals, $\radio_{\textrm{LO}}(\waveIdx)$ and $\radio_{\textrm{RF}}(\waveIdx)$.}
    \label{fig:mixer-schematic}
\end{figure*}
%% figure ends

We exploit the passive double-balanced diode mixer, Mini-Circuits ZEM-4300+~\cite{ZEM-4300+}, as the computing mixer in the implementation of {\name}, as shown in \autorefsubfig{fig:mixer-schematic}{a}.
The typical schematic of a double-balanced diode mixer is shown in \autorefsubfig{fig:mixer-schematic}{b}, whose core component is a four-diode bridge.
When performing signal frequency down-conversion, the LO and RF ports serve as the input ports, and the IF port is the output port.
Essentially, the waveform input to the LO port, $\radio_{\textrm{LO}}(\waveIdx)$, controls which two diodes are on while the other two diodes are off. The on/off status of this four-diode-bridge determines the current direction of the input waveform $\radio_{\textrm{RF}}(\waveIdx)$ at the RF port, and thus that of the output waveform $\radio_{\textrm{IF}}(\waveIdx)$ on the IF port.
Equivalently, this mixer modulates an on-off switching pattern on $\radio_{\textrm{RF}}(\waveIdx)$ based on $\radio_{\textrm{LO}}(\waveIdx)$; the output waveform on the IF port $\radio_{\textrm{IF}}(\waveIdx)$ can thus be formulated as
\begin{align}
    \radio_{\textrm{IF}}(\waveIdx) \propto \textsf{sgn}(\radio_{\textrm{LO}}(\waveIdx)) \cdot \radio_{\textrm{RF}}(\waveIdx), ~\textrm{where}~
    \textsf{sgn}(x) &=
    \begin{cases}
        -1, & \text{if}~x < 0, \\
        0, & \text{if}~x = 0, \\
        +1, & \text{if}~x > 0.
    \end{cases}
\end{align}
Such on-off switching can be treated as a low-resolution version of the ideal analog multiplication given by equation {\eqref{eq: frequency-mixer-multiplication}}, where $\radio_{\textrm{LO}}(\waveIdx)$ is quantized with an equivalent 1-bit resolution, i.e., ``ON'' or ``OFF''.
Fortunately, the waveform $\radio_{\textrm{LO}}(\waveIdx)$ is a narrowband signal modulated at a carrier frequency such that the quantization noise after the mixing process is mostly distributed across other carrier frequencies, which can be filtered out by an anti-aliasing filter. In our experiments, the residual noise still impacts the computing accuracy of {\name}. To mitigate this effect, a sourced analog multiplier, e.g., Gilbert cell~\cite{zhang20190, choi202419}, can be used for better analog computing performance at the cost of extra energy consumption as it is an active component.

Specifically, the ZEM-4300+ mixer supports an LO and RF frequency range of {0.3--4.3}\thinspace{GHz}, where $\waveWeight(\waveIdx)$ is modulated to the LO at {0.915}\thinspace{GHz} (within the ISM band) and $\waveInput(\waveIdx)$ is modulated to the RF at {1.2}\thinspace{GHz}. The mixer supports an IF frequency range of {0--1.0}\thinspace{GHz}, which includes the frequency {0.285}\thinspace{GHz} to which $\waveOutput(\waveIdx)$ modulated. Since the input waveform $\waveWeight(\waveIdx)$ to the LO port spans a bandwidth of {25}\thinspace{MHz}, different from the typical usage of a frequency mixer, i.e., a single tone signal, the frequency mixer's parameters (e.g., optimal input LO power, insertion loss, etc.) might deviate from that on the datasheet.

%% figure begins
\begin{figure*}[!t]
    \centering
    \includegraphics[width=1.0\columnwidth]{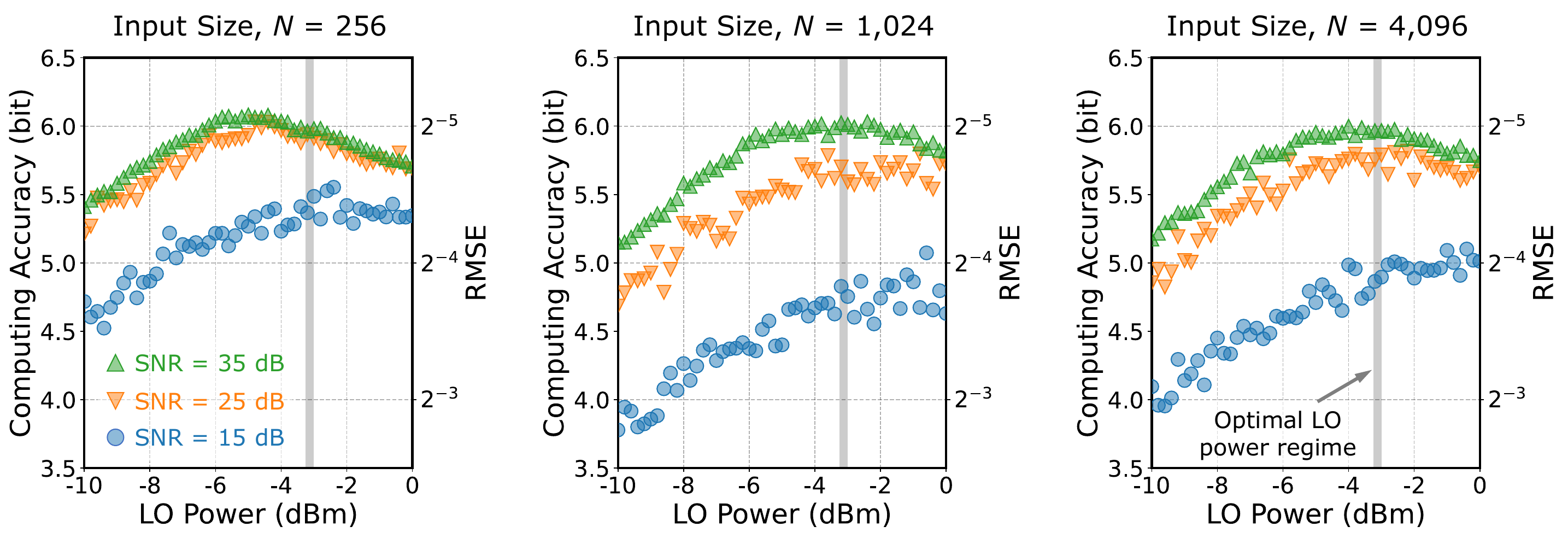}
    \caption{Experimental computing accuracy as a function of the input power level to the local oscillator (LO) port of the computing mixer, benchmarked using inner-product (IP) computations across varying input sizes $\inputSize = \{256, 1024, 4096\}$. Based on these results, we empirically select an LO power in the range of $-${3.00}\thinspace{dBm} to {$-$3.25}\thinspace{dBm}, which yields optimal computing accuracy.}
    \label{fig:mixer-linear}
\end{figure*}
%% figure ends

We benchmark the optimal LO power for the target in-physics computing tasks, employing a setup with wired transmissions of $\waveWeight(\waveIdx)$ to a single client without the channel calibration process, as shown in \autorefsubfig{fig:supp-setup}{b}.
We consider the same inner-product (IP) computation as described in {\autorefresults} with randomly generated complex-valued vectors $\mathbf{a}$ and $\mathbf{b}$ over the IP dimensions of $\inputSize \in \{256, 1024, 4096\}$. We follow the same carrier frequency configuration ({0.915}\thinspace{GHz}) and bandwidth ({25}\thinspace{MHz}) setup as described in {\autorefmethod}, and sweep the signal power of $\waveWeight(\waveIdx)$ input to the LO port between $[-10, 0]\thinspace\textrm{dBm}$ with a step size of {0.2}\thinspace{dB}.
We also consider three input power levels of $\{-63, -53, -43\}\thinspace\textrm{dBm}$ into the RF port ($\waveInput(\waveIdx)$), which correspond to three different SNR levels of $\textsf{SNR} \in \{15, 25, 35\}\thinspace\textrm{dB}$.
\autoreffig{fig:mixer-linear} plots the in-physics computing performance measured by the computing accuracy as a function of the LO power. 
Overall, the computing accuracy is slightly higher than that in {\autorefresults} under the wired channel setting. Given the same power of $\waveInput(\waveIdx)$, i.e., the same received SNR of $\waveOutput(\waveIdx)$, the LO power of $\waveWeight(\waveIdx)$ impacts the computing accuracy. Moreover, the optimal LO power that achieves the best computing accuracy, or the lowest RMSE, reduces as the RF power or SNR increases. For example, with an IP dimension of $\inputSize=4,096$, the optimal LO power is {$-$4.0}\thinspace{dBm} for $\textsf{SNR}={35}\thinspace\textrm{dB}$, corresponding to an RMSE of {0.031} or $\approx${6}-bit computing accuracy. On the other hand, the optimal LO power is {$-$0.4}\thinspace{dBm} for $\textsf{SNR}={15}\thinspace\textrm{dB}$, corresponding to an RMSE of {0.058} or $\approx${5}-bit computing accuracy.
This trend is plausible since a higher RF power input can compensate for the need for a higher LO power input that activates the frequency mixer into the optimal power regime.
To compromise across varying SNR values, we empirically select the LO power in the range of $[-3.25, -3.0]\thinspace\textrm{dBm}$ in our experiments.

Then, we measure the frequency mixer's insertion loss under the selected LO power, i.e., the power discrepancy of the input power to the RF port and the output power from the IF port. Note that the output power spans over a bandwidth of approximately $2\band$ after the convolution, as discussed in Supplementary Section~\ref{sec:supplementary-theory-vanilla-maft}. Under this setting, the measured insertion loss of the computing mixer is {11.4}\thinspace{dB}, i.e., an efficiency of $\efficiencyMixer=0.0724$.

\subsection{Tranceiver Radio Unit, USRP X310}

%% figure begins
\begin{figure*}[!t]
    \centering
    \includegraphics[width=1.0\textwidth]{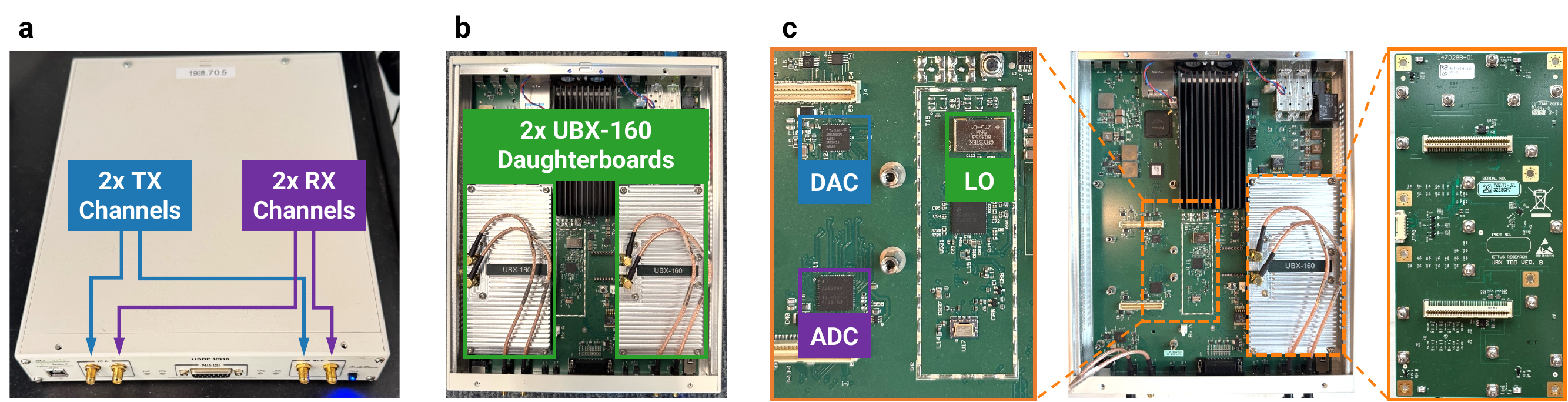}
    \caption{
    \textbf{A close view of the employed software-defined radio (SDR), USRP X310, in {\namebf}'s implementation.}
    \textbf{a--b}, The USRP X310 software-defined radio (SDR) with two UBX-160 daughterboards, supporting two transmit (TX) and two receive (RX) channels.
    \textbf{c}, Close-up view of the USRP X310's internal components, including the digital-to-analog converter (DAC), analog-to-digital converter (ADC), and local oscillator (LO).}
    \label{fig:supp-usrp}
\end{figure*}
%% figure ends 

We use USRP X310 equipped with a UBX-160 daughterboard as the basic transmitter/receiver radio unit, as shown in \autoreffig{fig:supp-usrp}.
Specifically, USRP X310 is a high-performance SDR that supports a carrier frequency of {10}\thinspace{MHz}--{6}\thinspace{GHz}. It is equipped with a 16-bit DAC and a 14-bit ADC per channel, both of which support a sampling rate of {0.196--200}\thinspace{MHz}. Note that the TX and RX path includes an LPF as the anti-aliasing filter in the baseband, whose cutoff frequency is default to half of the TX/RX sampling rate or {80}\thinspace{MHz}, whichever is smaller.
We use a Python-based interface built on top of GNU Radio for radio configuration and data streaming via a {10}\thinspace{Gbps} SFP+ interface to a host server.

The TX channel has a gain setting range of {0--31.5}\thinspace{dB} at a step size of {0.5}\thinspace{dB}, corresponding to a maximum transmitting power of $P_{\textrm{max}} \approx {+23}\thinspace\textrm{dBm}$ with $\abs{\samp[\sampIdx]}=1$, as discussed in Supplementary Section~\ref{sec:supplementary-theory-dac-adc}.
In practice, when transmitting $\wavePrecode(\waveIdx)$ on the central radio, we consider an average baseband I/Q waveform amplitude of $\sqrt{\mathbb{E}[\sampPrecode^2[\sampIdx]]}=0.2$, which corresponds to a peak-to-average power ratio (PAPR) of {14}\thinspace{dB} without saturation (see Supplementary Section~\ref{sec:supplementary-theory-dac-adc}). With the TX gain set to {31}\thinspace{dB}, the average transmit power of the central radio is {$\approx {9}\thinspace\textrm{dBm}$.
For the TX channel that generates $\waveInput(\waveIdx)$, the TX gain is set to {9}\thinspace{dB}, and a total of {30}\thinspace{dB} attenuators are employed to reduce the transmit power. We sweep the baseband I/Q waveform amplitude for different TX power/energy per MAC settings, whose upper bound is also $\sqrt{\mathbb{E}[\sampInput^2[\sampIdx]]}=0.2$, corresponding to a PAPR of {14}\thinspace{dB}.
Similarly, the RX channel has the same gain setting range of {0--31.5}\thinspace{dB} at the step of {0.5}\thinspace{dB}. We configure the RX gain at {20}\thinspace{dB}. Together with the frequency mixer, the RX noise figure is measured at {16.9}\thinspace{dB}, associated with an energy efficiency $\efficiencyRX=2.04 \times {10}^{-2}$.

\subsection{Embedded Anti-Aliasing Filter}

%% figure begins
\begin{figure*}[!t]
    \centering
    \includegraphics[width=0.80\columnwidth]{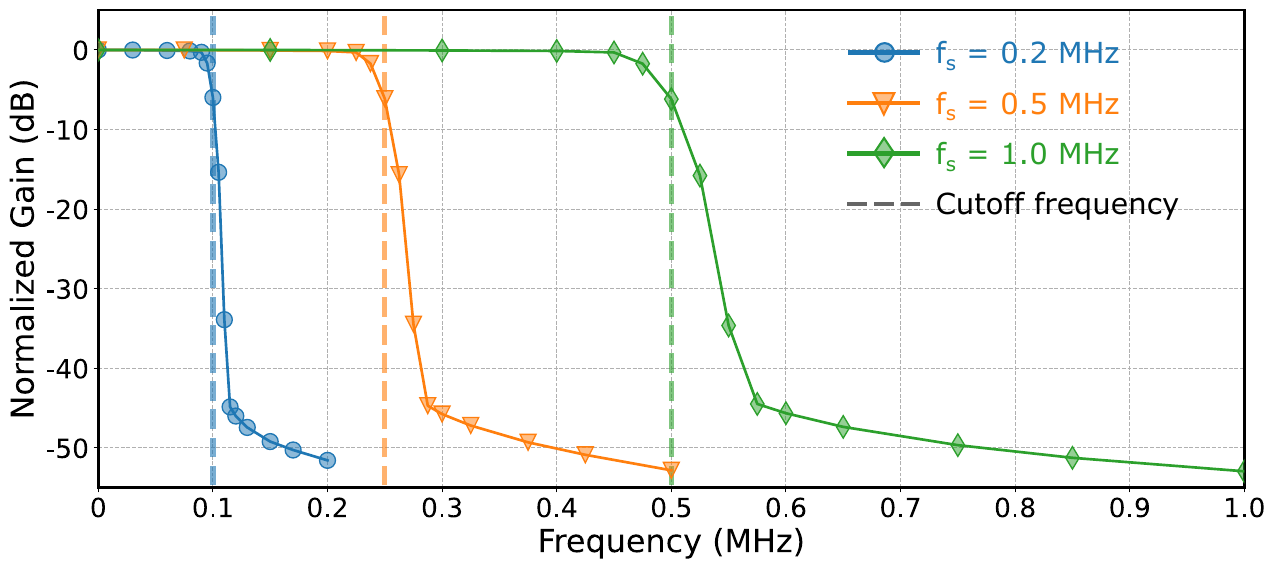}
    \caption{Measured normalized frequency response gain of the USRP X310's internal low-pass filter (LPF) at varying sampling rates, $\sampRate = \{0.2, 0.5, 1.0\}\thinspace\textrm{MHz}$, with cutoff frequencies of $\{0.1, 0.25, 0.5\}\thinspace\textrm{MHz}$ indicated by the dashed lines. A zero-subcarrier padding coefficient of $\overheadPad>0.11$ is sufficient to mitigate the roll-off effect of the LPF under these conditions.}
    \label{fig:anti-aliasing}
\end{figure*}
%% figure ends

We directly use the embedded anti-aliasing filter in USRP X310 as the LPF in {\name}, whose cutoff frequency is set to half of the sampling rate, i.e., $\freq_0 = \sampRate/2$.
To characterize the frequency response of this anti-aliasing filter, we transmit a continuous-wave (CW) signal at the power of {$-$30}\thinspace{dBm} using the signal generator function within the Keysight N9914B FieldFox Handheld RF Analyzer, and sweep its frequency $\freq$ around the carrier frequency of $\carrierOutput=${0.285}\thinspace{GHz}, at which $\waveOutput(\waveIdx)$ is received. Then, the amplitude of the frequency response of the anti-aliasing filter is calculated by the power of the received CW signal referred to the power when the CW signal's frequency is swept to exactly {0.285}\thinspace{GHz}.
Specifically, we consider three low sampling rates employed by the USRP RX $\sampRate = \{0.2, 0.5, 1.0\}\thinspace\textrm{MHz}$, corresponding to the anti-aliasing filter's cutoff frequency as $\freq_0=\{0.1, 0.25, 0.5\}\thinspace\textrm{MHz}$. The frequency of the CW tone, $\freq$, is swept with non-uniform step sizes with smaller step sizes around the cutoff frequency of $\sampRate/2$.

\autoreffig{fig:anti-aliasing} shows the gain of the frequency response of the anti-aliasing filter as part of the USRP X310 SDR, which is employed at the LPF for {\name}. Specifically, the frequency-gain descending slope is proportional to the cutoff frequency $\freq_0$ or the ADC sampling rate $\sampRate$.
In particular, the gain drops to below {$-$50}\thinspace{dB} on the stopband, which is sufficient enough to mitigate the frequency aliasing issue for an SNR of {30}\thinspace{dB} in our implementation. Also, the gain maintains over {$-$0.3}\thinspace{dB} at $\freq = 0.9 \cdot \freq_0$, which corresponds to the zero-subcarrier padding overhead coefficient of $\overheadPad=0.11$.
In our MVM implementation, we consider MVM decomposition with $\outputSizeNew=6$ and $\outputSizePad=1$, i.e., $\overheadPad={0.33}$. This configuration ensures that the padded zero subcarriers are sufficient to compensate for the edge effect of the anti-aliasing filter.
In the extreme case of IP computation with $\outputSizeNew=1$, we still need $\outputSizePad=1$, which leads to a large overhead coefficient of $\overheadPad=2$.

\subsection{Wireless Link Distance and Link Budget Analysis}

In this section, we examine the wireless link distance between the central radio and client that supports in-physics MVM computation via wireless broadcast of model weights, based on the optimized LO input power and \autoreffig{fig:mixer-linear}.
The link budget equation for a wireless link is given by
\begin{align}
    P_{\textrm{RX}} \textrm{[dBm]} &= P_{\textrm{TX}} \textrm{[dBm]} + G_{\textrm{TX}} \textrm{[dBi]} + BF_{\textrm{TX}} \textrm{[dB]} - L_{\textrm{TX}} \textrm{[dB]} - L_{\textrm{prop}} \textrm{[dB]} + G_{\textrm{RX}} \textrm{[dBi]} + BF_{\textrm{RX}} \textrm{[dB]} - L_{\textrm{RX}} \textrm{[dB]},
    \label{eq: link-distance-link-budget}
\end{align}
where $P_{\textrm{TX}}$ (resp. $P_{\textrm{RX}}$) denotes the TX (resp. RX) signal power, $G_{\textrm{TX}}$ (resp. $G_{\textrm{RX}}$) denotes the TX (resp. RX) antenna gain, $BF_{\textrm{TX}}$ (resp. $BF_{\textrm{RX}}$) denotes the TX (resp. RX) beamforming gain if an antenna array is employed for beamforming, $L_{\textrm{TX}}$ (resp. $L_{\textrm{RX}}$) denotes the insertion loss on the TX )(resp. RX) due to connectors and cables, etc., and $L_{\textrm{prop}}$ is the path loss of the wireless link.
In particular, we consider the free space path loss~\cite{friis1946note} given by
\begin{align}
    L_{\textrm{prop}} \textrm{[dB]} = 10 \cdot \log_{10} \left( \frac{4\pi d \carrier }{c} \right)^2
    = 20 \cdot \log_{10} \left( \frac{4\pi d \carrier }{c} \right),
    \label{eq: link-distance-path-loss}
\end{align}
where $d$ is the link distance between the TX and RX, $\carrier$ is the carrier frequency, and $c$ is the speed of light. 
Combining equations {\eqref{eq: link-distance-link-budget}} and {\eqref{eq: link-distance-path-loss}}, the link distance, $d$, can be written as
\begin{align}
    d \approx 10^{(P_{\textrm{TX}} - P_{\textrm{RX}} + G_{\textrm{TX}} + BF_{\textrm{TX}} - L_{\textrm{TX}} + G_{\textrm{RX}} + BF_{\textrm{RX}} - L_{\textrm{RX}})/20} \cdot \left(\frac{c}{4\pi \carrier}\right).
\end{align}
%%

%% figure begins
\begin{figure*}[!t]
    \centering
    \includegraphics[width=0.85\columnwidth]{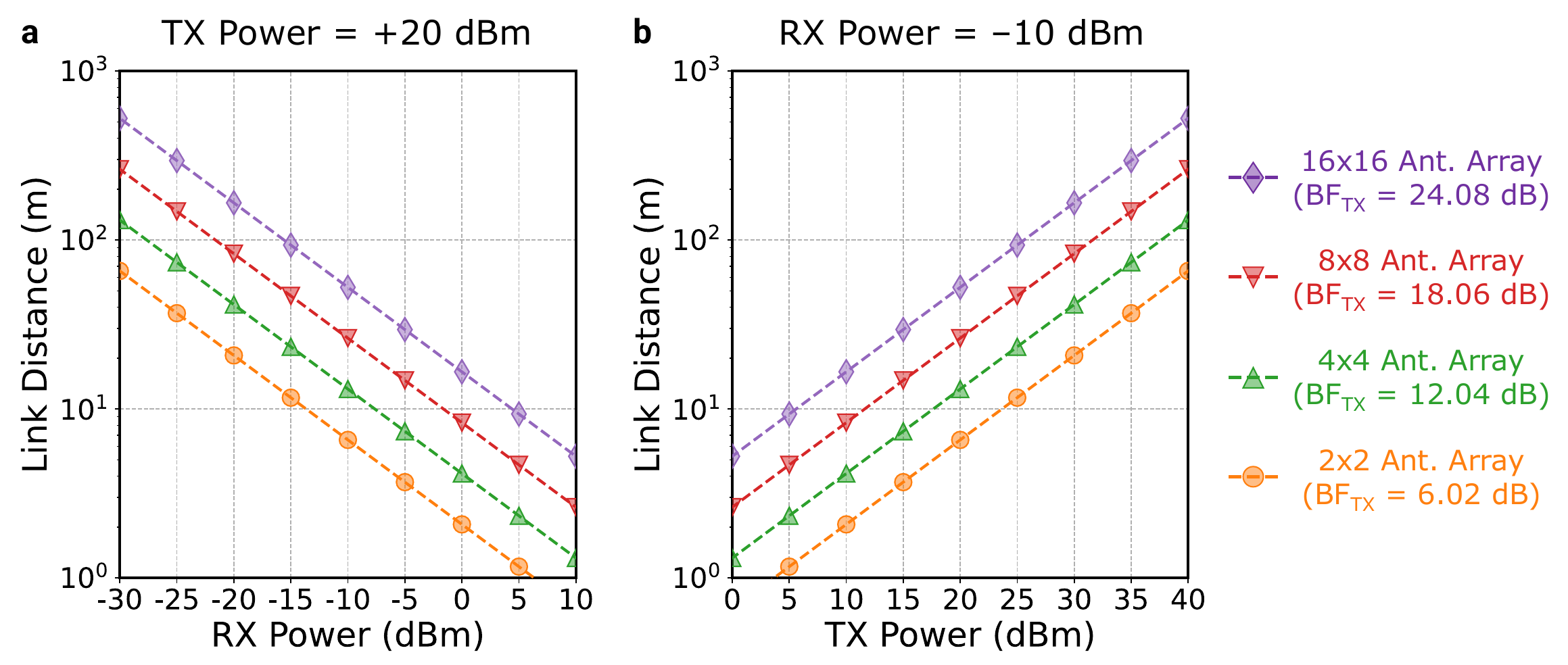}
    \caption{
    \textbf{The theoretical link distance analysis applying the free-space path loss model.}
    \textbf{a}, The link distance under varying TX power levels at the central radio, $P_{\textrm{TX}}$, where the central radio is equipped with an antenna array supporting TX beamforming, and each client is equipped with a signal antenna, and the antenna gain is $G_{\textrm{TX}} = G_{\textrm{RX}} = {6}\thinspace\textrm{dBi}$.
    \textbf{b}, The link distance under varying RX power input levels at the LO port of the computing mixer, $P_{\textrm{RX}}$, under the same setting as \textbf{a}.}
    \label{fig: link-distance}
\end{figure*}
%% figure ends

In our implementation, the USRP X310 supports an average transmit power of $P_{\textrm{TX}}={+9}\thinspace\textrm{dBm}$ with a PAPR of up to {14}\thinspace{dB}, the Yagi antenna provides an antenna gain of $G_{\textrm{TX}}=G_{\textrm{RX}}={9}\thinspace\textrm{dBi}$ with no beamforming ($BF_{\textrm{TX}}=BF_{\textrm{RX}}=0\thinspace\textrm{dB}$, and the insertion losses $L_{\textrm{TX}}=L_{\textrm{RX}}\approx{0}\thinspace\textrm{dB}$ due to short cable length and minimal connections. To feed the LO port with $P_{\textrm{RX}}={-3}\thinspace\textrm{dBm}$, the path loss $L_{\textrm{prop}}$ should be {30}\thinspace{dB}.
Plugging in the carrier frequency of $\waveWeight(\waveIdx) = {0.915}\thinspace\textrm{GHz}$, the wireless link distance is recommended to be around 1 meter, as employed in our experiments. This wireless link is determined by the relatively high input power at the computing mixer's LO power required to drive the frequency mixer of our choice. Such a high input power comes from the double-balanced diode architecture of this computing mixer.

To support wireless broadcast of model weight over larger wireless link distances, one can consider using a computing mixer that requires a lower LO input power (i.e., smaller values of $P_{\textrm{RX}}$). For example, RF mixers integrating an internal LO amplifier (e.g., the PE4152 UltraCMOS quad MOSFET mixer from pSemi) or analog multipliers based on integrated analog correlators~\cite{rashed2024scalable} can relax the input power constraint on $P_{\textrm{RX}}$. Another approach is to employ antennas with a higher antenna gain (i.e., larger values of $G_{\textrm{TX}}$ and/or $G_{\textrm{RX}}$), or beamforming using an antenna array (i.e., larger values o $BF_{\textrm{TX}}$ and/or $BF_{\textrm{RX}}$).
Specifically, when referred to a single antenna, a planar antenna array with $N_{\textrm{ant}} \times N_{\textrm{ant}}$ antenna elements with half-wavelength spacing between adjacent elements can provide a maximum beamforming gain of
\begin{align}
    BF_{\textrm{TX}}\textrm{[dB]} = BF_{\textrm{RX}}\textrm{[dB]} = 10 \log_{10} \left(N_{\textrm{ant}}^2 \right).
\end{align}
For example, the Argos massive MIMO radio~\cite{shepard2012argos} employs a {sub-7}\thinspace{GHz} antenna array with $N_{\textrm{ant}}=8$, which supports a beamforming gain $BF_{\textrm{TX}}$ of up to {18.06}\thinspace{dB} on the central radio. We illustrate the theoretical link distance in \autoreffig{fig: link-distance}, where the central radio employs an antenna array supporting TX beamforming. For example, as shown in \autorefsubfig{fig: link-distance}{a}, an {8}$\times${8} antenna array, which is commonly employed in modern cellular networks, can support a link distance of 100 meters with an improved TX power of $P_{\textrm{TX}}={+31.8}\thinspace\textrm{dBm}$.

%%%%%
%%%%%
\subsection{Time and Frequency Synchronization}

Generally, the central radio and each client are not naturally synchronized in the time or frequency domain. Specifically, the client is not aware of the starting point of the transmitted waveform from the central radio, and the LOs on both the central radio and clients may exhibit a carrier frequency offset (CFO), which can lead to inter-subcarrier interference, especially when the subcarrier spacing $\freqSub$ is small. Therefore, we insert preambles into $\waveInput(\waveIdx)$ and $\waveWeight(\waveIdx)$, which can be used for time and frequency synchronization between the central radio and each client.

Specifically, given a downsampling ratio $\inputSize$, the preamble as an I/Q sample sequence are defined by $\sampInputPreVec = [\sampInputPre[\sampIdx]] \in \mathbb{C}^{2\inputSize\preamLen}$ and $\sampWeightPreVec = [\sampWeightPre[\sampIdx]] \in \mathbb{C}^{2\inputSize\preamLen}$ for the baseband I/Q waveforms corresponding to $\waveInput(\waveIdx)$ and $\waveWeight(\waveIdx)$, respectively, where $\preamLen$ is recommended to be a prime number.
These two I/Q waveforms are streamed to the DACs operating at a sampling rate of $\sampRate$ to generate the analog waveform $\waveInputPre(\waveIdx)$ and $\waveWeightPre(\waveIdx)$.
These two preambles are composed of two identical sequences, each with $\inputSize\preamLen$ I/Q samples, each of which is generated with a constant amplitude $A$ and randomized phases $\phi_{x, \sampIdx}$ and $\phi_{w, \sampIdx}$ for $\sampInputPreVec$ and $\sampWeightPreVec$, respectively. Specifically, we consider a large amplitude $A$ close to $1$ to ensure a high output power and SNR without saturation; the phases are uniformly distributed within $[0, 2\pi]$ to ensure that the signal power is evenly distributed across the frequency band.
To sum up, the preamble generation can be written as
\begin{align}
    \sampInputPre[\sampIdx] & = \sampInputPre[\sampIdx + \inputSize\preamLen],\
    \sampWeightPre[\sampIdx] = \sampWeightPre[\sampIdx + \inputSize\preamLen],\ 
    \forall \sampIdx = 0, 1, \dots, \inputSize\preamLen-1. \\
    \sampInputPre[\sampIdx] & = A \cdot \eu^{\iu \phi_{x, \sampIdx}},\
    \sampWeightPre[\sampIdx] = A \cdot \eu^{\iu \phi_{w, \sampIdx}},\
    \forall \sampIdx = 0, 1, \dots, \inputSize\preamLen-1,\
    \text{where}~\phi_{x, \sampIdx}, \phi_{w, \sampIdx} \sim \mathcal{U}[0, 2\pi].
\end{align}
Based on equation {\eqref{eq: original-mvm-up-conversion-frequency}}, the received waveform $\waveOutputPre(\waveIdx)$ experiences a CFO, $\Delta\carrier$, between the TX and RX, i.e.,
\begin{align}
    \waveOutputPre(\waveIdx) \propto \waveInputPre(\waveIdx) \cdot \waveWeightPre(\waveIdx) \cdot \eu^{\iu \Delta\carrier \waveIdx}.
\end{align}
After the downsampling ratio of $\inputSize$, we denote the I/Q waveform corresponding to $\waveOutputPre(\waveIdx)$ as $\sampOutputPreVec = [\sampOutputPre[\sampIdx]] \in \mathbb{C}^{2\preamLen}$.
Assuming that the channel is stable within the transmission time of the preambles, the received $\sampOutputPreVec$ will include two identical sequences given by
\begin{align}
    \sampOutputPre[\sampIdx] = \sampOutputPre[\sampIdx + \preamLen] \cdot \eu^{\iu \Delta\carrier \cdot \frac{\sampIdx\inputSize}{\sampRate}},\
    \forall \sampIdx = 0, 1, \dots, \preamLen-1.
\end{align}
Therefore, the starting point of $\sampOutputPreVec$ can be detected by performing an auto-correlation with a copy of itself delayed by $\preamLen$ I/Q samples~\cite{bloessl2013ieee, gao2023swirls}, i.e.,
\begin{align}
    R_{y, \textrm{pre}} = \frac{\abs{\sum_{\sampIdx=0}^{\preamLen-1} \sampOutputPre[\sampIdx] \cdot \sampOutputPreConj[\sampIdx+\preamLen]}}{\sum_{\sampIdx=0}^{\preamLen-1} \sampOutputPre[\sampIdx] \cdot \sampOutputPreConj[\sampIdx]} \in [0, 1].
\end{align}
In practice, we calculate the auto-correlation, $R_{y, \textrm{pre}}$, for a sliding window containing $2\preamLen$ I/Q samples.
When the calculated $R_{y, \textrm{pre}}$ exceeds a threshold on a given sliding window as a local minimum (e.g., 0.8), a preamble $\sampOutputPreVec$ is considered to be detected, and the starting point of this sliding window is considered as the preamble's starting point.
When the starting point of $\sampOutputPreVec$ is detected, we can infer the starting point of the desired waveform $\waveOutput(\waveIdx)$ accordingly.
Generally, the starting point error of this auto-correlation-based detection algorithm is determined by the sliding window step~\cite{bloessl2013ieee}. As long as we calculate $R_{y, \textrm{pre}}$ for the sliding windows at the step size of every I/Q sample, a sub-symbol timing offset error can be achieved. Hence, one or two I/Q samples per OFDM symbol for the cyclic prefix is sufficient to ensure the desired synchronization performance.

In addition, the CFO can be estimated by
\begin{align}
    \widehat{\Delta\carrier} = \frac{\textsf{Angle}\Big(\sum_{\sampIdx=0}^{\preamLen-1} \sampOutputPre[\sampIdx] \cdot \sampOutputPreConj[\sampIdx+\preamLen] \Big)}{2\pi \inputSize \preamLen/\sampRate}.
\end{align}
To calibrate, we can either fine-tune the LO frequency $\carrierOutput$, or apply the estimated CFO on the I/Q sample sequence $\sampOutputVec$ in the digital domain~\cite{bloessl2013ieee, gao2023swirls}, i.e.,
\begin{align}
    \sampOutput^{\prime}[\sampIdx] = \sampOutput[\sampIdx] \cdot \eu^{-\iu \widehat{\Delta\carrier} \cdot \frac{\sampIdx\inputSize}{\sampRate}}.
\end{align}
To conclude, such a preamble-driven synchronization method allows a short cyclic prefix (e.g., $\fftSizeCP=1$) and a small subcarrier spacing $\freqSub$ for a large-scale of subcarrier assignment within the accessible bandwidth.

%%%%%
%%%%%
\section{Channel Calibration Schemes}
\label{sec:supplementary-experiment-ml-calibration}

In this section, we evaluate and compare the performance of {\name} across the three schemes: (\emph{i}) the basic scheme (Supplementary Section~\ref{sec:supplementary-theory-basic-scheme}), (\emph{ii}) the $\weightMat$-precoding scheme (Supplementary Section~\ref{sec:supplementary-theory-weight-precoding-scheme}), and (\emph{iii}) the $\inputVec$-precoding scheme (Supplementary Section~\ref{sec:supplementary-theory-input-precoding-scheme}).

\subsection{General MVM Computation}

%% figure begins
\begin{figure*}[!t]
    \centering
    \includegraphics[width=0.9\columnwidth]{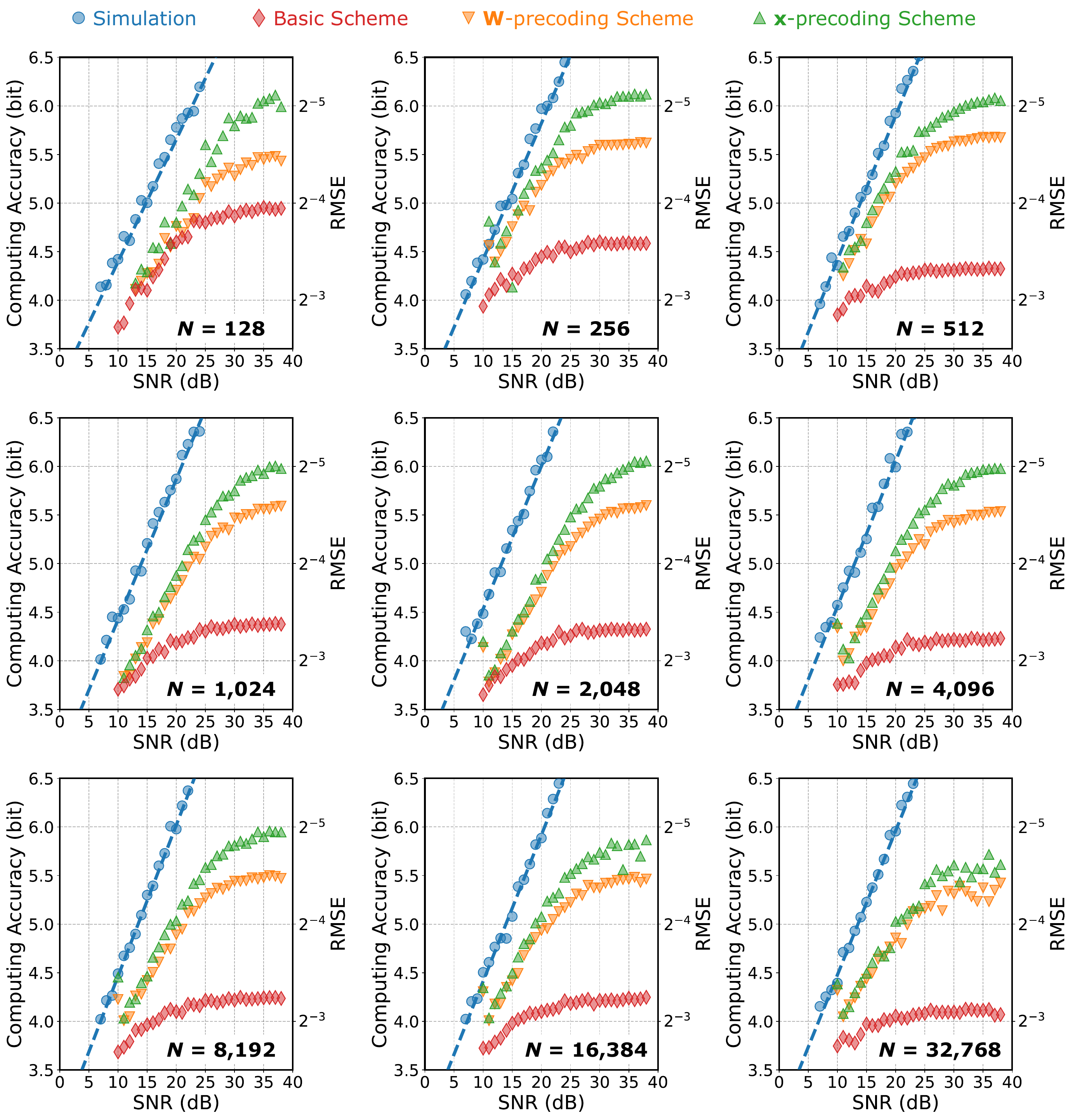}
    \caption{Benchmark computing accuracy achieved by {\name}'s difference schemes compared to simulations for matrix-vector multiplication (MVM) decomposed into inner-products (IPs), with randomized $\weightMat$ and $\inputVec$. Results are shown across varying input/output size, $\inputSize=\outputSize \in \{2^{7}, 2^{8}, \dots, 2^{15}\}$, and SNR values.}
    \label{fig:supplementary-snr-resol-all}
\end{figure*}
%% figure ends

%% figure begins
\begin{figure*}[!t]
    \centering
    \includegraphics[width=0.95\columnwidth]{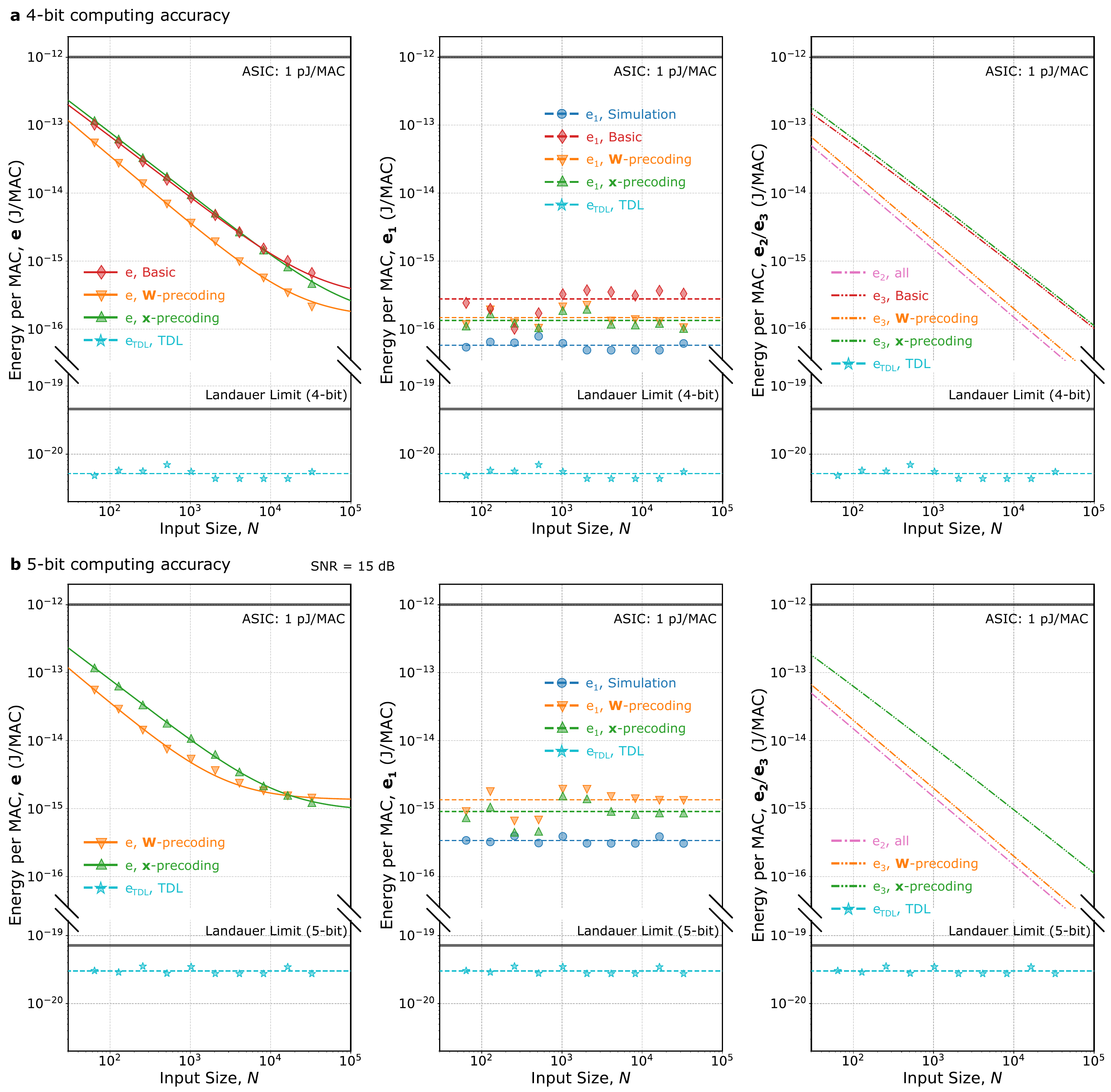}
    \caption{
    \textbf{The IP-based energy efficiency benchmarking of {\namebf}'s basic, $\weightMat$-precoding, and $\inputVec$-precoding schemes.}
    \textbf{a}, The minimum energy per MAC required by the basic, $\weightMat$-precoding, and $\inputVec$-precoding schemes to achieve $\textsf{RMSE} < 0.125$ (4-bit computing accuracy), with detailed breakdowns. In addition, the energy efficiency corresponding to the thermodynamic limit (TDL) is simulated and compared to the Landauer limit.
    \textbf{b}, The minimum energy per MAC for $\textsf{RMSE} < 0.0625$ (5-bit computing accuracy) of the $\weightMat$-precoding and $\inputVec$-precoding schemes. Note that the basic scheme cannot achieve the 5-bit computing accuracy and is thus not shown.}
    \label{fig:supplementary-inputsize-emac-all}
\end{figure*}
%% figure ends

We first benchmark {\name}'s three schemes for general complex-valued MVM computation, $\outputVec = \weightMat \cdot \inputVec$.
In particular, the IP-based MVM decomposition is considered with randomized $\inputVec$ and squared $\weightMat$ (i.e., $\inputSize=\outputSize$), where each element $\inputElem_{\inputIdx}$ and $\weightElem_{\outputIdx, \inputIdx}$ are independently randomized with uniformly distributed amplitudes $\abs{\inputElem},~\abs{\weightElem} \sim \mathcal{U}[0, 1]$ and uniformly distributed phases $\angle{\inputElem_{\inputIdx}},~\angle{\weightElem_{\outputIdx, \inputIdx}} \sim \mathcal{U}[0, 2\pi]$.
As discussed in Supplementary Section~\ref{sec:supplementary-theory-input-precoding-scheme}, the $\weightMat$-precoding scheme is independent of $\outputSize$, which means this MVM computation is equivalent to the IP computation in {\autorefresults}. This MVM computation is also suitable for the basic and $\inputVec$-precoding schemes that do not comply with the standalone IP computations.
Given the in-phyics MVM computing result ($\widehat{\outputVec}$) and ground truth ($\outputVec$), we define the RMSE~\cite{sludds2022delocalized, davis2022frequency} as 
\begin{align}
    \textsf{RMSE}=\sqrt{\mathbb{E}\left[ \frac{1}{\outputSize} \cdot \sum_{\outputIdx=0}^{\outputSize-1} \abs{\widehat{\outputElem}_{\outputIdx} - \outputElem_{\outputIdx}}^2 \right]},
\end{align}
and the computing accuracy can be derived as $-\log_2 (\textsf{RMSE}/2)$ [bit].

\autoreffig{fig:supplementary-snr-resol-all} compares the experimental computing accuracy of the three schemes with simulation results under perfect channel calibration and analog multiplication performed by an ideal computing mixer. The comparison is performed under varying MVM dimensions, where $\inputSize=\outputSize \in \{2^{7}, 2^{8}, \dots, 2^{15}\}$.
Overall, the experimental results show that the three schemes require approximately {5}\thinspace{dB} higher SNR than the simulations to achieve the same computing accuracy. The experimental computing accuracy is limited in the high SNR regime (e.g., $>${30}\thinspace{dB}) due to the on-off switching behavior of the double-balanced diode mixer, as discussed in Supplementary Section~\ref{sec:supplementary-experiment-setup}.
Across all values of $\inputSize$, the $\weightMat$-precoding and $\inputVec$-precoding schemes achieve higher computing accuracy than the basic scheme. For example, at {25}\thinspace{dB} SNR, the RMSE of the $\weightMat$-precoding scheme is {0.055/0.056} and RMSE of the $\inputVec$-precoding is {0.043/0.047} with $\inputSize = 4,096/32,768$, corresponding to $>$5-bit computing accuracy. In contrast, the RMSE of the basic scheme is {0.109/0.118}, corresponding to $\approx$4-bit computing accuracy. This performance gap highlights the effectiveness of the wireless channel calibration of the $\weightMat$-precoding and $\inputVec$-precoding schemes.
In addition, at higher SNR levels, the $\inputVec$-precoding scheme outperforms the $\weightMat$-precoding scheme, achieving an RMSE of {0.032/0.042} at {35}\thinspace{dB} SNR with $\inputSize=4,096/32,768$, equivalent to $\approx$6-bit computing accuracy. This is because the $\inputVec$-precoding scheme supports CSI estimation and calibration for individual clients.
Moreover, as $\inputSize$ increases, the computing accuracy achieved by all three schemes degrades, especially for the basic scheme. This degradation occurs since a larger value of $\inputSize$ results in reduced subcarrier spacing, $\freqSub$, as more subcarriers are packed within the same bandwidth, $\band$. In this case, the impact of inter-subcarrier interference becomes more significant, requiring finer frequency synchronization.

Next, we benchmark the energy efficiency of the three schemes across different input sizes, $\inputSize$, based on the energy efficiency analysis for IP-based MVM decomposition given by {\eqref{eq: basic-scheme-energy-mac-multiple}}, {\eqref{eq: weight-precoding-scheme-energy-per-mac-multiple}}, and {\eqref{eq: input-precoding-scheme-energy-mac-multiple}}, respectively.
The minimum energy per MAC required for the three schemes to achieve $\textsf{RMSE}<0.125$ and $\textsf{RMSE}<0.0625$ (4-bit and 5-bit computing accuracy, respectively) is shown in \autoreffig{fig:supplementary-inputsize-emac-all}. Note that the energy efficiency of the basic scheme to reach 5-bit computing accuracy is excluded since it cannot achieve this computing accuracy across all considered SNR values.
As summarized in \autoreftab{fig:supp-energy-table}, the energy efficiency ($\energyMAC$) of the basic and $\inputVec$-precoding schemes scales as $\complexity(\frac{1}{\inputSize}\log\inputSize)$ and the energy efficiency of the $\weightMat$-precoding scheme scales $\complexity(1/\inputSize)$. With large values of $\inputSize$, $\energyMAC$ converges to the energy efficiency corresponding to the waveform generation, $\energyMACTx$.
Specifically, with $\inputSize=4,096$, the $\weightMat$-precoding scheme has an energy efficiency of {0.99}\thinspace{fJ/MAC} and {2.37}\thinspace{fJ/MAC} ({1,014.20}\thinspace{TOPS/W} and {422.14}\thinspace{TOPS/W}) for achieving $\textsf{RMSE}<0.125$ (4-bit) and $\textsf{RMSE}<0.0625$ (5-bit), respectively, while the energy efficiency for the $\inputVec$-precoding scheme is {3.17}\thinspace{fJ/MAC} and {3.96}\thinspace{fJ/MAC} ({315.44}\thinspace{TOPS/W} and {252.26}\thinspace{TOPS/W}).
The energy saving of the $\weightMat$-precoding scheme comes from the elimination of the FFT-based encoding and precoding for wireless channel calibration, as discussed in Supplementary Section~\ref{sec:supplementary-theory-weight-precoding-scheme}, and thus a lower digital computing cost from $\energyMACDec={2.69}\thinspace\textrm{fJ/MAC}$ for the $\inputVec$-precoding scheme to $\energyMACDec={0.49}\thinspace\textrm{fJ/MAC}$ for the $\weightMat$-precoding scheme.
On the other hand, the extra energy for digital computing is averaged down as the $\inputSize$ and/or $\outputSize$ increases. For example, at $\inputSize={32,768}$, the energy efficiency becomes {0.21}\thinspace{fJ/MAC} and {1.43}\thinspace{fJ/MAC} ({4,710.54}\thinspace{TOPS/W} and {697.01}\thinspace{TOPS/W}) for the $\weightMat$-precoding scheme to reach $\textsf{RMSE}<0.125$ (4-bit) and $\textsf{RMSE}<0.0625$ (5-bit), respectively, which is {0.53}\thinspace{fJ/MAC} and {1.29}\thinspace{fJ/MAC} ({1,888.44}\thinspace{TOPS/W} and {774.47}\thinspace{TOPS/W}) for the $\inputVec$-precoding scheme.
By assuming an ideal channel calibration and perfect hardware with no overhead, we also simulate the TDL energy efficiency given by $\energyMACTDL$ in equations {\eqref{eq: basic-scheme-energy-per-mac-tdl}}, {\eqref{eq: weight-precoding-scheme-energy-per-mac-tdl}}, and {\eqref{eq: input-precoding-scheme-energy-per-mac-tdl}}, following the same forms over the three schemes.
The TDL energy efficiency averaging across all input sizes, $\inputSize$, is {5.15}\thinspace{zJ/MAC} and {30.15}\thinspace{zJ/MAC} ({194.17}\thinspace{EOPS/W} and {33.17}\thinspace{EOPS/W}) for the 4-bit and 5-bit computing accuracy, which is {9.1}$\times$ and {2.4}$\times$ lower than the corresponding 4-bit and 5-bit Landauer limit of {45.9}\thinspace{zJ/MAC} and {71.8}\thinspace{zJ/MAC}, respectively.

%% figure begins
\begin{figure*}[!t]
    \centering
    \includegraphics[width=0.95\columnwidth]{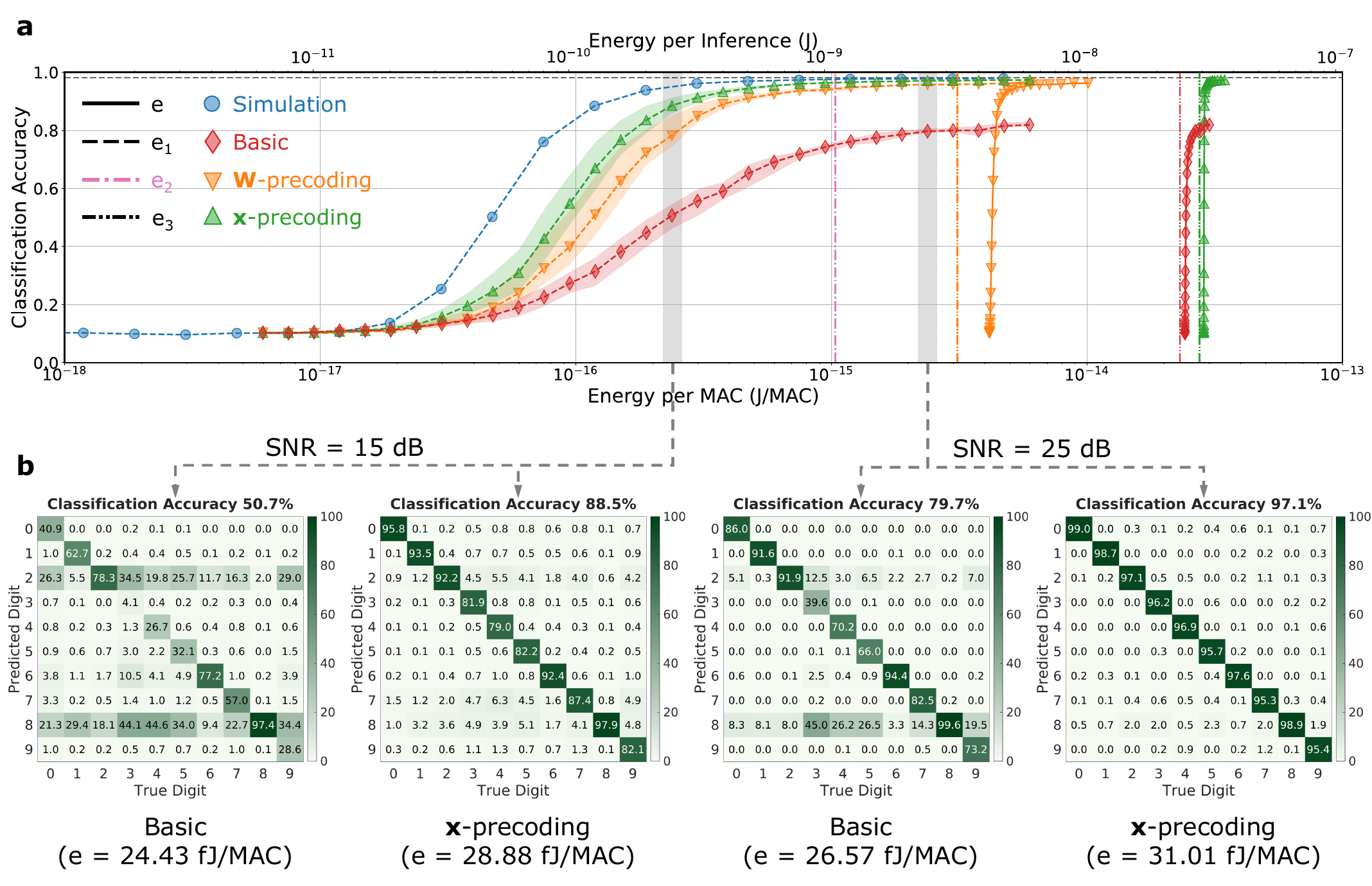}
    \caption{
    \textbf{The basic, $\weightMat$-precoding and $\inputVec$-precoding scheme comparison on the image classification task on the MNIST dataset.}
    \textbf{a}, Classification accuracy achieved by {\name}'s three schemes and simulation, shown as a function of the energy efficiency, $\energyMAC$, with a detailed breakdown into $\energyMACTx$, $\energyMACADC$, and $\energyMACDec$. The shaded area of $\energyMACTx$ indicates the accuracy variance across three clients under the same SNR.
    \textbf{b}, Confusion matrices of the basic scheme without channel calibration and the $\inputVec$-precoding scheme with per-client channel calibration at {15}\thinspace{dB} and {25}\thinspace{dB} SNR.}
    \label{fig:supplementary-scheme-mnist}
\end{figure*}
%% figure ends

\subsection{Image Classification on the MNIST Dataset}

We implement {\name} on a complex-valued model with three FC layers based on LeNet-300-100~\cite{lecun1998gradient} for handwritten digit image classification on the MNIST dataset.
\autorefsubfig{fig:supplementary-scheme-mnist}{a} shows the energy efficiency of {\name} on the MNIST dataset. For the MNIST dataset, the maximum input size is $\inputSize=784$, where the digital computing energy efficiency term $\energyMACDec$ dominates the total energy efficiency $\energyMAC$.
The energy efficiency required by the $\weightMat$-precoding scheme to achieve a classification accuracy of {90\%} is {4.62}\thinspace{fJ/MAC} ({216.35}\thinspace{TOPS/W}), which include $\energyMACTx=0.47\thinspace\textrm{fJ/MAC}$ for waveform generation and I/Q (de)modulation with {18.3}\thinspace{dB} SNR, $\energyMACADC={1.04}\thinspace\textrm{fJ/MAC}$ for I/Q sampling, and $\energyMACDec={3.11}\thinspace\textrm{fJ/MAC}$ for decoding performed in digital computing.
For the $\inputVec$-precoding scheme, the energy efficiency to achieve {90\%} classification accuracy is {28.94}\thinspace{fJ/MAC} ({35.55}\thinspace{TOPS/W}), with a breakdown of $\energyMACTx={0.30}\thinspace\textrm{fJ/MAC}$ (at {16.3}\thinspace{dB} SNR), $\energyMACADC={1.04}\thinspace\textrm{fJ/MAC}$, and $\energyMACDec={27.60}\thinspace\textrm{fJ/MAC}$.
Despite the degraded energy efficiency of the $\inputVec$-precoding scheme, it is still significantly better than that of digital computing using state-of-the-art ASICs at {1}\thinspace{pJ/MAC}~\cite{horowitz20141, abari201427, jouppi2017datacenter}. Note that the basic scheme, however, can only achieve a classification accuracy of up to {81.9\%} on the MNIST dataset.

Detailed confusion matrices of the classification accuracy achieved by the basic and $\inputVec$-precoding schemes at {15}\thinspace{dB} and {25}\thinspace{dB} are shown in \autorefsubfig{fig:supplementary-scheme-mnist}{b}. Due to the lack of CSI estimation and calibration, the basic scheme achieves the classification accuracy of only {50.7\%} and {79.7\%} under {15}\thinspace{dB} and {25}\thinspace{dB} SNR, respectively. In contrast, with proper CSI estimation and calibration, the $\weightMat$-precoding scheme achieves a classification accuracy of {78.2\%} and {95.7\%}, and the $\inputVec$-precoding scheme achieves a classification accuracy of {88.5\%} and {97.1\%} under the same SNR values. With an increased SNR of {29.3}\thinspace{dB}, the $\inputVec$-precoding scheme achieves a maximum classification accuracy of {97.4\%}, which is only {0.7\%} lower than the classification accuracy of {98.1}\% based on digital computing.

\subsection{Audio Signal Classification on the AudioMNIST Dataset}

%% figure begins
\begin{figure*}[!t]
    \centering
    \includegraphics[width=0.95\columnwidth]{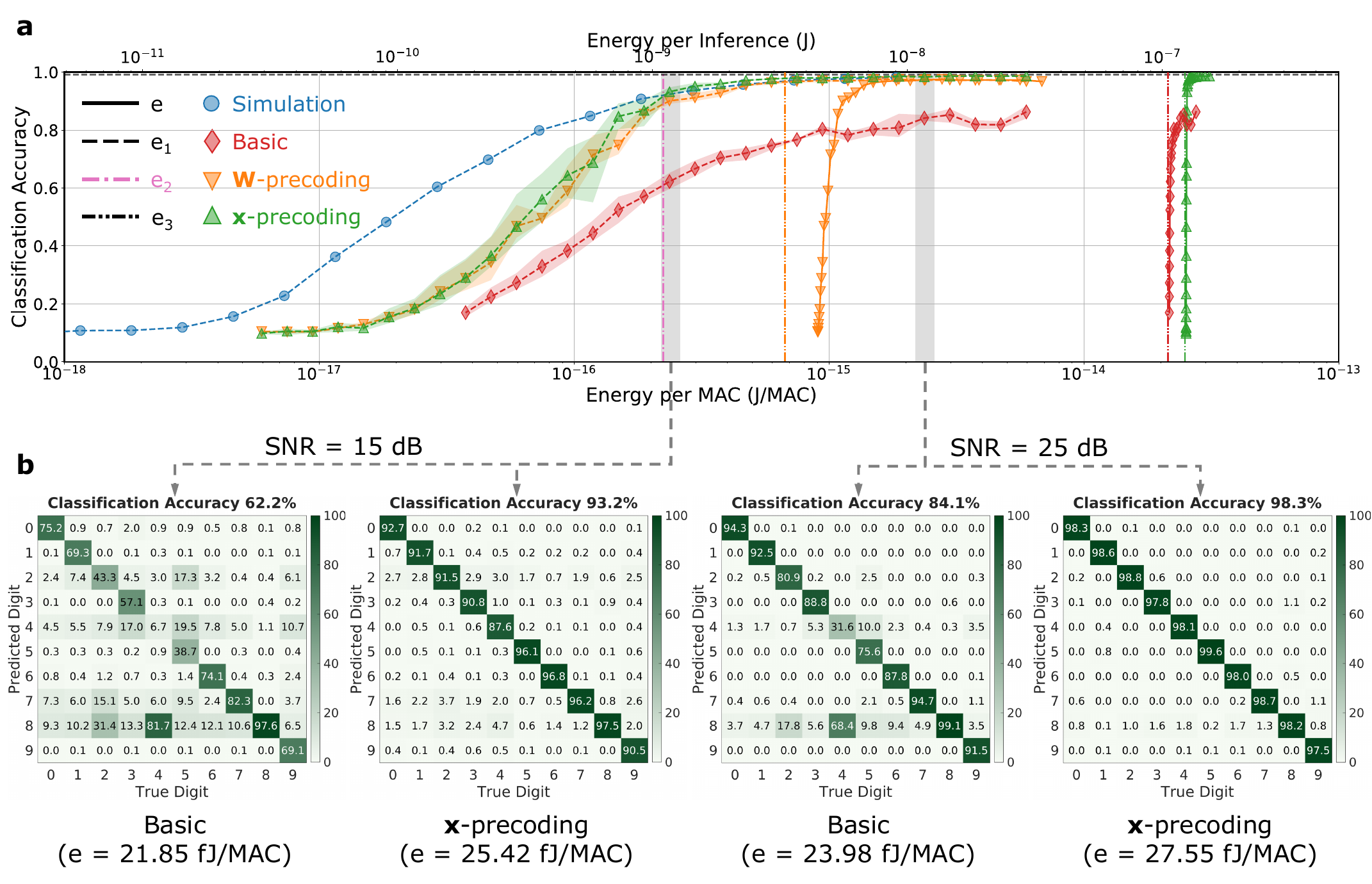}
    \caption{
    \textbf{The basic, $\weightMat$-precoding and $\inputVec$-precoding scheme comparison on the audio signal classification task on the MNIST dataset.}
    \textbf{a}, Classification accuracy achieved by {\name}'s three schemes and simulation, shown as a function of the energy efficiency with detailed breakdowns. The shaded area of $\energyMACTx$ indicates the accuracy variance across three clients under the same SNR.
    \textbf{b}, Confusion matrices of classification accuracy achieved by the basic scheme without channel calibration and the $\inputVec$-precoding scheme with per-client channel calibration at {15}\thinspace{dB} and {25}\thinspace{dB} SNR.}
    \label{fig:supplementary-scheme-audiomnist}
\end{figure*}
%% figure ends

We also evaluate the performance of {\name} on the AudioMNIST dataset~\cite{audiomnist2023} with spoken digits for audio signal classification, using a complex-valued model with three FC layers based on LeNet-300-100, with an input size of $\inputSize={4,000}$.
\autorefsubfig{fig:supplementary-scheme-audiomnist}{a} shows the energy efficiency of {\name} achieved by the three schemes. Overall, the energy efficiency of {\name} is lower on the AudioMNIST dataset compared to that on the MNIST dataset due to the large input size of $\inputSize={4,000}$.
Specifically, achieving an accuracy of {90\%} by the $\weightMat$-precoding scheme requires a minimum SNR of {15.3}\thinspace{dB} and an energy efficiency of {1.13}\thinspace{fJ/MAC} ({882.10}\thinspace{TOPS/W}), which includes $\energyMACTx={0.24}\thinspace\textrm{fJ/MAC}$, $\energyMACADC={0.22}\thinspace\textrm{fJ/MAC}$, and $\energyMACDec={0.67}\thinspace\textrm{fJ/MAC}$.
Moreover, the $\inputVec$-precoding scheme requires a minimum SNR of {15.3}\thinspace{dB} and an energy efficiency of {25.42}\thinspace{fJ/MAC} ({33.34}\thinspace{TOPS/W}), which can be decomposed into $\energyMACTx={0.24}\thinspace\textrm{fJ/MAC}$, $\energyMACADC={0.22}\thinspace\textrm{fJ/MAC}$, and $\energyMACDec={24.96}\thinspace\textrm{fJ/MAC}$. Similarly, the degraded energy efficiency of the $\inputVec$-precoding scheme results from the encoding and precoding performed in digital computing, which dominates the total energy efficiency, given the problem size of the AudioMNIST dataset.
In this case, the $\inputVec$-precoding scheme achieves an energy efficiency gain of approximately ${40}\times$ compared to the {1}\thinspace{pJ/MAC} energy efficiency by the state-of-the-art ASICs.
Beyond, as the MVM scales up on the future DL tasks, e.g., Llama-2-7b~\cite{touvron2023llama} with $\inputSize={11,008}$, the energy efficiency of both the $\weightMat$-precoding and $\inputVec$-precoding schemes can be further improved.

Detailed confusion matrices of the classification accuracy achieved by the basic and $\inputVec$-precoding schemes at {15}\thinspace{dB} and {25}\thinspace{dB} are shown in \autorefsubfig{fig:supplementary-scheme-audiomnist}{b}. It can be seen that the basic scheme achieves a classification accuracy of {62.2\%} and {84.1\%} under {15}\thinspace{dB} and {25}\thinspace{dB} SNR, respectively, and is bounded by {86.3\%} with further increased SNR values. 
On the other hand, the $\inputVec$-precoding scheme that exploits the per-client CSI estimation and calibration archives a classification accuracy of {93.2\%} and {98.3\%} under {15}\thinspace{dB} and {25}\thinspace{dB} SNR, outperforming the $\weightMat$-precoding scheme that achieves a classification accuracy of {90.1\%} and {97.2\%}. Further, it achieves a maximum accuracy of {98.6}\%, only {0.6}\% lower compared to the classification accuracy of {99.2}\% based on digital computing.

%%%%%
%%%%%
\section{MVM Decomposition into IPs}
\label{sec:supplementary-experiment-ml-inner}

%% figure begins
\begin{figure*}[!t]
    \centering
    \includegraphics[width=0.95\columnwidth]{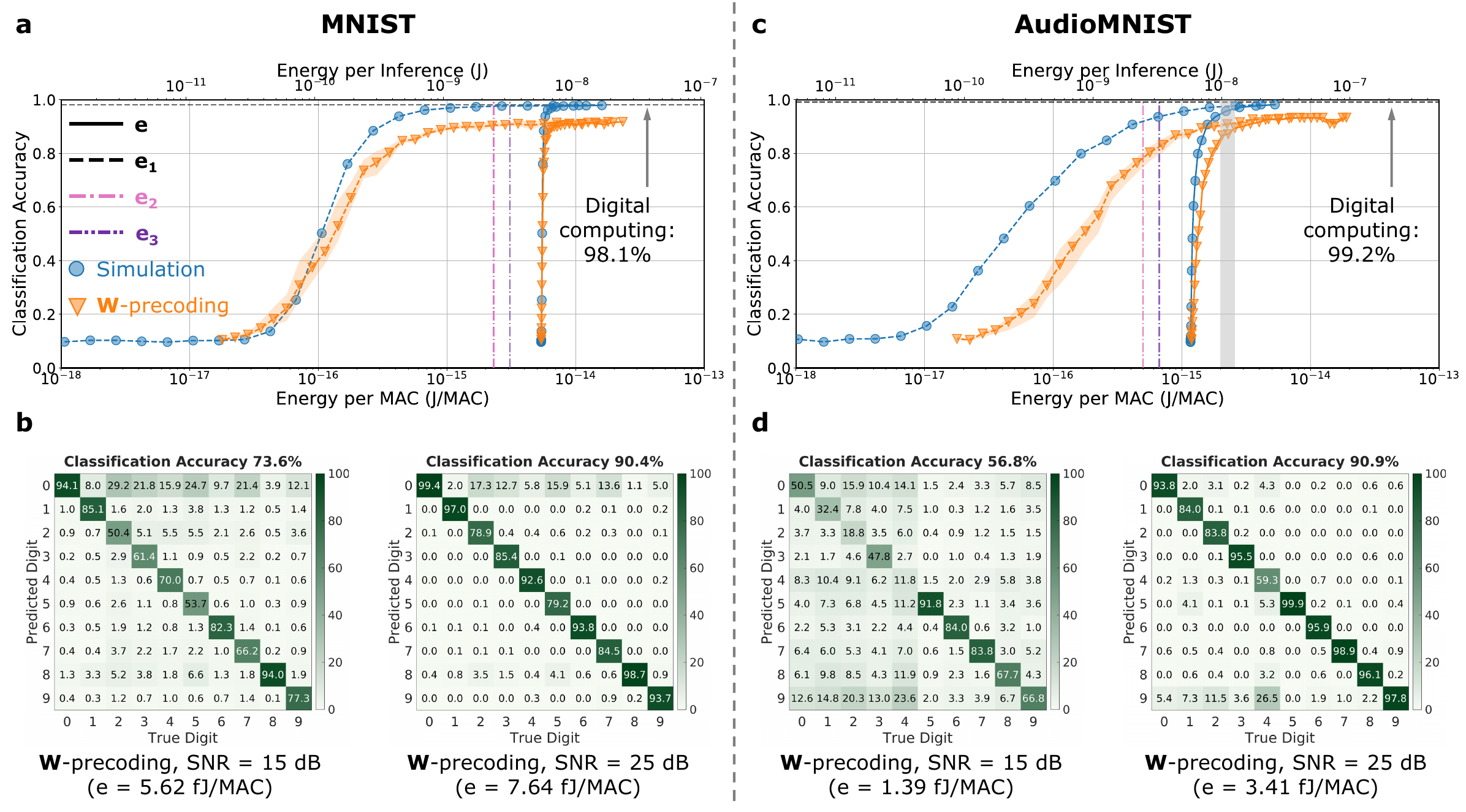}
    \caption{
    \textbf{The DL inference performance on the MNIST and AudioMNIST by {\namebf}'s $\weightMat$-precoding scheme with IP-based MVM decomposition ($\outputSizeNew=1$).}
    \textbf{a}, Classification accuracy on the MNIST and AudioMNIST datasets as a function of the energy efficiency with detailed breakdowns. The shaded area of $\energyMACTx$ indicates the accuracy variance across three clients under the same SNR.
    \textbf{b}, Confusion matrices of classification accuracy achieved by the $\weightMat$-precoding scheme at {15}\thinspace{dB} and {25}\thinspace{dB} SNR.}
    \label{fig:supplementary-inner}
\end{figure*}
%% figure ends

%% figure begins
\begin{figure*}[!t]
    \centering
    \includegraphics[width=0.95\columnwidth]{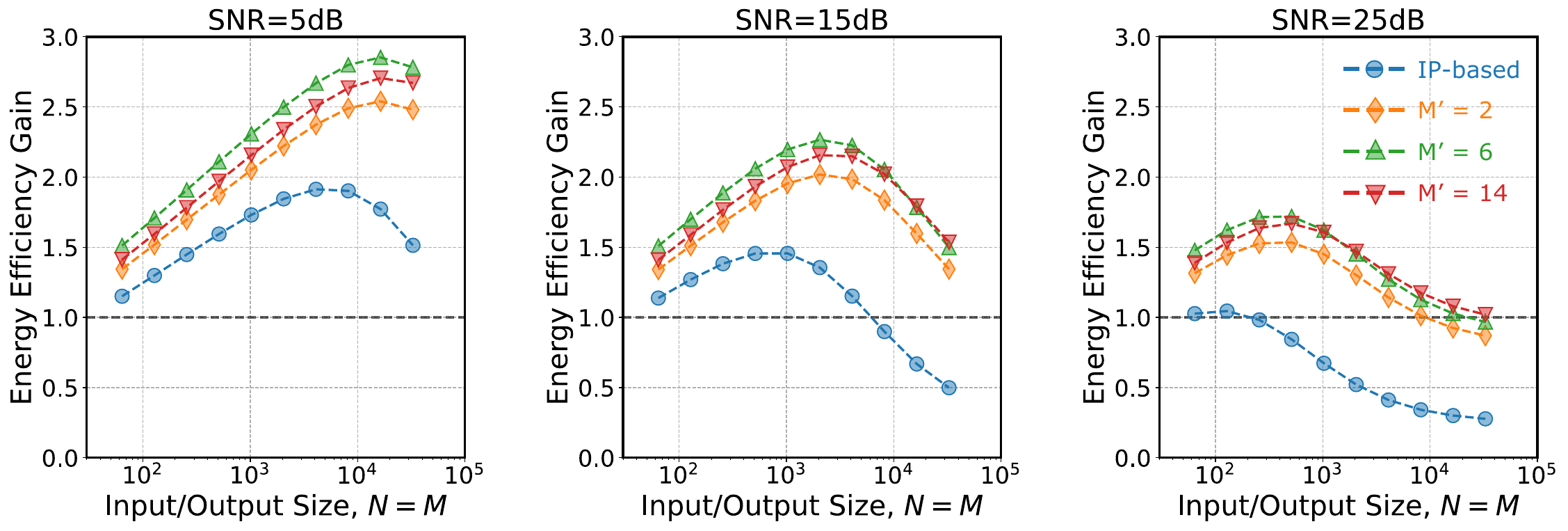}
    \caption{Energy efficiency gain achieved by {\name}'s $\weightMat$-precoding scheme across varying MVM decompositions, $\outputSizeNew = \{1, 2, 6, 14\}$, compared to the baseline without MVM decomposition. The zero padding overhead is set to $\outputSizePad=1$ ($\overheadPad=\frac{2}{\outputSizeNew}$, and the cyclic prefix overhead $\overheadCP$ is selected assuming a single I/Q sample $\fftSizeCP=1$ as the cyclic prefix, i.e., $\overheadCP=\frac{1}{\outputSizeNew+2}$.}
    \label{fig:supplementary-decomposition-selection}
\end{figure*}
%% figure ends

For the $\weightMat$-precoding scheme, we now consider the case where each MVM is decomposed into $\outputSize$ IPs, i.e., $\outputSizeNew=1$, with minimal FFT size at the cost of a slightly higher overhead of $\overheadPad=2$ and $\overheadCP=0.33$, as discussed in Supplementary Section~\ref{sec:supplementary-theory-weight-precoding-scheme}. This IP-based MVM decomposition has a lower computation throughput of $\throughput={75}\thinspace\textrm{MOPS}$ across three clients, and its energy consumption $\energyMACIPMult$ is given by equation {\eqref{eq: weight-precoding-scheme-energy-per-mac-multiple}}.
\autorefsubfig{fig:supplementary-inner}{a} shows the confusion matrices on the MNIST dataset under $\textsf{SNR}={15/25}\thinspace\textrm{dB}$, whose classification accuracies are {73.6\%/90.4\%} for MNIST, lower than the performance with $\outputSizeNew=6$ as {\name}'s default choice in {\autorefresults}.
This performance degradation comes from the potentially higher PAPR on $\waveInput(\waveIdx)$, $\waveWeight(\waveIdx)$, and $\waveOutput(\waveIdx)$, which incurs a relatively higher saturation level on the DACs/ADCs.
\autorefsubfig{fig:supplementary-inner}{b} shows the energy efficiency of this IP-based MVM decomposition. 
Specifically, to achieve a classification accuracy of {90\%} on MNIST, the energy efficiency is {$\energyMACIPMult={7.64}\thinspace\textrm{fJ/MAC}$ ({130.85}\thinspace{TOP/W})}, including the breakdown of $\energyMACTx={2.25}\thinspace\textrm{fJ/MAC}$ for an SNR of {25.1}\thinspace{dB}, $\energyMACADC={2.31}\thinspace\textrm{fJ/MAC}$, and $\energyMACDec={3.08}\thinspace\textrm{fJ/MAC}$.
The same experiments are repeated on the AudioMNIST.
\autorefsubfig{fig:supplementary-inner}{c} shows the classification accuracy of {56.8\%} and {90.9\%} under the SNR of {15}\thinspace{dB} and {25}\thinspace{dB}.
The energy efficiency analysis is further shown in \autorefsubfig{fig:supplementary-inner}{d}, where a minimum energy efficiency $\energyMACIPMult={3.41}\thinspace\textrm{fJ/MAC}$ ({293.17}\thinspace{TOPS/W}) is needed to achieve a classification accuracy of {90\%}. This energy efficiency corresponds to $\energyMACTx={2.25}\thinspace\textrm{fJ/MAC}$, $\energyMACADC={0.50}\thinspace\textrm{fJ/MAC}$, and $\energyMACDec={0.67}\thinspace\textrm{fJ/MAC}$.

We further investigate the optimal value of $\outputSizeNew$ for the MVM decomposition based on energy efficiency.
Specifically, the energy efficiency is simulated based on equations {\eqref{eq: weight-precoding-scheme-energy-per-mac}} and {\eqref{eq: weight-precoding-scheme-energy-per-mac-multiple}}, which is normalized by the energy efficiency without the MVM decomposition, as shown in \autoreffig{fig:supplementary-decomposition-selection}.
For the IP-based decomposition, despite the smallest energy for the term $\energyMACDec$, the overhead of $\overheadPad$ and $\overheadCP$ results in degraded energy efficiency on the term $\energyMACTx$, especially in the high SNR regime. As a result, for $\inputSize=4,096$, the energy efficiency gain is {1.92}$\times$, {1.16}$\times$, and {0.41}$\times$ to under an SNR value of {5}\thinspace{dB}, {15}\thinspace{dB} and {25}\thinspace{dB}, respectively.
We consider three levels of MVM decompositions with $\outputSizeNew=\{2, 6, 14\}$, which correspond to the FFT sizes of $\{4, 8, 16\}$ after attaching the padded zero-subcarriers with $\outputSizePad=1$.
Among these three decomposition levels, $\outputSizeNew=6$ achieves the highest energy efficiency gain of {2.70}$\times$, {2.22}$\times$, and {1.27}$\times$ for the three SNR levels for $\inputSize={4,096}$. This is due to a smaller overhead of $\overheadPad$ and $\overheadCP$ compared to $\outputSizeNew=2$, and a smaller FFT size compared to $\outputSizeNew=14$.
To conclude, we empirically select $\outputSizeNew=6$ for the MVM decomposition used by the $\weightMat$-precoding scheme, as described in {\autorefmethod}. Similar conclusions can also be extended to the basic scheme and the $\inputVec$-precoding scheme.

%%%%%
%%%%%
\section{A Case Study of {\namebf} on a Three-Layer DL Model}
\label{sec:supplementary-experiment-case-study-three-layer}

%% figure begins
\begin{figure*}[!t]
    \centering
    \includegraphics[width=0.95\columnwidth]{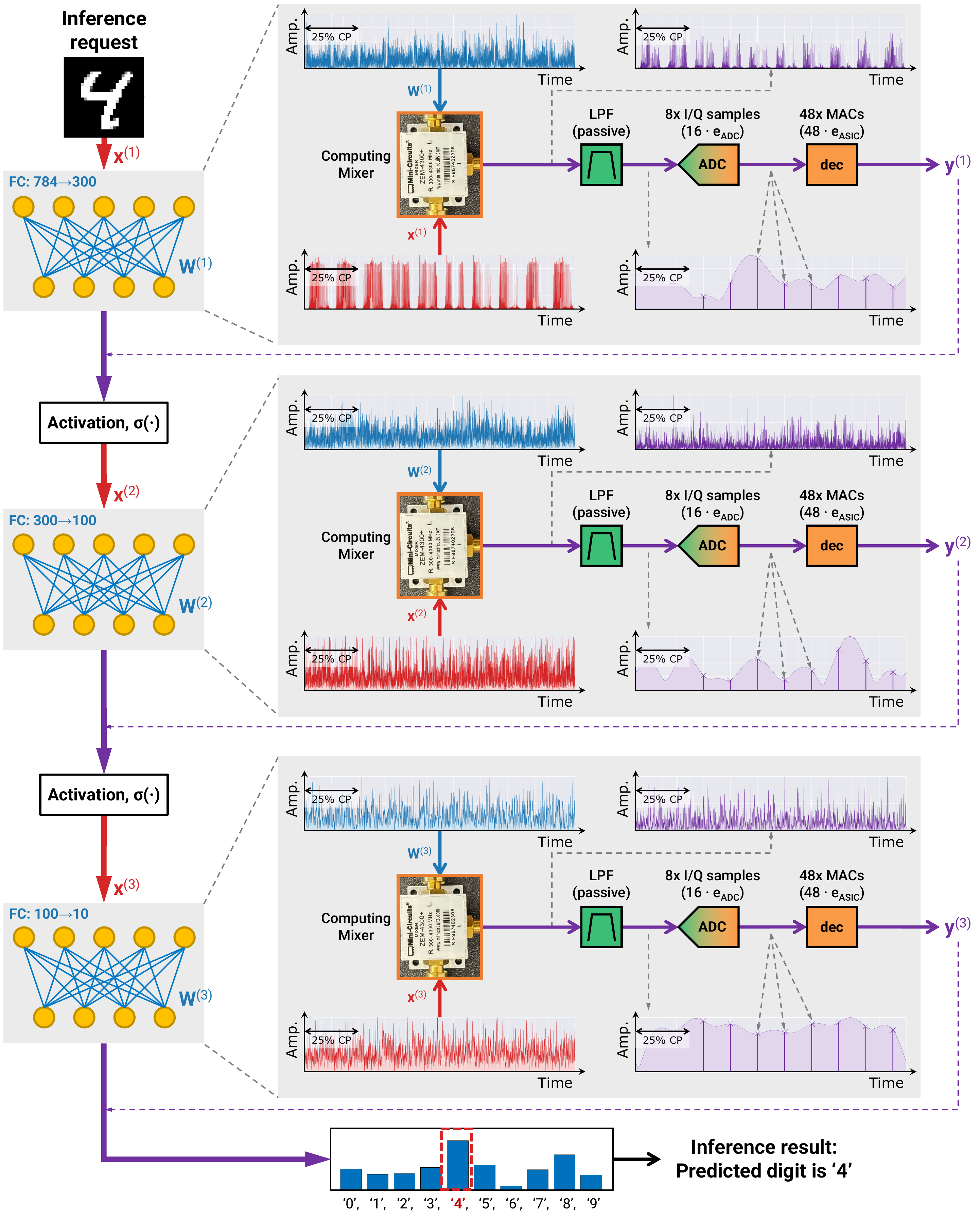}
    \caption{Example workflow of {\name}: A complex-valued model with three FC layers processes an inference request to predict the digit `4' from a handwritten image. For each FC layer, we show the amplitudes of the time domain waveform $\waveInput(\waveIdx)$, $\waveWeight(\waveIdx)$, $\waveOutput(\waveIdx)$ before/after the LPF, the sampled I/Q samples after ADC $\sampOutputVecLow$, of a single decomposed MVM.}
    \label{fig:dataset-waveform}
\end{figure*}
%% figure ends

We show a detailed workflow of how {\name} performs inference on an image of a handwritten digit `4' in MNIST on a 3-FC layer DL model, shown in \autoreffig{fig:dataset-waveform}.
In particular, the {28}$\times${28}-pixel image of digit `4' is formed as a {28}$\times${28} real-valued matrix, and is then flattened into a {784}-element vector. This vector is then modulated with a {784}-point Zadoff-Chu (ZC) phase sequence $\zadoffVec$ as defined in {\autorefmethod}, which yields a {784}-element complex-valued vector, $\inputVec^{(1)} \in \mathbb{C}^{784}$, as the input to the first FC layer.

The first FC layer has an input size of $\inputSize^{(1)}=784$ and an output size of $\outputSize^{(1)}=300$. The MVM decomposition technique described in Supplementary Section~\ref{sec:supplementary-theory-basic-scheme} first decomposes the entire MVM with $\weightMat^{(1)} \in \mathbb{C}^{300 \times 784}$ into 50 smaller MVMs, each of which has a smaller output size of $\outputSizeNew={6}$. We employ a zero-subcarrier padding overhead of $\overheadPad=0.33$ with $\outputSizePad=1$, which extends the output dimension per decomposed MVM to $(1+\overheadPad)\outputSizeNew = 8$.
For each decomposed MVM, the time-encoded input sequence $\sampInputVec$ contains eight duplicated $\inputVec^{(1)}$ in the time domain, with a total number of 6,272 I/Q samples.
Finally, we employ a cyclic prefix overhead coefficient of $\overheadCP=0.25$, which further appends two copies of $\inputVec^{(1)}$ to the front of $\sampInputVec$. The resulting I/Q waveform streamed to the DACs has 7,840 I/Q samples. With a DAC sampling rate of $\sampRate={25}\thinspace\textrm{MHz}$, the generated waveform $\waveInput(\waveIdx)$ has a duration of {0.314}\thinspace{ms}. The waveform $\waveInput(\waveIdx)$ is then I/Q modulated to the carrier frequency of $\carrierInput = {1.2}\thinspace\textrm{GHz}$.
Similarly, at the central radio, the model weights for each decomposed MVM are encoded into I/Q waveform $\waveWeight(\waveIdx)$, which is then I/Q modulated to the carrier frequency of $\carrierWeight = {0.915}\thinspace\textrm{GHz}$. The model weights are then broadcast wirelessly to the client.

On the client side, $\waveInput(\waveIdx)$ is mixed with the received waveform $\waveWeight(\waveIdx)$ and filtered by the LPF with a cutoff frequency of $\freq_0={15.9}\thinspace\textrm{kHz}$, the output waveform $\LPF{\waveOutput(\waveIdx)}$ is sampled by two I/Q ADCs operating at {31.9}\thinspace{kHz} to obtain $\sampOutputVecLow \in \mathbb{C}^{8}$, where the first {25}\% waveform is excluded as the cyclic prefix.
Then, an 8-point FFT is performed on $\sampOutputVecLow$ via digital computing, which yields the subcarrier symbols $\specOutputVecLow \in \mathbb{C}^{8}$ of eight complex-valued symbols. 
Finally, the middle six symbols in $\specOutputVecLow$ are considered as the output $\outputVecNew \in \mathbb{C}^{6}$ of the decomposed MVM; concatenating $\outputVec$ from all $\outputSize/\outputSizeNew=50$ decomposed MVMs yields the final output of the first FC layer, $\outputVec^{(1)} \in \mathbb{C}^{300}$.
This output is passed through the activation function, $\activation_{300}(\cdot)$, including the absolute function and phase modulation with a {300}-point Zadoff-Chu sequence $\zadoffVec$, which generates the input to the second FC layer, $\inputVec^{(2)}$.

This process is repeated three times, one for each layer, to obtain the output of the last FC layer, $\outputVec^{(3)} \in \mathbb{C}^{10}$. The amplitude of the final output, $\abs{\outputVec^{(3)}}$ represents the probability of the input image being one of the ten digits `0' to `9'. In this example, the $5^{\textrm{th}}$ element has the highest amplitude, corresponding to the classification result of digit `4'.

%%%%%
%%%%%
\section{A Fully Analog Linear Regression Model}
\label{sec:supplementary-experiment-ml-regression}

%% figure begins
\begin{figure*}[!t]
    \centering
    \includegraphics[width=0.95\columnwidth]{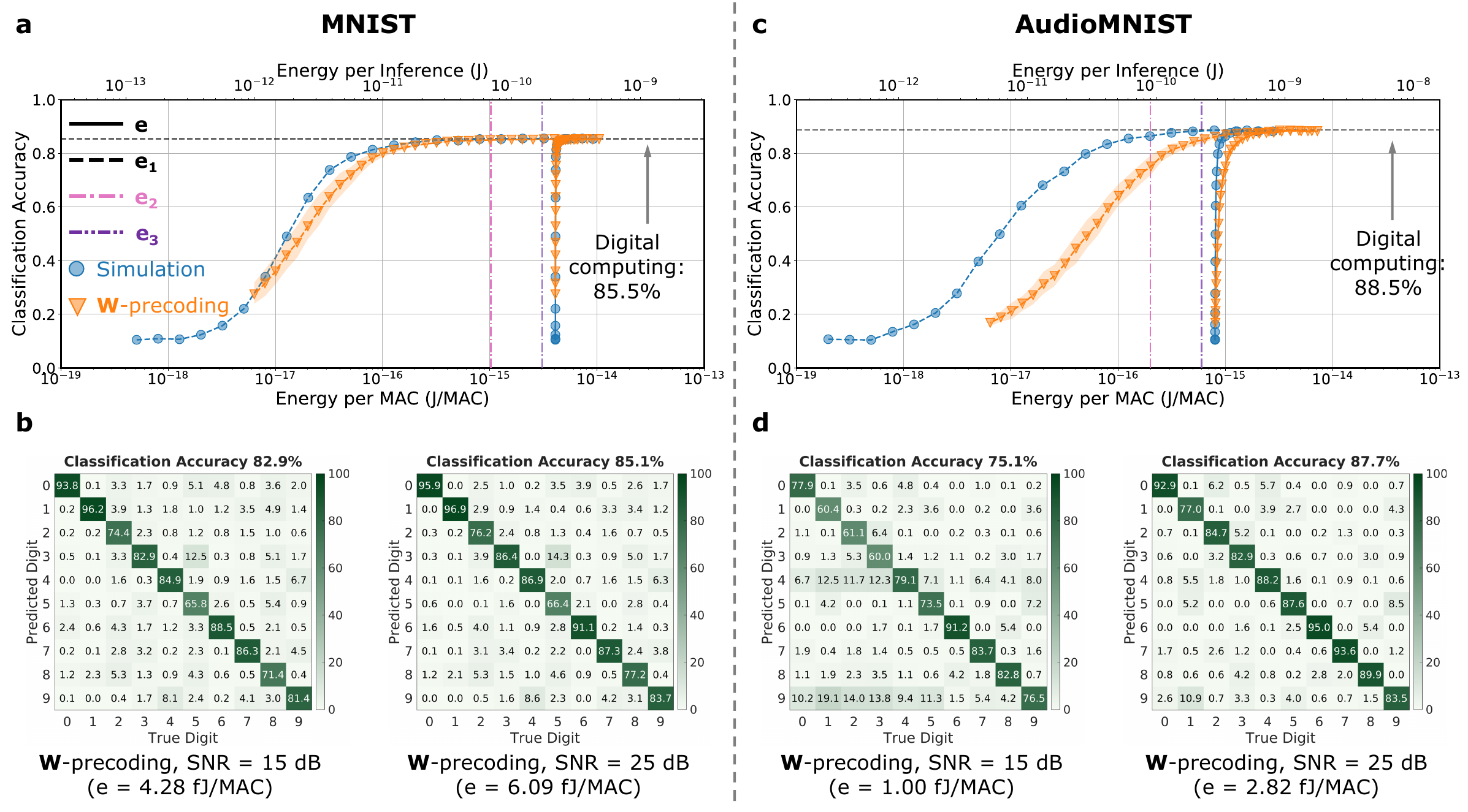}
    \caption{
    \textbf{The DL inference performance on the MNIST and AudioMNIST by a fully analog linear regression model, which skips the digital computing-based activation functions in the middle layers, and the $\weightMat$-precoding scheme is applied}
    \textbf{a}, The classification accuracy on MNIST under different energy efficiency $\energyMAC$ of the fully analog linear regression model. The shadow area indicates the accuracy variance over the three users. 
    \textbf{b}, Under {15/25}\thinspace{dB} SNR, the confusion matrices on the MNIST dataset by fully analog linear regression model.
    \textbf{c--d}, The energy efficiency analysis and the confusion matrices on the AudioMNIST dataset, respectively.}
    \label{fig:supplementary-model}
\end{figure*}
%% figure ends

We also consider a small linear regression model~\cite{montgomery2021introduction} with a single complex-valued FC layer.
For the MNIST and AudioMNIST datasets, only one FC layer of {784/4,000}$\times${10} complex-valued parameters transfers the {784/4,000}-element input $\inputVec$ into the {10}-element output $\outputVec$ for the likelihood of the input being one of the ten digits. Since there is only one FC layer, no nonlinear activation function with absolute function and Zadoff-Chu phase sequence is applied. The absolute function after the FC layer can be realized by directly measuring the absolute power of each subcarrier in $\specOutputVecLow$.
Therefore, this linear regression model only requires a single transmission without digitally performing absolute functions. On the other hand, the one-time-FFT-based decoding is still required to extract $\outputVec$ from the time-domain waveform $\waveOutput(\waveIdx)$, which can be done either in digital computing or internally when being received by a spectrum analyzer.
Similar to the LeNet-300-100 models, we train the linear regression model using the Adam optimizer~\cite{kingma2014adam} with a learning rate of $1.0 \times 10^{-3}$ over 100 epochs and cross-entropy as the loss function.

Using digital computing, this linear regression model achieves a classification accuracy of {85.5\%} on the MNIST dataset.
\autorefsubfig{fig:supplementary-model}{a} shows the energy efficiency of the linear regression model on the MNIST dataset.
Specifically, this linear regression model achieves a classification accuracy of {80\%} at energy efficiency of $\energyMAC={4.18}\thinspace\textrm{fJ/MAC}$ ({239.23\thinspace{TOP/W}}), with a breakdown of $\energyMACTx={0.10}\thinspace\textrm{fJ/MAC}$, $\energyMACADC={1.02}\thinspace\textrm{fJ/MAC}$, and $\energyMACDec={3.06}\thinspace\textrm{fJ/MAC}$.
This linear regression model only involves {31,360} MACs, which corresponds to the total energy consumption of {131.08}\thinspace{pJ} per inference.
Compared to the LeNet-300-100 model at a classification accuracy of {80\%}, this linear regression model consumes {36.4}$\times$ less energy per inference.
As shown in \autorefsubfig{fig:supplementary-model}{b}, {\name} achieves a classification accuracy of {82.9\%} and {85.1\%} under {15}\thinspace{dB} and {25}\thinspace{dB}, respectively. The accuracy gap between digital computing and the in-physics computing of {\name} of is only {0.4\%}, which is smaller compared to that of the LeNet-300-100 model described in {\autorefresults}. This is due to the shallow model architecture, where errors introduced during the in-physics computing process do not accumulate across layers.

On the AudioMNIST dataset, the classification accuracy of this linear regression model with full-precision digital computing is {88.5\%}.
\autorefsubfig{fig:supplementary-model}{c} shows the energy efficiency of the in-physics computing by {\name} and the corresponding classification accuracy.
To achieve {80\%} classification accuracy, the energy efficiency of {\name} is $\energyMAC={1.12}\thinspace\textrm{fJ/MAC}$ ({892.86\thinspace{TOPS/W}}), which includes $\energyMACTx={0.32}\thinspace\textrm{fJ/MAC}$, $\energyMACADC={0.22}\thinspace\textrm{fJ/MAC}$, and $\energyMACDec={0.67}\thinspace\textrm{fJ/MAC}$.
Given the total number of {160,000} MACs involved in the model, this energy efficiency corresponds to an energy consumption of {179.35}\thinspace{pJ/MAC} per inference, which is only {29.8}$\times$ lower compared to that of the LeNet-300-100 model at the same classification accuracy.
Moreover, as shown in \autorefsubfig{fig:supplementary-model}{d}, the linear regression model achieves a classification accuracy of {75.1\%} and {87.7\%} at {15}\thinspace{dB} and {25}\thinspace{dB} SNR.

\section{{\namebf} over Wired Channels}
\label{sec:supplementary-experiment-ml-wired}

%% figure begins
\begin{figure*}[!t]
    \centering
    \includegraphics[width=0.95\columnwidth]{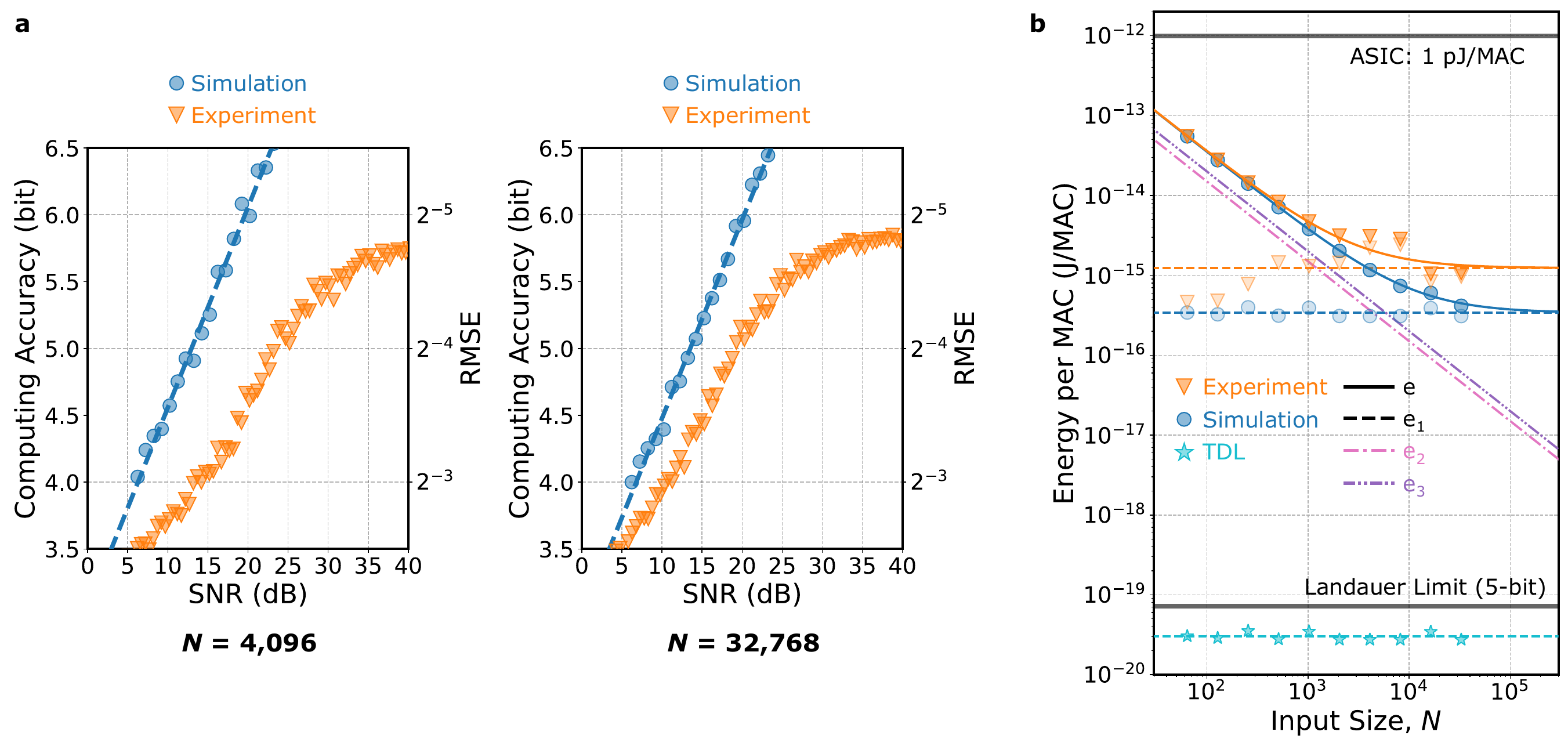}
    \caption{
    \textbf{General MVM computing accuracy achieved by {\namebf}'s basic scheme compared to simulations over a {100}\thinspace{MHz} wired channel.}
    \textbf{a}, Computing accuracy across varying SNR levels with input sizes $\inputSize=\{4096, 32768\}$.
    \textbf{b}, Energy efficiency required to achieve $\textsf{RMSE} < 0.0625$ (5-bit computing accuracy) across varying input sizes, $\inputSize$, with its breakdown.}
    \label{fig:supplementary-wired-inner}
\end{figure*}
%% figure ends

To comply with the ISM band regulations, our wireless experiments are conducted using a bandwidth of {25}\thinspace{MHz}, which limits the computation throughput $\throughput$.
On the other hand, a larger bandwidth $\band$ can proportionally increase $\throughput$ without increasing the energy consumption $\energyMAC$.
In this section, we consider a wired channel as a substitute for the wireless transmission of $\waveWeight(\waveIdx)$, where the central radio's TX port is connected to the computing mixer's LO port using an SMA cable, as shown in \autorefsubfig{fig:supp-setup}{b}.
Such a wired setting ensures the flat channel response so that the precoding process can be skipped, while the time-encoded $\inputVec$ to waive the encoding energy can still be applied. On the other hand, only a single client is supported at a time.
In the wired experiment, we employ a bandwidth of as {100}\thinspace{MHz} for $\waveWeight(\waveIdx)$ and $\waveInput(\waveIdx)$, which is the maximum sampling rate at which the USRP X310 can maintain stable data streaming. 
According to equation {\eqref{eq: computation-throughput-throughput}}, the per-client computation throughtput is increased from {60}\thinspace{MOPS} to {240}\thinspace{MOPS}, due to the {4}$\times$ increase in the signal bandwidth.

%% figure begins
\begin{figure*}[!t]
    \centering
    \includegraphics[width=0.95\columnwidth]{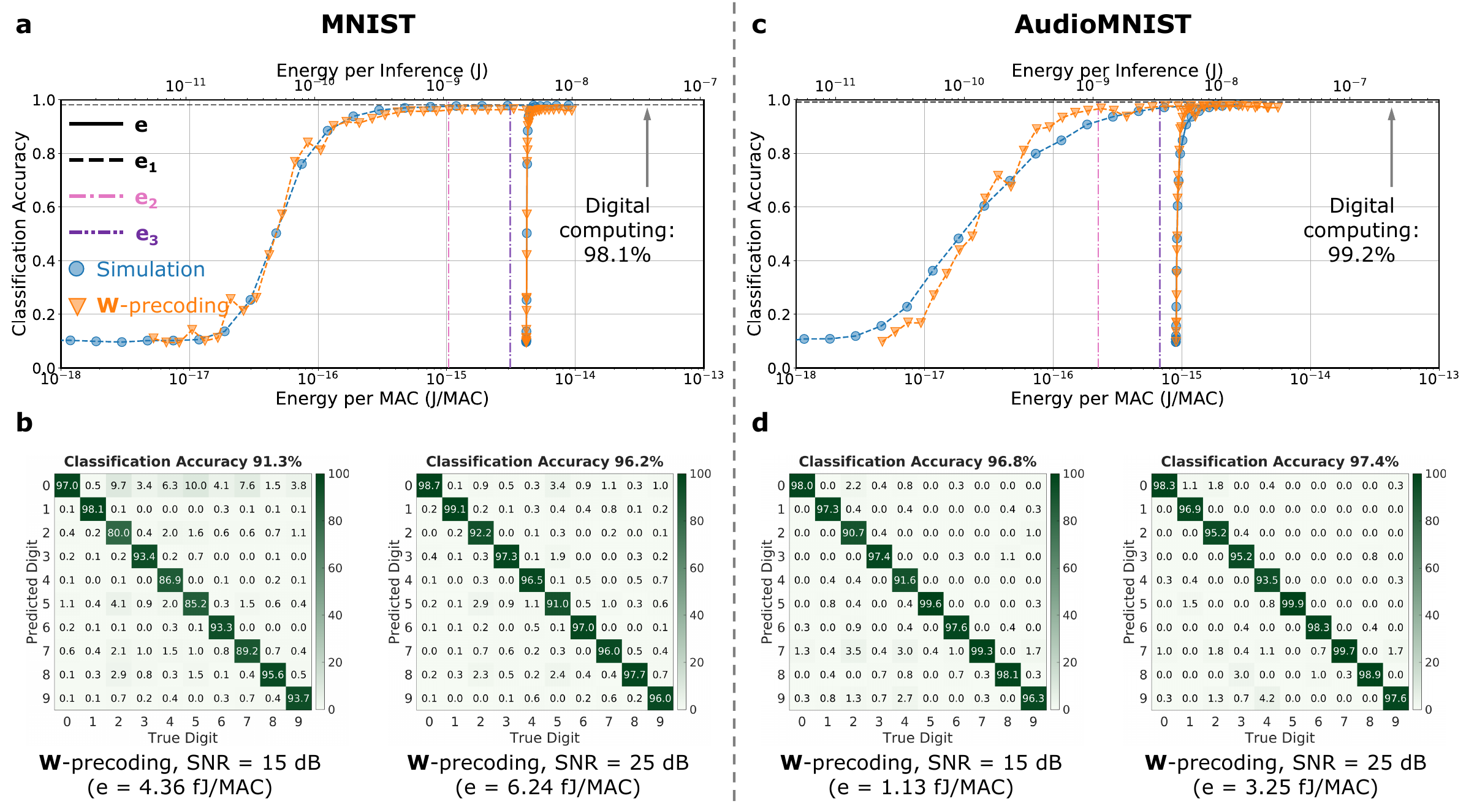}
    \caption{
    \textbf{The DL-inference performance on the MNIST and AudioMNIST by the {\namebf}'s basic scheme over a {100}\thinspace{MHz} wired channel.}
    \textbf{a}, The energy efficiency analysis on the MNIST dataset. 
    \textbf{b}, The confusion matrices on the MNIST dataset of the wired channel. 
    \textbf{c--d}, The energy efficiency analysis and the confusion matrices on the AudioMNIST dataset, respectively.}
    \label{fig:supplementary-wired}
\end{figure*}
%% figure ends

Under the {100}\thinspace{MHz} wired channel, we first showcase the performance of general MVM computation under the same settings as described in Supplementary Section~\ref{sec:supplementary-experiment-ml-calibration}. As shown in \autorefsubfig{fig:supplementary-wired-inner}{a}, this wired channel enables slightly higher computing accuracy than the wireless {\name} under higher SNRs. For example, under {30}\thinspace{dB} SNR, the basic scheme without channel calibration achieves an RMSE of {0.045} and {0.038} with $\inputSize=4,096$ and $32,768$.
In addition, \autorefsubfig{fig:supplementary-wired-inner}{b} shows the minimum energy efficiency required to achieve $\textsf{RMSE}<0.0625$, which is similar to the $\weightMat$-precoding scheme presented in {\autorefresults}.
For example, given $\inputSize=4,096$ and $\inputSize=32,768$, the energy efficiency required for {\name} is {3.07}\thinspace{fJ/MAC} and {1.07}\thinspace{fJ/MAC} ({325.73}\thinspace{TOPS/W} and {934.58}\thinspace{TOPS/W}), respectively.

On the MNIST dataset, \autorefsubfig{fig:supplementary-wired}{a} presents the energy efficiency achieved by {\name} in the wired setup. Note that the flat channel response of the wired setup reduces the gap between the simulation and experimental results. To achieve {90\%} classification accuracy on the MNIST dataset, the energy efficiency of {\name} is $\energyMAC={4.28}\thinspace\textrm{fJ/MAC}$ ({233.64}\thinspace{TOPS/W}), with a breakdown of $\energyMACTx={0.13}\thinspace\textrm{fJ/MAC}$, $\energyMACADC={1.04}\thinspace\textrm{fJ/MAC}$, and $\energyMACDec={3.11}\thinspace\textrm{fJ/MAC}$.
As shown in \autorefsubfig{fig:supplementary-wired}{b}, at SNR values of {15}\thinspace{dB} and {25}\thinspace{dB}, the experimental classification accuracy on MNIST is {91.3\%} and {96.2\%}, respectively, which closely matches the simulation results of {93.8\%} and {97.7\%}.
As for the AudioMNIST dataset, \autorefsubfig{fig:supplementary-wired}{c} shows that to achieve {90\%} classification accuracy, the energy efficiency of {\name} is $\energyMAC={1.01}\thinspace\textrm{fJ/MAC}$ ({985.61}\thinspace{TOPS/W}), with a breakdown of $\energyMACTx={0.12}\thinspace\textrm{fJ/MAC}$, $\energyMACADC={0.22}\thinspace\textrm{fJ/MAC}$, and $\energyMACDec={0.67}\thinspace\textrm{fJ/MAC}$.
Moreover, in \autorefsubfig{fig:supplementary-wired}{d}, the experimental classification accuracies are {96.8\%} and {97.4\%} under {15}\thinspace{dB} and {25}\thinspace{dB} SNR, respectively, compared to simulation results of {90.8\%} and {98.2\%}.
These results showcase the promising performance of {\name}'s basic scheme in the wired setup, demonstrating its scalability toward higher computation throughput.